\documentclass[12pt,a4paper]{article}

\usepackage{textcomp}
\usepackage[latin1]{inputenc}
\usepackage{pstricks,pst-node,pst-text}
\usepackage{amsmath}
\usepackage{graphicx}
\usepackage{axodraw}

\newcommand{\be}{\begin{equation}}
\newcommand{\ee}{\end{equation}}
\newcommand{\bea}{\begin{eqnarray}}
\newcommand{\eea}{\end{eqnarray}}
\newcommand{\beqn}{\begin{eqnarray}}
\newcommand{\eeqn}{\end{eqnarray}}
\newcommand{\ba}{\begin{array}}
\newcommand{\ea}{\end{array}}

\newcommand{\ben}{\begin{enumerate}}
\newcommand{\een}{\end{enumerate}}

\newcommand{\crn}{\nonumber \\}

\newcommand{\noi}{\noindent}

\newcommand{\la}{\lambda}

\newcommand{\eps}{\varepsilon}

\newcommand{\RE}{\operatorname{Re}}

\newcommand{\Trace}{\operatorname{Tr}}

\newcommand{\fr}{\frac}
\newcommand{\bc}{\begin{center}}
\newcommand{\ec}{\end{center}}

\newcommand{\ra}{\rightarrow}

\newcommand{\beq}{\begin{equation}}
\newcommand{\eeq}{\end{equation}}

\newcommand{\pr}{Phys.\,Rev.\,}

\newcommand{\zp}{Z.\,Phys.\,}

\newcommand{\hs}{\hspace*{3mm}}

\newcommand{\Img}{\operatorname{Im}}
\newcommand{\GeV}{\textrm{GeV}}
\newcommand{\sign}{\operatorname{sign}}

\def\slashc{c\kern -.400em {/}}
\def\slashp{p\kern -.400em {/}}
\def\slashq{q\kern -.450em {/}}
\def\slashL{L\kern -.450em {/}}
\def\slashcl{\cl\kern -.600em {/}}
\def\slashr{r\kern -.450em {/}}
\def\slashk{k\kern -.500em {/}}
\def\slashep{\epsilon\kern -.450em {/}}
\def\slashpbar{\bar{p}\kern -.450em {/}}
\def\slashepi{\epsilon_i\kern -.720em {/}}
\def\slashpi{p_i\kern -.600em {/}}

\begin{document}

\begin{titlepage}

\vspace*{0.1cm}\rightline{LAPTH-1256/08}

\vspace{1mm}
\begin{center}

{\Large{\bf $b\bar{b}H$ production at the LHC: Yukawa
corrections and the leading Landau singularity}}

\vspace{.5cm}

F.~Boudjema and LE Duc Ninh

\vspace{4mm}

{\it LAPTH, Universit\'e de Savoie, CNRS,  \\
BP 110, F-74941
Annecy-le-Vieux Cedex, France}

\vspace{10mm} \abstract{At tree-level Higgs production in
association with a $b$-quark pair proceeds through the small
Yukawa bottom coupling in the Standard Model. Even in the limit
where this coupling vanishes, electroweak one-loop effects,
through the top-Higgs Yukawa coupling in particular, can still trigger
this reaction. This contribution is small for Higgs masses around
$120$GeV but it quickly picks up for higher Higgs masses
especially because the one-loop amplitude develops a leading
Landau singularity and new thresholds open up. These effects  can
be viewed as the production of a pair of top quarks which
rescatter to give rise to Higgs production through $WW$ fusion. We
study the leading Landau singularity in detail. Since this
singularity is not integrable when the one-loop amplitude is
squared, we regulate the cross section by taking into account the
width of the internal top and $W$ particles. This requires that we
extend the usual box one-loop function to the case of complex masses.
We show how this can be implemented analytically in our
case. We study in some detail the cross section at the LHC as a
function of the Higgs mass and show how some distributions can be
drastically affected compared to the tree-level result.}

\end{center}


\normalsize

\end{titlepage}

\section{Introduction}
The LHC will soon start running and collecting data. Although one
expects surprises, discovering the Higgs is the highest priority.
A lot of effort has gone in calculating the rate of production of
this particle, within the Standard Model and beyond,  for a host
of channels and signatures, see \cite{djouadi_H1,Higgs_LH07}   for
a review.

Higgs production in association with a pair of bottom quarks is
not, especially in the Standard Model, a discovery channel since
the coupling of the Higgs to the bottom quark is given by the
small, ${\cal O}(m_b/v)$ bottom-Higgs Yukawa coupling, where $m_b$
is the bottom mass and $v\sim 246$GeV the scale of electroweak
symmetry breaking. Nonetheless given the special role that can
play the third generation of fermions in the mechanism of symmetry
breaking and in particular  the top-bottom quark doublet, a
reconstruction of this Higgs coupling to bottom quarks is
important. This reconstruction and interpretation of the
measurements requires theoretical predictions that go beyond the
tree-level approximation. \noi Many of these calculations, most of
which concern the important QCD corrections, have already been
performed\cite{bbh_LHC_all}. Usually one expects the electroweak
corrections to be small and not compete with the QCD corrections.
However, one should bear in mind that the top Yukawa coupling
${\cal O}(m_t/v)$ is of order the strong coupling constant. If
this coupling takes part in the electroweak corrections the latter
may not necessarily be small. Other Yukawa couplings that are not
negligible are the Higgs Yukawa coupling\footnote{As this paper is
on the Yukawa corrections neglecting corrections of order the
electroweak gauge coupling , we use the terminology {\em Higgs
Yukawa coupling} for the Higgs self-coupling which in the Standard
Model is not a gauge coupling.}. Both these couplings are involved
when one considers the electroweak corrections to $b \bar b H$
production at the LHC. Another important property of the
electroweak effects is that this cross section can be triggered
off by one-loop corrections involving the top quark and $W$ gauge
boson (or Goldstone) loops even for vanishing $b \bar b H$ (or
$m_b=0$) coupling, where the Born cross section vanishes.

We\cite{fawzi_bbH} have, very recently, studied the effects of
the leading (Yukawa-type) electroweak corrections for $b \bar b H$
production at the LHC in a situation where both $b$'s are tagged,
requiring somewhat large $p_T$ $b$, as would be relevant for a
measurement of the  $b \bar b H$ couplings and a complete
identification of this channel. The study we performed
concentrated on a Higgs with a mass below $150$GeV not only
because this range is preferred by the precision electroweak data
but also because the cross section decreases much with increasing
Higgs mass.   It was found that, after all, the Next-to-Leading
Order (NLO) corrections were small and could be safely neglected.
In the limit where the  $b \bar b H$ coupling vanishes and where
the cross section is induced solely through electroweak loops, we
found that this effect is much larger than the NLO correction and
increases rapidly with the Higgs mass. We pointed out that, for
this contribution, as $M_H \ge 2 M_W$ our perturbative calculation
becomes unreliable since the loop integrals start showing
numerical instabilities. We had identified this behaviour as a
leading Landau singularity (LLS)\cite{Landau,book_eden} which is a
pinch singularity of the loop integral. This, in part, has an
interesting physical origin: the on-shell production and
rescattering of  the top quarks into on-shell $W$ bosons, the
latter giving rise to Higgs production through $WW$ boson fusion.
This LLS of the one-loop four-point function is not integrable
when one considers the square of the loop amplitude as needed for
vanishing $b \bar b H$ coupling. The NLO contribution, on the
other hand, is integrable.

The aim of this paper is to extend the study we made in
\cite{fawzi_bbH} to higher Higgs masses. The emphasis will be on
the LLS problem and the pure  one-loop contribution in the limit
of vanishing $m_b$ since this is the major hurdle. For
completeness we will also give results for the NLO contribution
for Higgs masses not covered in our previous calculation. Beyond
the phenomenological impact of the LLS for the case at hand, the
study of the LLS in this paper should be of interest for other
situations considering that one rarely encounters such
singularities, as compared to the inverse (vanishing) Gram
determinant which is not a genuine physical singularity but an
artifact of the reduction of the tensorial integrals. {Some
of the few examples in the relatively recent literature where some
aspect of a Landau singularity shows up include $ZZ \ra
ZZ$\cite{denner_4z} and the $6$-photon amplitude\cite{6_photons}
in the Standard Model both with massless particles in the internal
states involving a four-point function. Beyond the Standard Model
we can mention loop corrections to sfermion pair production in
supersymmetry\cite{freitas} and Higgs production from the decay of
a fourth generation b-like quark\cite{stuart}, both these examples
involve heavy instable particles in a three-point function.
In\cite{freitas} no special treatment of the singularity is
required since the study is made at the NLO level where this
singularity is integrable. In\cite{stuart} the width of the
internal unstable particle is called for. In $ZZ \ra ZZ$, the
study\cite{denner_4z} keeps away from the region of the LLS, while
it is argued that the LLS should disappear if one considers a more
inclusive cross section where the $Z$ boson would decay or the
initial $Z$ are grafted to light stable fermions. For the case of the
$6$-photon amplitude the situation is quite subtle. The QED
dynamics is such that the LLS disappears at the level of the total
gauge invariant amplitude after summing on individual diagrams.}
The LLS issue can also be relevant for the nascent {\em cut
techniques} of computing loop amplitudes, for a recent review
see\cite{Bern:2008ef}.   This is the reason we devote a fair part
of this paper to the study and solution of the LLS. Our solution
to the problem of the LLS for Higgs production through $gg \ra b
\bar b H$ is to endow the resonating internal particles, namely
the top quark and $W$ gauge boson with a width.
The extension of the usual loop libraries, such as {\tt
FF}\cite{FF} of {\tt LoopTools}\cite{looptools}, to the case of
complex masses is not trivial especially if one insists on an
analytical implementation. We will show how the case at hand lends
itself to a fairly manageable implementation of complex masses for
the four-point function which is computer-time effective. The
introduction of the widths avoids all numerical instabilities and
smooths out the cross section when we enter the phase space region
of the LLS. It rests that this effect can still give large
corrections particularly for some specific distributions, like for
example the
$p_T$ distribution of the bottom quark or the Higgs boson.\\

The plan of the paper is as follows.  In the next section we set
the framework for our calculation with a reminder on the $SU(3)$
(QCD) gauge invariant classes of the electroweak contributions and
the helicity properties of the amplitudes. We then briefly uncover
the class and type of diagrams that contain a potential leading
Landau singularity. Section~3 follows with a general discussion on
the Landau singularities first exposing the conditions under which
such singularities can show up for the scalar $N$-point function.
We then carefully extract the nature of the singularity before
moving into a detailed investigation of the scalar $4$-point
function at the origin of the LLS in our case, for $gg \ra b \bar
b H$. Section~4 discusses how this singularity can be regulated
through taking into account the width of the unstable particles
running in the loop. Section~5 describes how these widths are
implemented through a modification of the loop integrals that
should be defined for complex masses of the loop particles. In
particular we describe our analytical implementation of the
complex masses suitable for our problem. We will also discuss the
various checks we made to insure the correctness of the
implementation. Section~6 gives briefly our input parameters and
cuts and describe how the cross section at the $pp$ level is
obtained. Section~7 gives our main results for the cross section
$pp \ra b \bar b H$ at the LHC in the limit of vanishing Higgs
coupling to $b$-quarks. In this case the cross section is induced
at one-loop and we need, in particular, to integrate the square of
the $4$-point loop integral over the kinematical phase space. This
calls for our new implementation of the box one-loop functions. We
will discuss the behaviour of the cross section as a function of
the Higgs mass and study a few distributions. Section~8 turns to
the NLO result for $M_H>150$GeV, completing therefore the study we
made in \cite{fawzi_bbH}. Section~9 summarises our findings. The
paper contains also three appendices. In the first we give the
details of our derivation of the nature of the singularity while
the second appendix gives technical details about the handling of
complex masses in one-loop scalar box functions. The third
appendix details the singularities of the 3-point function. Many
key issues about the LLS are unravelled in this case which help in
better understanding the issues in the 4-point function.
\section{A quick reminder and general considerations of the one-loop electroweak structure}
\label{section_general}

At LHC energies the exclusive $b \bar b H$ production with both
$b$-quarks tagged is dominated, by far, by the gluon gluon
initiated subprocess. We therefore only consider, as we have done
in \cite{fawzi_bbH}, the gluon-gluon initiated subprocess
$g(p_1,\la_1)+g(p_2,\la_2)\rightarrow
b(p_3,\la_3)+\bar{b}(p_4,\la_4)+H(p_5)$. $\la_i=\pm$ and $p_i$
with $i=1,2,3,4$ stand for the helicity  for the momentum of the
particle. The corresponding helicity amplitude will be denoted as
${\cal A}(\la_1,\la_2;\la_3,\la_4)$.
\begin{figure}[h]
\begin{center}
\includegraphics[width=12cm]{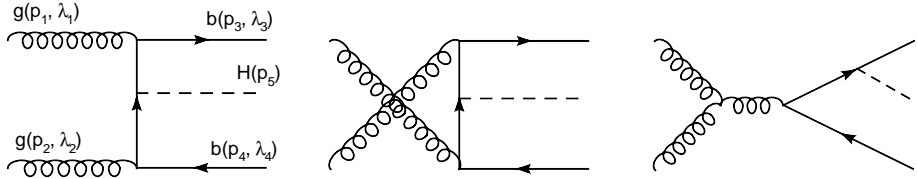}
\caption{\label{diag_gg_LO}{\em All the eight Feynman diagrams can
be obtained by inserting the Higgs line to all possible positions
in the bottom line.}}
\end{center}
\end{figure}

At tree-level the process is given by Higgs radiation off the
$b$-quark line, see Fig.~\ref{diag_gg_LO}. The tree-level
amplitude, ${\cal A}_0(\la_1,\la_2;\la_3,\la_4)$,  is therefore
proportional to $\la_{bbH}$ the Higgs coupling to $b$.  As has
been done in previous
analyses~\cite{dawson_bbH,LH03_bbh,fawzi_bbH}, for the exclusive
$b\bar{b}H$ final state, we will require the outgoing $b$ and
$\bar{b}$ to have  transverse momenta
$|\textbf{p}_{T}^{b,\bar{b}}|\ge 20$GeV and pseudo-rapidity
$|\eta^{b,\bar{b}}|<2.5$. These kinematical cuts reduce the total
rate of the signal but also greatly reduce the QCD background. As
pointed in~\cite{dittmaier_bbH} these cuts also stabilise the
scale dependence of the QCD NLO corrections compared to the case
where no cut is applied. In the approximation of neglecting the bottom mass the whole
contribution vanishes, since the Higgs coupling to $b$ vanishes.
The massless bottom limit can also be taken, but by keeping
$\la_{bbH}$ as an independent parameter with a non zero value. In
this limit the tree-level contribution consists of only the
amplitude ${\cal A}_0(\la_1,\la_2;\la,-\la)$\footnote{The helicity
amplitude method and the convention we use in this paper for the
definition of the helicity state are based on \cite{fawzi_bbH,
kleiss_stirling}.}. This turns out to be a very good approximation
with the cuts we have taken, see\cite{fawzi_bbH}.

\begin{figure}[hbt]
\begin{center}
\includegraphics[width=16cm]{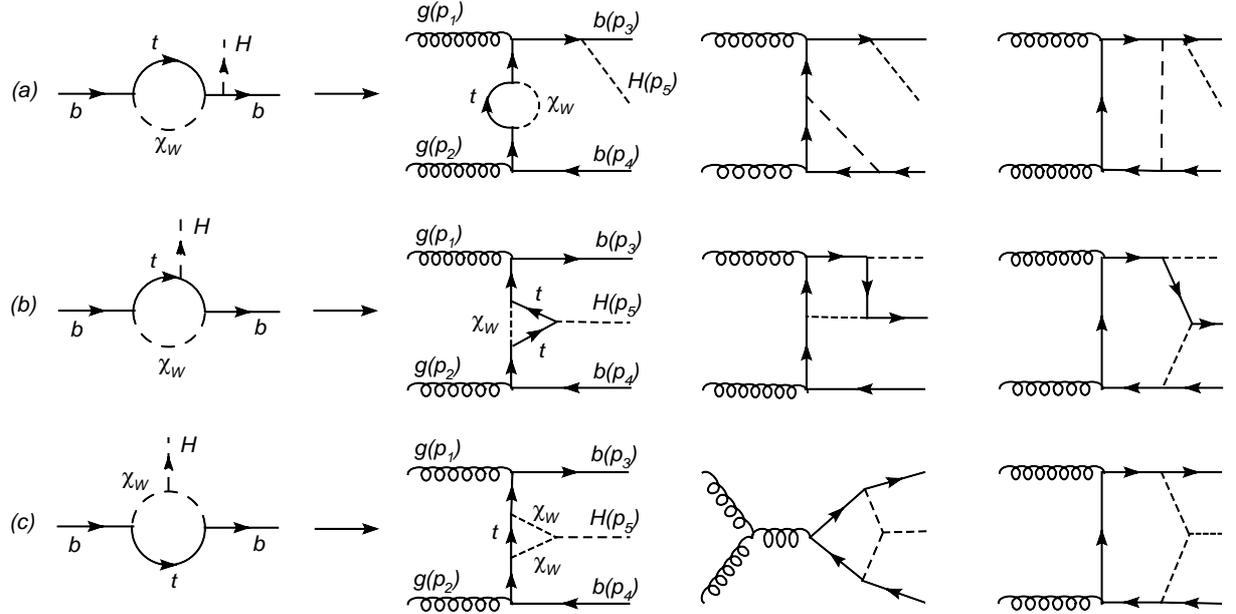}
\caption{\label{diag_3group}{\em All the diagrams in each group
can be obtained by inserting the two gluon lines or one triple
gluon vertex (shown in class (c)) to all possible positions in the generic
bottom line, which is the first diagram on the left. }}
\end{center}
\end{figure}

At the one-loop level the electroweak effects introduce a rich
structure even in the limit where one takes the leading Yukawa
(top and Higgs)  couplings that are most easily given by the
contribution of the top/charged Goldstones contribution in the
Feynman gauge\cite{fawzi_bbH}, see Fig.~\ref{diag_3group}.
At one-loop, the diagrams are classified into three QCD gauge
invariant classes as displayed in Fig.~\ref{diag_3group}. The
Higgs couples to the bottom quark in  class (a), to the top quark
in the class (b) and to the charged Goldstone boson in class (c).
As shown in Fig.~\ref{diag_3group} each class can be efficiently
reconstructed from the one-loop vertex $b \bar b H$, depending on
which leg one attaches the Higgs, by then grafting the gluons in
all possible ways. The difference in the coupling structure is
another indication that each group forms a QCD gauge independent
subset, see \cite{fawzi_bbH} for details. The analysis of
\cite{fawzi_bbH} reveals that the contribution of class (a) at NLO
is about $-0.1\%$ and thus can be totally neglected.  Class (a)
contribution naturally vanishes in the limit $\la_{bbH}=0$ as does
the tree-level. In this limit the process is loop induced and
triggered by diagrams in classes (b) and (c). Moreover in the
limit $m_b \ra 0$ with $\la_{bbH} \neq 0$, the one-loop
corrections induce new helicity structures compared to those found
at tree-level in this limit.

When trying to extend the study we have performed in
\cite{fawzi_bbH}  for $M_H > 2 M_W$ we encountered severe
numerical instabilities for the cross section involving the square
of the one-loop induced  amplitude, which is the only remaining
contribution in the limit $\la_{bbH} \ra 0$. At the level of the
NLO which involves the interference term between the tree-level
and one-loop amplitudes no instability was present. On close
inspection it was found that the instabilities were only due to
the   contribution from class (c) in particular to the box
diagrams, including the box obtained from the reduction of the
pentagon diagrams as displayed in Fig.~\ref{landau_box}). At the
partonic gluon-gluon level it was found there is  no instability
for $\sqrt{s_{gg}}< 2 m_t$  and that independently of $M_H$ and
$\sqrt{s_{gg}}$ the result was completely stable for $m_t=M_W$.
These threshold conditions were a sign for the possible existence
of a leading Landau singularity for the box diagrams whose square
is not integrable. The pentagon diagram in  class (c) has no LLS
but contains a sub-leading Landau singularity which is exactly the
same as the LLS of the box diagram, obtained through the reduction
of the pentagon to boxes. Some triangle diagrams of class (c) have
also LLS (see Appendix~\ref{appendix_3pt}) but they are integrable
hence do not cause any numerical instability. Since such
singularities are little known nowadays and hardly encountered
though we have referred to a few examples from the relatively
recent literature in the introduction, we will discuss the issue
of the LLS, their location and the condition on their appearance
in the next section.

Before that, let us remind the reader that, to calculate the cross
sections, we use the same helicity amplitude method as the one
used and explained in \cite{fawzi_bbH}. Details of the
renormalisation scheme, for the NLO,  and the optimization
implemented in our code are the same as in \cite{fawzi_bbH}. To
check the amplitudes and the cross sections we perform (QCD) gauge
invariance tests and verify that our results are ultraviolet
finite, see \cite{fawzi_bbH} for details of implementing these
checks.
\section{Landau singularities}
\label{section_landau}

Part of the discussion in this section has been summarised in
\cite{Bern:2008ef} and relies  on \cite{Landau,book_eden} although
a few results are new.

\subsection{Conditions for a Landau singularity and the nature of the singularity}
\begin{figure}[h]
\begin{center}
\includegraphics[width=0.3\textwidth]{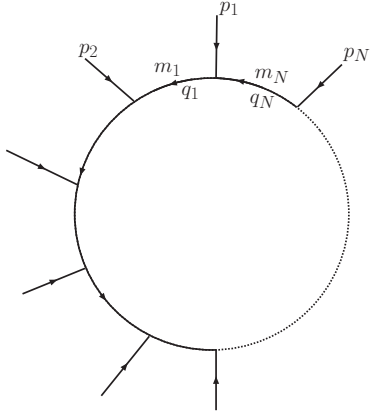}
\caption{\label{diagram_Npt}{\em One-loop Feynman diagram with $N$
external particles.}}
\end{center}
\end{figure}
Consider the one-loop process $F_1(p_1)+F_2(p_2)+\ldots
F_N(p_N)\to 0,$ where $F_i$ stands for either a scalar, fermion or
vector field with momentum $p_i$ as in Fig.~\ref{diagram_Npt}. The
internal momentum for each propagator is  $q_i$ with $i=1,\ldots
N$. Each momentum $q_i$ is associated with one real Feynman
parameter $x_i$ respectively. The scalar N-point loop integral in
$D$ space-time dimension reads
\bea
T^{N}_{0}&\equiv & \int\fr{d^Dq}{(2\pi)^Di}\fr{1}{D_1D_2\cdots
D_{N}},
\nonumber \\
D_i&=&q_i^2-m_i^2+i\eps \;\; {\rm with}\;\;\eps > 0, \,\,\, q_i=q+r_i\;\; {\rm and}\;\; q_i=q_i^*, \nonumber \\
r_i&=&\sum_{j=1}^{i}p_j,\,\,\, i=1,\ldots,N,
\eea
$q_i=q_i^*$ comes from  the fact that the $q$-integration
hypercontour is along the real axis, according to the
(infinitesimal) $i\eps$ prescription. The Feynman parameter
representation reads
\bea
T^{N}_{0}=\Gamma(N)\int_0^\infty dx_1\cdots
dx_N\delta(\sum_{i=1}^{N}x_i-1)\int\fr{d^Dq}{(2\pi)^Di}\fr{1}{(x_1D_1+x_2D_2+\cdots x_ND_{N})^N}.
\label{eq_TN0}
\eea
Because of the Dirac delta function, the integration boundary in
the Feynman parameter space are $x_i=0$, $i=1,\ldots ,N$. Thus the
only important condition on $x_i$ is that they are real and {\em
not negative.} The singularities are given by the Landau
conditions \cite{Landau,polkinghorne_1960,book_eden}
\bea
\left\{
\begin{array}{ll}
\forall i \,\,\, x_i(q_i^2-m_i^2)=0,\\
\sum_{i=1}^Nx_iq_i=0,
\\
q_i=q_i^*. \label{landau_eqN}
\end{array}
\right.
\eea
If Eq.~(\ref{landau_eqN}) has a solution $x_i>0$ for every
$i\in\{1,\ldots ,N\}$, {\it i.e. all particles in the loop are
simultaneously on-shell}, then  the integral $T^N_{0}$ has a
leading Landau singularity (LLS). If a solution exists but with
some $x_i=0$ while the other  $x_i$'s are positive, the Landau
condition corresponds to a sub-leading Landau singularity. To
keep the analysis general let us therefore assume that Eq.
(\ref{landau_eqN}) admits a solution with $x_i=0$ for
$i=M+1,\ldots,N$ with $1\le M\le N$ and $x_i>0$ for every
$i\in\{1,\ldots ,M\}$. Eq. (\ref{landau_eqN}) would read
\bea
\left\{
\begin{array}{lll}
x_i=0\,\,\, \text{for}\,\, i=M+1,\ldots,N,\\
q_i^2=m_i^2 \,,\; x_i > 0 \,\, \text{for}\,\, i=1,\ldots,M,\\
\sum_{i=1}^Mx_iq_i=0.\label{landau_eqM}
\end{array}
\right.
\eea
For $M=N$ one has a leading singularity, otherwise if $M<N$ this
is a subleading singularity. Multiplying the third equation in
Eq.~(\ref{landau_eqM}) by $q_j$ leads to a system of $M$ equations
\bea
\left\{
\begin{array}{ll}
Q_{11}x_1 + Q_{12}x_2 + \cdots Q_{1M}x_M& =0,\\
Q_{21}x_1 + Q_{22}x_2 + \cdots Q_{2M}x_M& =0,\\
\vdots \\
Q_{M1}x_1 + Q_{M2}x_2 + \cdots Q_{MM}x_M& =0,\label{landau_Meqs}
\end{array}
\right.
\eea
where the $Q$ matrix is defined as
\bea
Q_{ij}=2q_i.q_j=m_i^2+m_j^2-(q_i-q_j)^2=m_i^2+m_j^2-(r_i-r_j)^2;\,\,\, i,j\in\{1,2,\ldots,M\},\label{def_Qij}
\eea
and use is made of the on-shell constraint,{\it i.e.} the second
equation in (\ref{landau_eqM}). Note that in Eq.~(\ref{landau_Meqs})
$x_i>0$.

The necessary conditions for the appearance of a Landau
singularity can be summarized as follows
\bea
\left\{
\begin{array}{ll}
\det(Q)=0\\
x_i>0\\
q_i^2=m_i^2\\
q_i=q_i^*
\end{array}
\right.
\label{landau_cond0}
\eea
for $i=1,\ldots,M$.  The last condition, already encoded in
Eq.~(\ref{landau_eqN}), will prove to be useful, as we shall see.

It has been shown by Coleman and Norton \cite{Coleman:1965xm} that
if the matrix $Q_{ij}$ has {\em only one} zero eigenvalue then
these equations are necessary and sufficient conditions for the
appearance of a singularity in the physical region.

\noi If some internal (external) particles are massless like in the
case of six photon scattering\cite{6_photons}, then some $Q_{ij}$
are zero, the above conditions can be easily checked. However, if
the internal particles are massive then it is difficult to check
the second  condition in Eq.~(\ref{landau_cond0})  explicitly,
especially if $M$ is large. In this case, we can rewrite the
second condition as follows
\bea
x_j=\det(\hat{Q}_{jM})/\det(\hat{Q}_{MM})>0,\,\,\,
j=1,\ldots,M-1,\label{landau_cond1}
\eea
where $\hat{Q}_{ij}$ is obtained from $Q$ by discarding row $i$
and column $j$ from $Q$ and
$\det(\hat{Q}_{jM})=d[\det(Q)]/(2dQ_{jM})$,
$\det(\hat{Q}_{MM})=d[\det(Q)]/dQ_{MM}$. If $\det(\hat{Q}_{MM})=0$
then  condition Eq.~(\ref{landau_cond1}) becomes
$\det(\hat{Q}_{jM})=0$ with $j=1,\ldots,M-1$.

The condition of vanishing Landau determinant means that $Q$ has
at least one zero eigenvalue. In general, $Q$ has $N$ real
eigenvalues  $\la_1$, \ldots, $\la_N$. Consider the case where $Q$
has {\em only one} (non degenerate) very small eigenvalue
$\la_N\ll 1$, which is what is occurring in our present
calculation for $gg \ra b \bar b H$. To leading order
\bea
\la_N=\fr{a_0}{a_1},\,\,\, a_{1}=\la_1\la_2\ldots \la_{N-1}\neq 0,
a_0=\det(Q).
\eea
With $V=\{x_1^0,x_2^0,\ldots,x_N^0\}$  the eigenvector
corresponding to $\la_N$, we define $\upsilon^2=V.V$.  We will
assume that $\la_i>0$ for $i=1,\ldots,K$ and $\la_j<0$ for
$j=K+1,\ldots,N-1$ with $0\le K\le N-1$. It can then be shown that
in $D$-dimension (see Appendix~\ref{appendix_landau})

\bea
(T^{N}_{0})_{div}&=&\frac{1}{\pi} \frac{(-1)^{N+1}}{2^{(N+3)/2}}
\; \fr{e^{i\pi\alpha_{K}}\upsilon}{\sqrt{(-1)^{2 \alpha_{K}}a_1}}
\; \fr{(4
\pi)^{\alpha_{D}}\Gamma(\alpha_{D})}{(\frac{1}{2}\la_N\upsilon^2-i\eps)^{\alpha_{D}}}
\nonumber \\ \alpha_K&=&\frac{N-K+1}{2} \;\;\;\;\;\;
\alpha_D=\frac{N-D+1}{2}. \label{ccnature_landau_N}
\eea
This result holds provided $a_1\neq 0 $ or in other words that the matrix
$Q$ does not have a degenerate zero
eigenvalue. A similar result for the nature of the singularity has
been derived in \cite{polkinghorne_1960} in the general case of a
multi-loop diagram including the behaviour of the non-leading
singularity. The extraction of the overall, regular, factor  which
is the $K$-dependent part in Eq.~(\ref{ccnature_landau_N}) is more
transparent in our derivation. As stressed earlier the above
result holds provided $a_1 \neq 0$. This general result has been
derived with the assumption that formally $N-D+1>0$, however
unlike in \cite{polkinghorne_1960}  we can trivially analytically
continue the result by using dimensional regularisation with
$D=4-2\epsilon$ so that we can easily derive the nature of the
singularity from Eq.~(\ref{ccnature_landau_N}) even for the case of
$N\leq 3$ in $D=4$. For  the box in $4$-dimension, $N=4$, $D=4$,
$a_0\to 0$ and $a_1\neq 0$ we get
\bea
(T_0^4)_{div}=\fr{e^{i\pi(3-K)/2}}{4\sqrt{(-1)^{3-K}\det(Q_4)-i\eps}}.\label{eq_T04h}
\eea
This shows that $(T_0^4)_{div}$ is integrable but its square is
not. \noindent In the case $N=3$ (the triangle), $D=4$, one gets
(see Appendix~\ref{appendix_landau} for an alternative derivation
not based on dimensional regularisation but along the one followed
in \cite{polkinghorne_1960})
\bea
(T^{3}_{0})_{div}=\fr{e^{i\pi(2-K)/2}\upsilon}{8\pi\sqrt{(-1)^{2-K}\la_1\la_2}}\ln(\la_3\upsilon^2-i\eps).
\label{nature_landau_vertex}
\eea
$T_0^3$ and its square are therefore integrable.

The situation becomes more complicated when $Q$ has a degenerate
zero eigenvalue which happens in the case of the box diagram
obtained in the case  of the $6$ photon amplitude or $gg\to
W^+W^-$\cite{ggww_landau} with massless internal particles. In
$D=4$ and for $N\geq 6$ a leading Landau singularity does not
obtain, see for example p.~115 of \cite{book_eden}. We leave some
of these  issues for another publication though and will
concentrate here only on our process.

\subsection{Application to $gg \ra b\bar{b}H$}
\begin{figure}[htb]
\begin{center}
\includegraphics[width=0.4\textwidth]{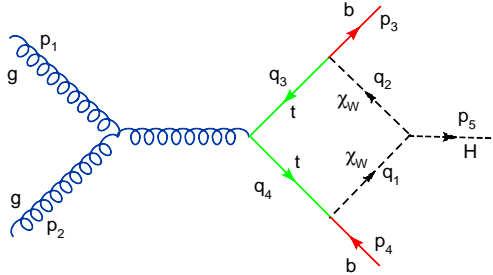}
\caption{\label{landau_box}{\em A box diagram contributing to
$gg \ra b \bar b H$  that can develop a Landau singularity for
$M_H\ge 2M_W$ and $\sqrt{s}\ge 2m_t$, i.e. all the four particles
in the loop can be simultaneously on-shell.}}
\end{center}
\end{figure}

Having set the stage for the occurrence of the Landau
singularities we now turn to check that the numerical
instabilities found in $gg \ra b \bar b H$ are indeed due a Landau
singularity. We concentrate on the box diagram in
Fig.~\ref{landau_box} which can contribute a leading Landau
singularity. The leading singularity of the 3-point function
relevant for our problem is studied in Appendix~\ref{appendix_3pt}
and serves as good starting point for the discussion to follow.
The associated $5-$point function where both external gluons
attach to the internal top quark has no leading Landau singularity
but rather a sub-leading Landau singularity which is exactly the
same as the leading singularity that appears in the box diagram in
Fig.~\ref{landau_box}. It is thus enough to study, in detail, the
structure and the singularity behaviour of this box diagram. We
will keep the bottom quark massless unless otherwise stated.

Defining the invariants $s=s_{gg}=(p_1+p_2)^2$,
$s_1=(p_3+p_5)^2,\,\, s_2=(p_4+p_5)^2$, and the on-shell
conditions $p_1^2=p_2^2=p_3^2=p_4^2=0$, $p_5^2=M_H^2$, the
kinematically allowed phase space region leads to the constraint
\bea
M_H^2\le  s_{1} \le s,\;\; M_H^2\fr{s}{s_1}\le s_2  \le
M_H^2+s-s_1.\label{phys_region_ggbbH}
\eea

We need to keep these constraints in mind as the solution of the
Landau equations may fall outside the phase space.

In terms of these invariants, the scalar box integral depicted in
Fig.~\ref{landau_box} writes, in the nomenclature of {\tt
LoopTools} for example, as
\bea
T_0^4(s_1,s_2)=D_0(M_H^2,0,s,0,s_1,s_2,M_W^2,M_W^2,m_t^2,m_t^2).
\eea

\subsubsection{On-shell and real conditions on the internal momenta $q_i$}

For the leading Landau singularity of the box in
Fig.~\ref{landau_box}, the on-shell conditions on the internal
particles read as $q_1^2=q_2^2=m_1^2=m_2^2=M_W^2$,
$q_3^2=q_4^2=m_3^2=m_4^2=m_t^2$. The condition of real
$q_i=(q_i^0,\textbf{q}_i)$ means that $\textbf{q}_i^2\ge 0$. At
each vertex, one has
\bea
\lambda(M_i^2,m_i^2,m_{i+1}^2)&=& \biggl( M_i^2-(m_i+m_{i+1})^2
\biggr) \biggl( M_i^2-(m_i-m_{i+1})^2 \biggr) \ge 0,\hs \nonumber
\\ M_i^2&=&(q_i-q_{i+1})^2, \label{cond_real}
\eea
with the usual $\lambda$ kinematical function,
\noi vertex $i$ is identified as the the vertex to which the
vector $q_i$ points according to Fig.~\ref{landau_box}, $M_i$ is
the invariant mass of the external leg at vertex $i$. Applying the condition of Eq.~(\ref{cond_real}) for
the cases $i=1,3$ we get
\bea
M_H\ge 2M_W \hs \text{and} \hs \sqrt{s}\ge 2m_t.
\label{cond_MH_s_0}
\eea
This requires that the normal thresholds for top quark production and
Higgs decay into a $W$ pair be opened.

Condition $\sum_{i=1}^Nx_iq_i=0$ in Eq.~(\ref{landau_eqN}) in the
case of the leading Landau singularity with $x_i>0$ is nothing
else but the addition of $N$ vectors, $x_i q_i$, with norm $|x_i
m_i|$. This says for example that not all time components $q_i^0$
can be positive or negative. For $N=4$, either one vector has
$\sign(x_iq_i^0)$ opposite to all the other three or there
are 2 vectors $x_iq_i^0$ with positive signs while the others have
a negative sign. In our case it is easy to see that we can only take
$q_{1,4}^0>0, q_{2,3}^{0} <0$. These simple considerations furnish
additional inequalities that are constraints on the appearance of
a LLS. Applied at the four vertices, for example in the rest frame
of one of the internal on-shell particle\cite{Ninh_thesis}, these
give the additional normal thresholds of this 4-point function
\bea
\label{thresh_mtgmw} m_t &> & M_W    \\
\label{thresh_s1s2} s_1\ge (m_t+M_W)^2\hs &\textrm{and} & \hs
s_2\ge (m_t+M_W)^2.
\eea
These strong requirements  on the opening up of the normal
thresholds will delimit the region where a LLS will occur, as
given by the vanishing of the Landau determinant. These normal thresholds
are also normal thresholds of the reduced diagrams, 3-point and 2-point
functions, obtained from $x_i=0$ and are necessary condition for a
LLS for these integrals, see Appendix~\ref{appendix_3pt}.

The on-shell and real conditions on the internal momenta $q_i$
with $\sum x_iq_i=0, \hs x_i>0$ have been
given a beautiful pictorial physical interpretation by Coleman and
Norton\cite{Coleman:1965xm}. Each $q_i$ can be regarded as the
physical momentum of a physical particle, we can associate to the
Feynman diagram a space-time graph of a process with on-shell
classical particles moving forward in time, $x_i m_i$ can be
regarded as the proper time of particle $i$. The vertices are regarded as
space-time points. $\Delta X_i=x_iq_i$ (no sum over $i$) is a space-time separation.

\subsubsection{Landau determinant}
The necessary conditions given by the inequalities above having to
do with the opening up of normal thresholds need to be
supplemented by the requirements of a vanishing Landau
determinant. The reduced matrix, $S^{(4)}$, which is equivalent in
this case to the $Q$ matrix for studying the Landau singularity,
is given by
\bea S_{4}=\left( \begin{array}{cccc}
1 & \fr{2M_W^2-M_H^2}{2M_W^2} & \fr{m_t^2+M_W^2-s_1}{2M_Wm_t} & \fr{M_W^2+m_t^2}{2M_Wm_t}\\
\fr{2M_W^2-M_H^2}{2M_W^2} & 1 & \fr{M_W^2+m_t^2}{2M_Wm_t} & \fr{m_t^2+M_W^2-s_2}{2M_Wm_t}\\
\fr{m_t^2+M_W^2-s_1}{2M_Wm_t} & \fr{M_W^2+m_t^2}{2M_Wm_t} & 1 & \fr{2m_t^2-s}{2m_t^2}\\
\fr{M_W^2+m_t^2}{2M_Wm_t} & \fr{m_t^2+M_W^2-s_2}{2M_Wm_t} & \fr{2m_t^2-s}{2m_t^2} & 1\\
\end{array}\right), \;\; S_{4}^{ij}=\frac{Q_4^{ij}}{2m_i m_j}.\eea
With $s$ and $M_H$ fixed one can study the behaviour of the
determinant as a function of the invariant $s_1$ and $s_2$. The
determinant is a polynomial of order $2$ in each of these
variables. In terms of $s_2$ for example it reads
\bea
\det(Q_4)&=&16M_W^4m_t^4 \det(S_4)=as_2^2+bs_2+c=a \left\{
\left(s_2-s_{2}^{0}\right)^2+\bar{\Delta}(s_1)\right\},\crn
a&=&\la(s_1,m_t^2,M_W^2)=[s_1-(m_t+M_W)^2][s_1-(m_t-M_W)^2],\crn
b&=&2\left\{-s_1^2(m_t^2+M_W^2)+s_1[(m_t^2+M_W^2)^2-(s-2m_t^2)(M_H^2-2M_W^2)]
\right. \nonumber \\ & & \left. +sM_H^2(m_t^2+M_W^2)\right\},
\quad s_{2}^0=-b/2a, \crn
c&=&s_1^2(m_t^2-M_W^2)^2+2M_H^2s(m_t^2+M_W^2)s_1\crn
&+&sM_H^2[(s-4m_t^2)(M_H^2-4M_W^2)-4(m_t^2+M_W^2)^2],\crn
\bar{\Delta}(s_1)&=&-\fr{b^2-4ac}{4a^2}. \label{det_abc}
\eea

Writing $\det Q_4$ as perfect square in $s_2$, like above for
example, and a remainder which is the discriminant of the
quadratic form that does not depend on $s_2$ can be revealing. In
our case we find
\beqn
\det(Q_4)&=&-\det Q_2(s_1;m_t^2,M_W^2)  \biggl( (s_2-s_2^0)^2
\nonumber \\
& & \;\;\;\;\; - \fr{\det Q_3(s_1,M_H^2,0;m_t^2,M_W^2,M_W^2)}{\det
Q_2(s_1;m_t^2,M_W^2)}\; \fr{\det Q_3(s_1,s,0;M_W^2,m_t^2,m_t^2)}{\det
Q_2(s_1;m_t^2,M_W^2)} \biggr) \label{landau_extreme}
\eeqn
$\det Q_3$'s are the Landau determinants of the 3-point function
sub-diagrams obtained from the original 4-point function by
shrinking one internal line  to a point, forming sub-diagrams
where the invariant $s_1$ is an argument of these 3-point
functions. Likewise for $\det Q_2$ obtained by further shrinking
one of the triangles. The corresponding 2- and 3-point functions
are shown in Fig.~\ref{fig_gbbH_sub-LLS}. Our convention for $\det
Q_{3,2}$ as concerns its arguments is given in
Appendix~\ref{appendix_3pt}. The factorisation in
Eq.~(\ref{landau_extreme}) can be derived\cite{Tarski} for
symmetric matrices based on the Jacobi ratio theorem for
determinants\cite{determinants_wonders}. Each sub-determinant of
the reduced three-point function  can be further reduced into
exactly such a factorised form, see Appendix~\ref{appendix_3pt}.
\begin{figure}[htb]
\begin{center}
\includegraphics[width=0.9\textwidth]{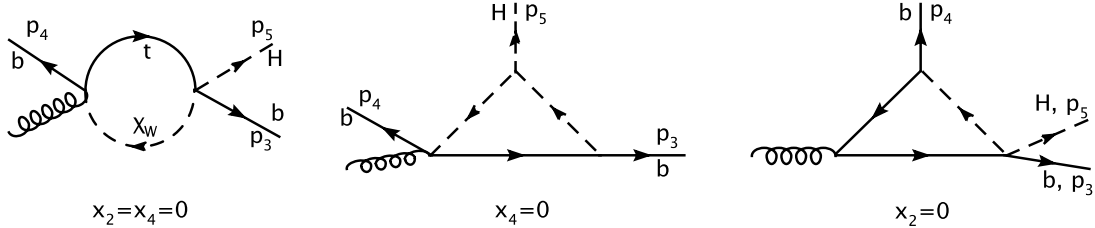}
\caption{\label{fig_gbbH_sub-LLS}{\em Three reduced diagrams of
the box diagram in Fig.~\ref{landau_box}, that contain $s_1$ as an
invariant of the 2- and 3-point functions and whose Landau
determinants  are given in Eq.~(\ref{landau_extreme}). The
self-energy diagram has a normal threshold. The two triangle
diagrams contain anomalous thresholds. Note that the singularity
structure of the second diagram, in the $s_1,M_H$ variables, is
the same as the triangle studied in much detail in Appendix~3 but
with $s_1 \ra s_2$, see Fig.~\ref{3pt-LLS}. The singularities of
the second triangle can be obtained from the first one by $s
\leftrightarrow M_H^2, m_t \leftrightarrow M_W$.}}
\end{center}
\end{figure}
This makes the identification of the sub-leading singularities
very transparent. For example, $\det Q_2(s_1;m_t,M_W)=0$
corresponds to a normal threshold, see Eq.~(\ref{thresh_s1s2}). It
occurs for $\sqrt{s_1}=m_t+M_W$ ($\sqrt{s_1}=m_t-M_W$ is outside
the physical region for Higgs masses of interest). Obviously we
could have written the quadratic form in any of the variables
$s_1,s_2,M_H^2,s$, the completion of the determinant will  be the
product of the determinant of two sub-diagrams.
\subsubsection{Numerical investigation of the four-point function and the Landau determinant}
\begin{figure}[htb]
\begin{center}
\mbox{
\includegraphics[width=0.45\textwidth]{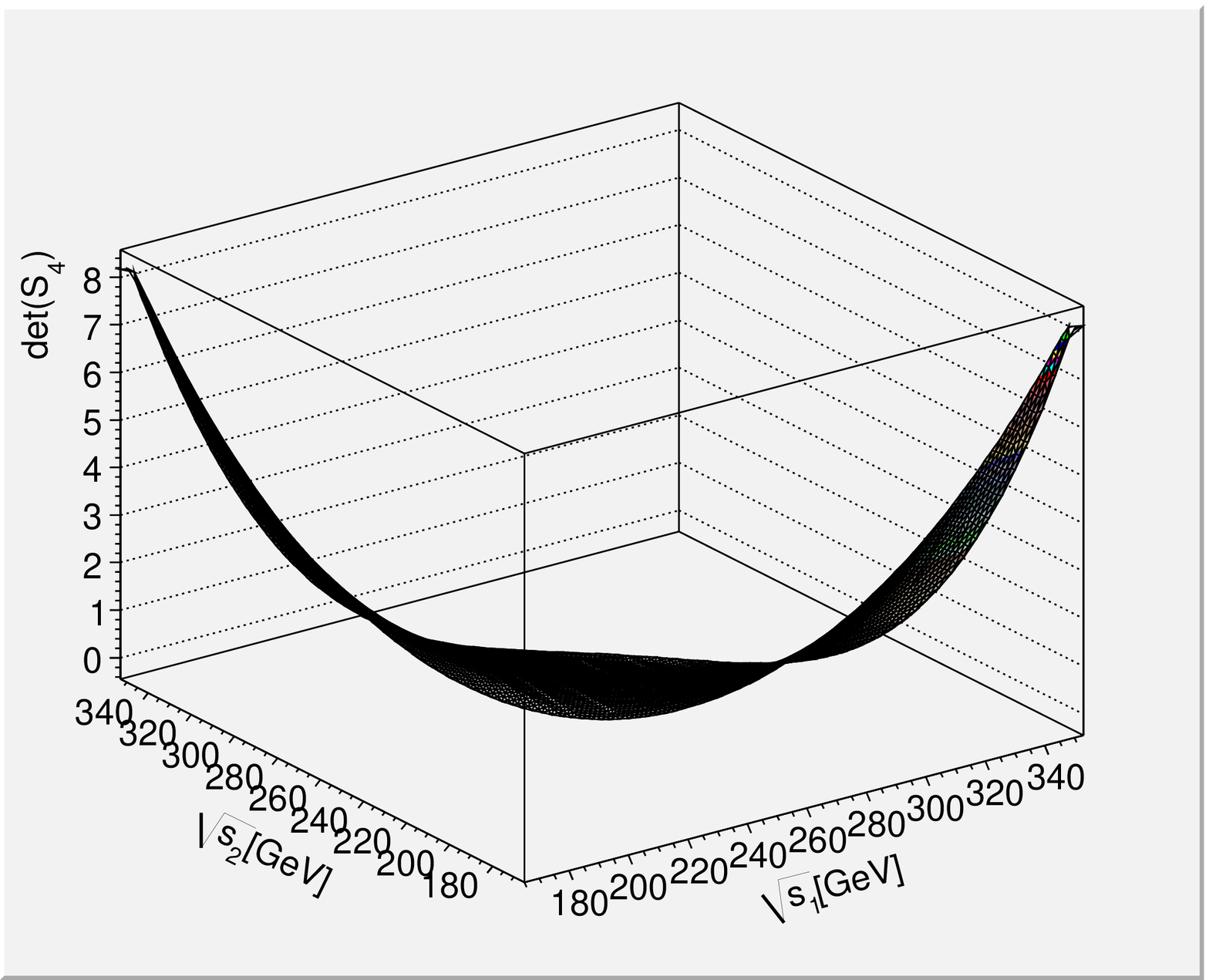}}
\mbox{\includegraphics[width=0.45\textwidth]{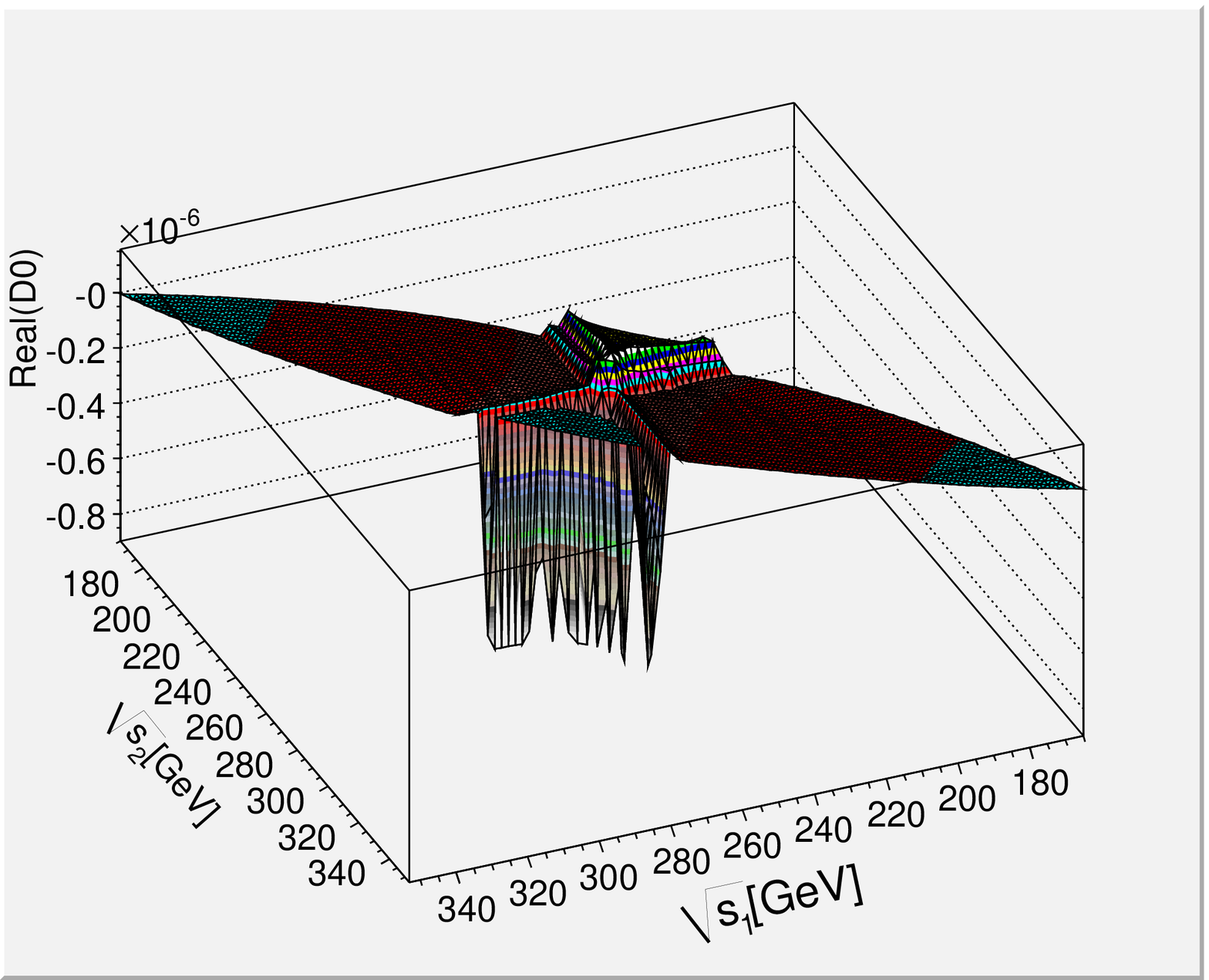}
\hspace*{0.075\textwidth}
\includegraphics[width=0.45\textwidth]{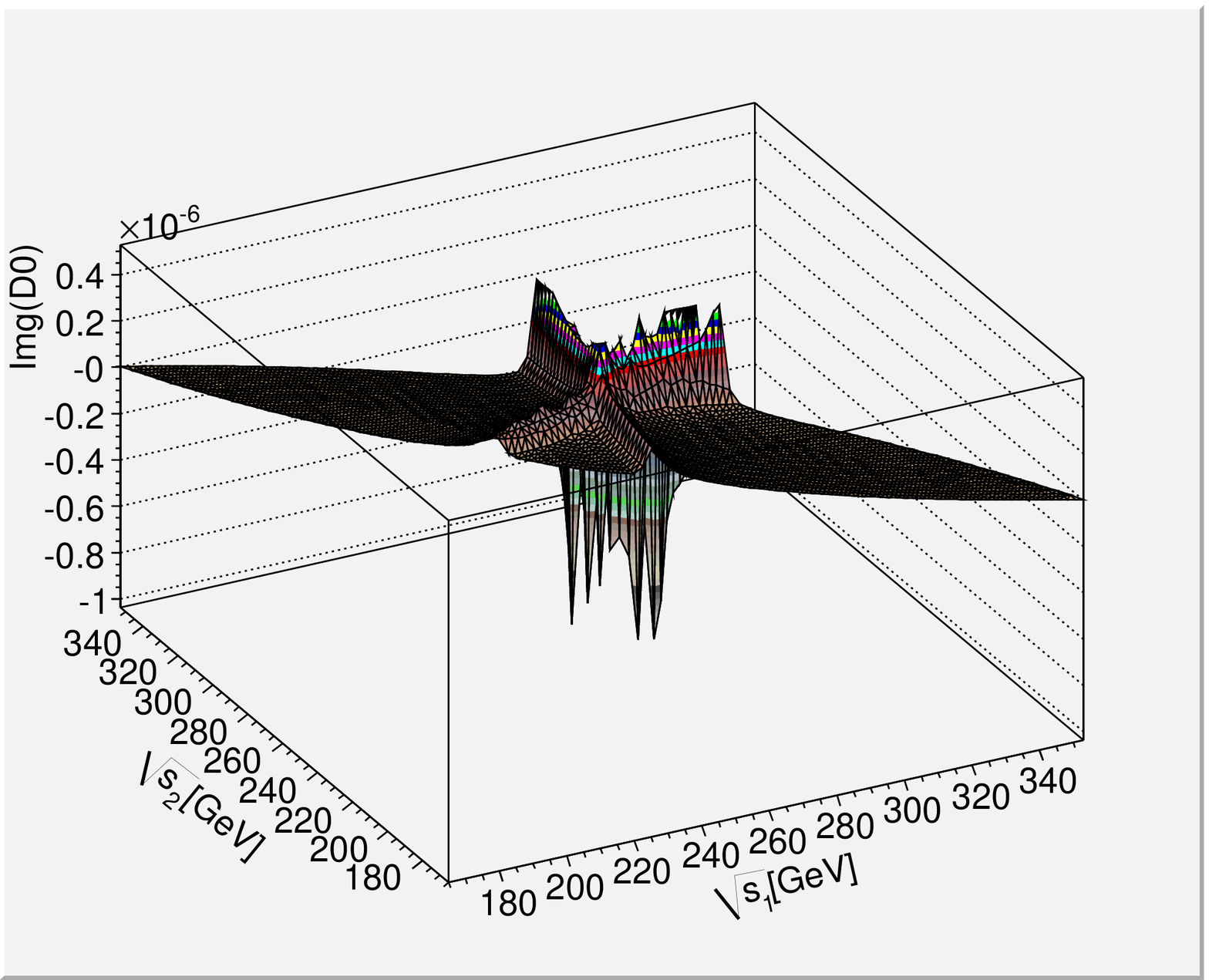}}
\caption{\label{box_diag_3D_plots}{\em The Landau determinant as a
function of $s_1$ and $s_2$ (upper figure). The real and imaginary
parts of $D_0$ as a function of $s_1$ and $s_2$. The figure for
the real part of $D_0$ has been rotated since the structure is
best seen with this view.}}
\end{center}
\end{figure}

We will always take $m_t=174\;$GeV and $M_W=80.3766\;$GeV. Our
investigation starts by taking $\sqrt{s}=353\;$GeV,
$M_H=165\;$GeV. The behaviour of the Landau determinant, the real
and imaginary parts of the $4-$point function $T_0^4$ are
displayed in Fig.~\ref{box_diag_3D_plots} as a function of $s_1$,
$s_2$ within the phase space. We clearly see that the Landau
determinant vanishes inside the phase space and leads to regions
of severe instability in both the real and imaginary parts of the
scalar integral.

To  investigate the structure of the singularities in more
detail let us fix $\sqrt{s_1}=\sqrt{2(m_t^2+M_W^2)}\approx
271.06\;$GeV, such that the properties are studied for the single
variable $s_2$. This will also exhibit the sub-leading Landau
singularities related to the reduced diagrams. In the variables
$s_2$ these are exactly the same as the ones we uncovered through
Eq.~(\ref{landau_extreme}). They are represented in
Fig.~\ref{fig_gbbH_sub-LLS} allowing for $s_1 \ra s_2$ (and $x_2
\ra x_1 , x_4 \ra x_3$).

\begin{figure}[h]
\begin{center}
\includegraphics[width=0.78\textwidth]{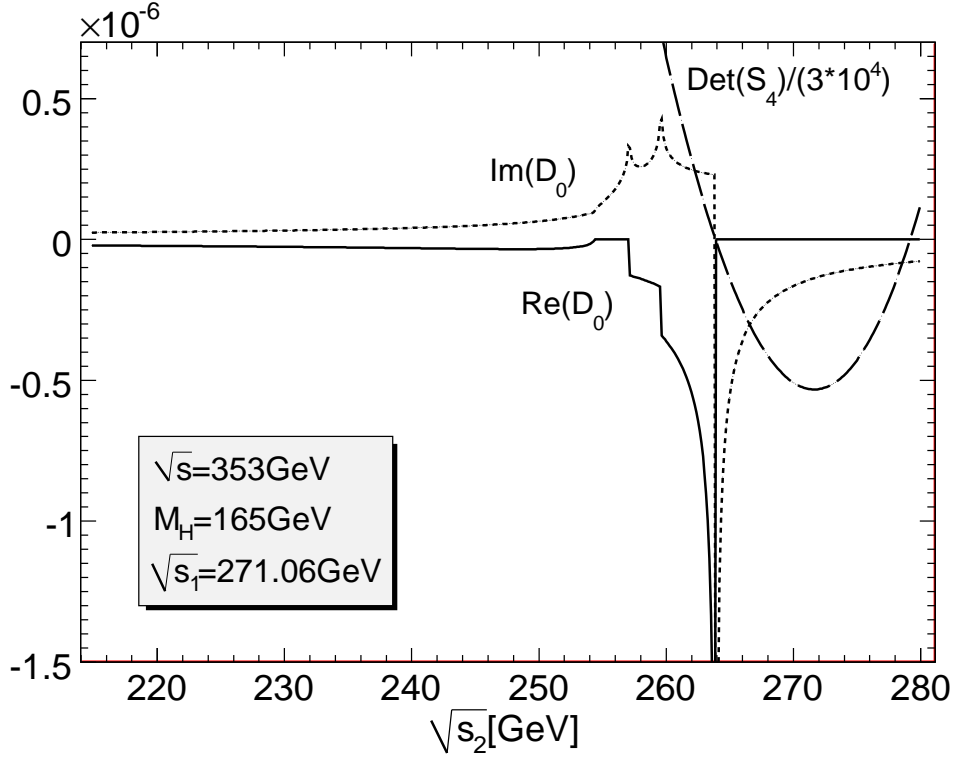}
\caption{\label{gbbH_img_real_det0_2D}{\em The imaginary, real
parts of $D_0$ and the Landau determinant as functions of $\sqrt{s_2}$.}}
\end{center}
\end{figure}
Fig.~\ref{gbbH_img_real_det0_2D} is very educative. We see that
there are four discontinuities in the function representing the
real part of the scalar integral in the variable $\sqrt{s_2}$.
\begin{itemize}
\item
As $s_2$ increases we first encounter a discontinuity  at the
normal threshold $\sqrt{s_2}=\sqrt{s_2^{{\rm
tW}}}=m_t+M_W=254.38\;$GeV, representing $H b \ra W t$. This
corresponds to the solution (for the Feynman parameters)
$x_{1,3}=0$ and $x_{2,4}>0$ of the Landau equations and can be
associated to a leading Landau singularity for the 2-point scalar
integral.
\item
The second discontinuity occurs $\sqrt{s_2}=257.09\;$GeV. This
corresponds to an anomalous threshold of a reduced triangle
diagram. This corresponds to the solution $x_{3}=0$ and
$x_{1,2,4}>0$ of the Landau equations, (see
Fig.~\ref{fig_gbbH_sub-LLS}).  The singularity structure of this
diagram is studied in more detail in Appendix~\ref{appendix_3pt}.
We can explicitly check that $\sqrt{s_2}=257.09\;$GeV corresponds
to the condition of vanishing determinant. One solution of this
equation does not satisfy the sign condition,
Eq.~(\ref{landau_cond1}) and is not even inside of phase space. As
shown in Appendix~\ref{appendix_3pt} only one solution, see also
Eq.~(\ref{s2_3pt}), is acceptable with
\bea
\label{landaupole3_x3}
s_{2}^H&=&\fr{1}{2M_W^2}\left(M_H^2(M_W^2+m_t^2)-(m_t^2-M_W^2)M_H\sqrt{M_H^2-4M_W^2}\right)
 \\
&=&2 (m_t^2+M_W^2) +(M_H^2-4
M_W^2)\Biggl(1+\frac{m_t^2-M_W^2}{2M_W^2}  \biggl(1-
\frac{1}{\sqrt{1-4M_W^2/M_H^2}} \biggr)  \Biggr) ,
 \nonumber
\eea
which gives $\sqrt{s_2^H}=257.09\;$GeV. Note that one of the
necessary conditions for this anomalous threshold to occur in the
physical region is $M_H\ge 2M_W$. At this {\em normal} threshold
the value of $s_2^H$ is $s_2^H=2 (m_t^2+M_W^2)$, see
Eq.~(\ref{landaupole3_x3}).
\item
The third discontinuity at $\sqrt{s_2}=259.58\;$GeV corresponds to
the anomalous threshold of the reduced three point function
obtained from the box diagram by contracting to a point the $x_1$
line, see the third diagram in Fig.~\ref{fig_gbbH_sub-LLS}, so
that $\det Q_3(s_1,s,0;M_W^2,$ $m_t^2,$ $m_t^2)$ $=0$. Analogously
$\sqrt{s_2}=259.58\;$GeV is given by
\bea
\label{landaupole3_x3b}
s_2^s&=&\fr{1}{2m_t^2} \left(
s(m_t^2+M_W^2)- \sqrt{s}\sqrt{s-4m_t^2}(m_t^2-M_W^2) \right),
\\
&=&2 (m_t^2+M_W^2) +(s-4 m_t^2)\Biggl(1+\frac{m_t^2-M_W^2}{2m_t^2}
\biggl(1- \frac{1}{\sqrt{1-4m_t^2/s}} \biggr)  \Biggr).
 \nonumber
\eea
\item
The last singular discontinuity is the leading Landau singularity.
The condition $\det(S_4)=0$ for the box has two solutions which
numerically correspond to $\sqrt{s_2}=263.88\;$GeV or
$\sqrt{s_2}=279.18\;$GeV. Both values are inside the phase space,
see Fig.~\ref{gbbH_img_real_det0_2D}. However after inspection of
the corresponding sign condition, only $\sqrt{s_2}=263.88\;$GeV
(with  $x_1 \approx 0.53, x_2 \approx 0.75, x_3 \approx
0.77$)  qualifies as a leading Landau singularity.
$\sqrt{s_2}=279.18\;$GeV has $x_1\approx -0.74, x_2 \approx
-0.75, x_3 \approx 1.07$ and is outside the physical
region.
\end{itemize}

The nature of the LLS in Fig.~\ref{gbbH_img_real_det0_2D} can be
extracted  by using the general formula (\ref{eq_T04h}). With the
input parameters given above, the Landau matrix has only one
positive eigenvalue at the leading singular point, {\it i.e.}
$K=1$. The leading singularity behaves as\footnote{The singularity of the $3$-point function is logarithmic, see Eq.~(\ref{nature_landau_vertex}). Fig.~\ref{gbbH_img_real_det0_2D}
shows two $3$-point singularities which look as if better behaved within {\tt LoopTools}.}
\bea D_0^{div}=-\fr{1}{16M_W^2m_t^2\sqrt{\det(S_4)-i\eps}}.\label{d0_detS}
\eea
When approaching the singularity from the left, $\det(S_4)>0$,
the real part turns singular. When we cross the leading
singularity from the right, $\det(S_4)<0$, the imaginary part of
the singularity switches on, while the real part vanishes. In this
example, both the real and imaginary parts are singular because
$\det(S_4)$ changes  sign when  the leading singular point is
crossed.\\

\subsubsection{The leading Landau singularity region in  the $(M_H$, $\sqrt{s})$ plane}
\label{landau_region_blue} In practice we will have to integrate
over the $s_1$ and $s_2$ variables to obtain  the total cross
section at the partonic level. We will also have to integrate over
$s=s_{gg}$ to arrive at the cross section at the $pp$ level.
Moreover, we would like to study the behaviour of the cross
section by varying $M_H$. It is therefore important to quickly
localise the range or region in the $(\sqrt{s},M_H)$ plane where
the leading Landau singularity occurs. This approach should in
fact be followed in more general cases to check if one might
encounter a potential problem prior to carrying the full phase
space integration procedure with the full matrix elements.
\begin{figure}[httb!]
\begin{center}
\includegraphics[width=0.6\textwidth]{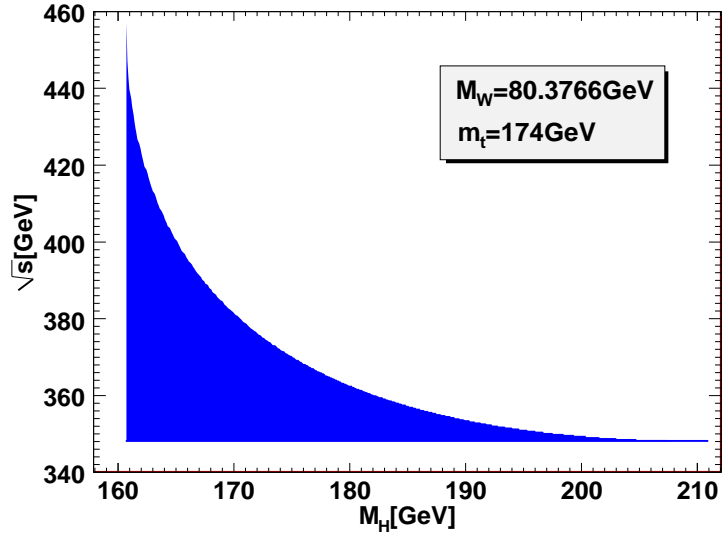}
\caption{\textit{The region of the leading Landau singularity in
the variables $\sqrt{s}=\sqrt{s_{gg}},M_H$.}} \label{landau_range}
\end{center}
\end{figure}
Necessary (but not sufficient) conditions on $M_H$ and $\sqrt{s}$
to have a LLS correspond to the opening of normal thresholds as
given in Eq.~(\ref{cond_MH_s_0}). These are easy to guess and are
contained in the last two equations of Eq.~(\ref{landau_cond0}).
We have however to solve all of Eq.~(\ref{landau_cond0}) together
with the constraint that one is inside  the phase space
Eq.~(\ref{phys_region_ggbbH}). This, in general, is too
complicated to be done analytically in a situation like ours with
$4$ variables ($M_H$, $s$, $s_1$, $s_2$) and $2$ parameters
($M_W$, $m_t$). However numerically the algorithm that goes
through all the conditions is quite simple to implement. For
instance one can start with the Landau determinant written as a
quadratic form in $s_2$ by first computing the discriminant of the
quadratic equation and check whether the latter is positive or
negative, assuming the solutions are in the physical region. If
the discriminant is positive one checks if the corresponding
solution does not conflict with the positivity solution as
implemented in Eq.~(\ref{landau_cond1}). If this condition is
satisfied then there is a LLS. In our case the result is shown in
Fig.~\ref{landau_range}. We conclude that the LLS occurs when
$2M_W\le M_H< 211$GeV and $2m_t\le \sqrt{s}< 457$GeV. The range of
the LLS region depends on $M_W$ and $m_t$. If $m_t/M_W\le 1$ then
the first two conditions in Eq.~(\ref{landau_cond0}) can never be
satisfied. In particular, if $m_t/M_W=1$ then the Landau
determinant can vanish but the sign condition cannot be realised.
When $M_H>210$GeV or $\sqrt{s}>456$GeV the Landau determinant
$\det(Q_4)$ can vanish inside the phase space but the sign
condition $x_i>0$ cannot be fulfilled. \\

\noi {The region of the leading Landau singularity in
Fig.~\ref{landau_range} is a surface of singularities in the plane
of the kinematical variables $\sqrt{s}=\sqrt{s_{gg}},M_H$.  This
is bounded by three curves. It is important to stress again that
the horizontal and vertical lines or boundaries correspond to the
normal thresholds. These lines are also {\em tangent} to the upper
curve delimiting the surface of LLS. We will get back to this
property later.} \\

\noi  The algorithm we have just outlined is very easy to implement.
The importance of the sign condition is crucial in determining the
boundary of the leading Landau singularity region which occurs
when $x_i \ra 0$. We will come back to this point shortly. Before
doing so, it is worth coming back to the behaviour of $D_0$ as a
function of $s_2$ like what we have shown in
(Fig.~\ref{gbbH_img_real_det0_2D}) and see how the location of the
leading Landau singularity and the other discontinuities (related
to other thresholds)  move as $M_H$ is varied.

\begin{figure}[h]
\begin{center}
\includegraphics[width=0.8\textwidth]{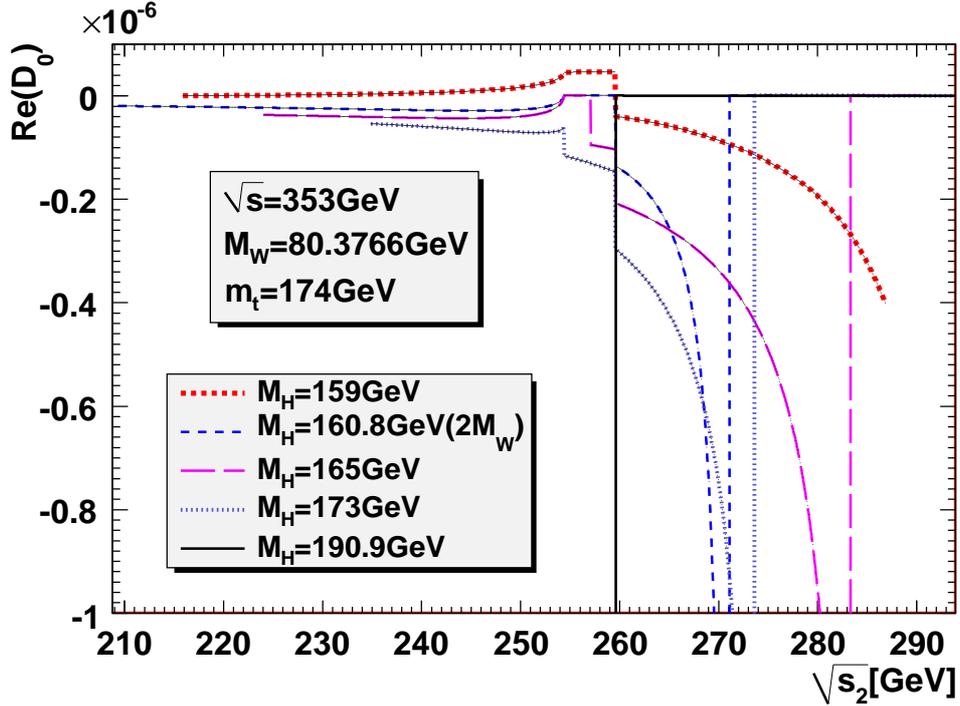}
\caption{\label{ggbbH_d0_width0}{\em The real part of $D_0$ as a function of $\sqrt{s_2}$ for various values of $M_H$.
For $M_H=2M_W$ we have taken
$s_1=2(m_t^2+M_W^2)$. For the other cases, we take $s_1=260$GeV.}}
\end{center}
\end{figure}

As in (Fig.~\ref{gbbH_img_real_det0_2D}) we fix $\sqrt{s}=353$GeV
and $\sqrt{s_1}$ but with $\sqrt{s_1}=260$GeV  for
$M_H=159,165,190$GeV. All the curves will therefore show the
two-point function discontinuity (normal threshold) at
$\sqrt{s_2^{{\rm tW}}}=254.38$GeV and three-point function
discontinuity $\sqrt{s_2^s}=259.58$GeV, see
Eq.~(\ref{landaupole3_x3b}). The other three-point function
discontinuity at $\sqrt{s_2^H}$ and the leading Landau
singularity, if at all there, will of course move. The results are
shown in Fig.~\ref{ggbbH_d0_width0}.
\begin{itemize}
\item
For $M_H=159$GeV only the normal threshold at $\sqrt{s_2^{{\rm tW}}}$, and the $\sqrt{s_2^s}$ discontinuity show
up as expected since $M_H < 2M_W$.
\item For $M_H=165$GeV ($M_H>2M_W$), the other three-point singularity
shows at $\sqrt{s_2^H}=257.09$GeV, together with the LLS  at
$\sqrt{s_2^{{\rm LLS}}}\approx 283.5$GeV. As $\sqrt{s_1}$ is
increased the LLS moves to smaller values of $s_2$, closer to the
three-point function singularity as can be seen by comparing with
Fig.~\ref{gbbH_img_real_det0_2D} for the same value of $M_H$ but
higher value of $\sqrt{s_1}$. This will be a common feature with
the other cases with $M_H> 2M_W$, till the LLS disappears from the
physical region. For $\sqrt{s_1} < 260$GeV, no LLS develops. We
have the ordering $\sqrt{s_2^{{\rm tW}}}<\sqrt{s_2^H}<
\sqrt{s_2^s}<\sqrt{s_2^{{\rm LLS}}}$.
\item
For $M_H\approx 173$GeV, $\sqrt{s_2^H}=\sqrt{s_2^{{\rm
tW}}}=254.38$GeV {\it i.e} the $s_2^H$ threshold coincides with
the normal threshold. The LLS starts showing up at
$\sqrt{s_2}\approx 274$GeV when $\sqrt{s_1}=260$GeV and moves to
smaller values of $s_2$ as $\sqrt{s_1}$ increases. We have the ordering
$\sqrt{s_2^{{\rm tW}}}=\sqrt{s_2^H}< \sqrt{s_2^s}<\sqrt{s_2^{{\rm
LLS}}}$. The coincidence $\sqrt{s_2^{{\rm tW}}}=\sqrt{s_2^H}$
signals the termination a leading singularity in the 3-point
function, see Appendix~\ref{appendix_3pt}. As we increase $M_H$
the LLS moves to smaller values of $s_2$ and the $s_2^H$ discontinuity disappears from the physical region.
\item  For the special value $M_H=190.88$GeV, the  $s_2^H$ singularity has
moved out of the physical region but now the LLS coincides with
the location of the  $s_2^s$ three-point function singularity. We
therefore have $\sqrt{s_2^{{\rm tW}}}<
\sqrt{s_2^s}=\sqrt{s_2^{{\rm LLS}}}$. For $M_H>190.88$GeV the
LLS disappears from the physical region.
\item Finally, we consider the special case of the threshold $M_H=2M_W$
where $\sqrt{s_2^H}=271.06$GeV. One has to change $s_1$ in the
range defined by Eq.~(\ref{phys_region_ggbbH}) with the condition
$s_1\ge (m_t+M_W)^2$ to make the LLS appear. It is easy to find
out that the LLS only occurs when
$\sqrt{s_1}=\sqrt{2(m_t^2+M_W^2)}=271.06$GeV and the LLS position
coincides with the position of the three-point singularity
$\sqrt{s_2^H}$. We
have the ordering $\sqrt{s_2^{{\rm tW}}}<
\sqrt{s_2^s}<\sqrt{s_2^H}=\sqrt{s_2^{{\rm LLS}}}$.
\end{itemize}

For future reference, it is worth noting that the LLS region opens
up rather sharply when the normal thresholds open up and the bulk
of the region is concentrated around these thresholds. Already for
$M_H \geq 200$GeV the region squeezes into
a very thin line.
\subsubsection{The leading Landau singularity region: analytical insight}
We will take two approaches. The first one is based on the
observation that the boundary of the singularity region
corresponds to a coincidence of a leading Landau singularity with
a sub-leading singularity, this is the termination of the  LLS
\cite{Cunningham,Tarski,Ninh_thesis}.  The second approach starts
directly from the constraint or equation given by the  vanishing
of the Landau determinant. The extrema of this equation with
respect to a particular choice of kinematical variables will
define the termination of the LLS. Interpreting the equation as
that defining a  surface or a hypercurve, the extrema are tangents
to the surface and are parallel to the corresponding coordinate
variables. This will become clearer when we expose the derivation.

{\bf i)} \noi A study of the LLS in the 3-point scalar integral
relevant to our problem is quite simple since this function does,
for fixed $m_t,M_W$, involve a very small number of variables. Yet
the study, see Appendix~\ref{appendix_3pt}, reveals some very
general features. There is an LLS region, or curve, that is
bounded by the normal threshold. This  a manifestation of the fact
that at the boundary, the leading singularity moves to the
sub-leading singularity\cite{Cunningham,Tarski,Ninh_thesis}. This
is also a phenomenon we
observed in section~\ref{landau_region_blue}. \\
\noi Let us now analytically derive the surface shown in
Fig.~\ref{landau_range}, or rather the curve representing its
boundaries in the ($M_H,s$) range. The lower bounds are just given
by the normal thresholds of the two-point function so that $M_H
\ge 2 M_W$ and $\sqrt{s} \ge 2 m_t$, see Eq.~(\ref{cond_MH_s_0}).
For each value of ($M_H,s$) there is a  curve of LLS's defined by
${\cal {F}}(s_1,s_2,|M_H,s)$ which is constrained by the vanishing
of $\det Q_4(s_1,s_2)$ and subject to the sign conditions. For
this discussion about the ($M_H,s$) range it is sufficient to only
keep the $(s_1,s_2)$ dependence of $\det Q_4$. As we scan over
($M_H,s$) we span a surface of LLS's. The key observation is that
the curves terminate at a point corresponding to a sub-leading
singularity, in this case a leading singularity of one of the
3-point function sub-diagrams which itself will terminate at the
2-point singularity, {\it i.e.} the normal threshold. For
instance, writing $\det(Q_4)$ as a quadratic polynomial of $s_2$
as we did in Eq.~(\ref{landau_extreme}), there are $2$ three-point
sub-LLSs given by each $\det Q_3$ in Eq.~(\ref{landau_extreme})
vanishing. The solutions of the latter are  given, respectively,
by Eq.~(\ref{landaupole3_x3}) and Eq.~(\ref{landaupole3_x3b}). Let
us take for definiteness the sub-leading singularity corresponding
to $s_2^H$ in Eq.~(\ref{landaupole3_x3}). The argument works just
as well with the other 3-point singularity $s_2^s$. The
coincidence constraint implies, for $s_2$ for example, a solution
$\hat {s}_2=s_2=s_2^H$ and  $\det Q_4(s_1,{s}_2 )=0$ (with the
proviso about the sign condition). Exactly the same argument can
be put but now solving for the variable $s_1$ and exploiting the
fact that our problem is symmetric in $s_1 \leftrightarrow s_2$.
The coincidence problem or the constraint we are looking for
translates into
\bea
\label{eq_s2_s2s} s_2&=&s_2^H\hs \text{and} \hs \det
Q_4(s_1,s_2)=0, \nonumber \\
 s_1&=&s_2^H \hs \text{and} \hs \det
Q_4(s_1,s_2)=0, \quad \implies \nonumber \\
\det Q_4(\hat{s}_2,\hat{s}_2)&=&0 \hs \text{and} \hs
\hat{s}_2=s_2^H
\eea
Only one solution to $\det Q_4(\hat{s}_2,\hat{s}_2)=0$ passes the
LLS sign conditions, with
\bea
\hat{s}_2=2(m_t^2+M_W^2)-\sqrt{(s-4m_t^2)(M_H^2-4M_W^2)}.
\label{s2-root_LLS}
\eea
Equating Eq.~(\ref{s2-root_LLS}) with Eq. \ref{landaupole3_x3}, we
arrive at the equation of the termination curve
\bea
\sqrt{(s-4m_t^2)}&=&\frac{1}{2 M_W^2} \bigg(M_H
(m_t^2-M_W^2)-(m_t^2+M_W^2)\sqrt{(M_H^2-4M_W^2)} \biggr).
\label{curve_s_MH_max}
\eea
Observe that this equation shows, in a very transparent way, that
all thresholds:
 $$m_t > M_W, M_H\ge 2 M_W, \sqrt{s} \ge 2 m_t$$  need
to be open simultaneously. We can invert
Eq.~(\ref{curve_s_MH_max}) to write the solution in terms of
$M_H$. To arrive at the same result, it is more judicious however
to go through exactly the same steps but choosing $s_2^s$
instead of $s_2^H$. We derive
\bea
\sqrt{(M_H^2-4 M_W^2)}&=&\frac{1}{2 m_t^2} \bigg(\sqrt{s}
(m_t^2-M_W^2)-(m_t^2+M_W^2)\sqrt{(s-4 m_t^2)} \biggr).
\label{curve_MH_s_max}
\eea
The maximum value of $M_H$ ($\sqrt{s}$) is obtained by setting
$\sqrt{s}=2m_t$ ($M_H=2M_W$), {\it i.e.} when the LLS, the two
$3$-point sub-LLSs and the normal threshold coincide. We have
\bea
&&4M_W^2\le M_H^2\le 4M_W^2+\fr{(m_t^2-M_W^2)^2}{m_t^2},\crn
&&4m_t^2\le s\le 4m_t^2+\fr{(m_t^2-M_W^2)^2}{M_W^2}.
\label{bounds_MH_s}
\eea
or numerically,
\bea
348.00\text{GeV}\le \sqrt{s}\le 457.05\text{GeV}\hs \text{and}\hs
160.75\text{GeV}\le M_H\le 211.13\text{GeV}.
\eea
Of course, these analytical formulae reproduce exactly the curve
in Fig.~\ref{landau_range} that was obtained numerically.  For
example, we have arrived at the same, unique, solution by taking
$s_{1,2}=s_2^H$ and $s_{1,2}=s_2^s$ in turn. This also means that
the curve is also given by
\bea
s_2^s=s_2^H. \label{eq_s2s_s2H}
\eea
This constraint gives directly the equation for the bounding curve
and avoids having to solve for $s_1$ or $s_2$ as is done as an
intermediate step in Eq.~(\ref{s2-root_LLS}).

{\bf ii)} Another interesting interpretation of the bounding curve
which also leads to Eq.~(\ref{eq_s2s_s2H}) is based on the
following. The leading Landau singularity in the $(s_1,s_2)$ plane
is a solution of $\det Q_4(s_1,s_2)=0$ supplemented by the sign
conditions. With fixed values of the internal masses, the
constraint $\det Q_4(s_1,s_2,s,M_H^2)=0$ is a constraint on the
kinematical invariants for which a LLS can occur. This therefore
defines a surface of LLS singularities, which one may want to
visualise in the plane $(s_1,s_2)$ or $(s,M_H^2)$. Within the
plane $(s_1,s_2)$, the extrema of this surface are given by the
tangents to this surface which are parallel to the coordinate
variables, in this case ${s_1,s_2}$\cite{Tarski}, therefore
\beqn
\frac{\partial\det Q_4(s_1,s_2)}{\partial s_2}=0 \quad {\rm with}
\quad \det Q_4(s_1,s_2)=0 \quad {\rm and} \nonumber \\
\frac{\partial\det Q_4(s_1,s_2)}{\partial s_1}=0 \quad {\rm with}
\quad  \det Q_4(s_1,s_2)=0. \label{extr-detq4}
\eeqn
These conditions are best exploited by using the quadratic form of
$\det Q_4(s_1,s_2)$ in $s_2$ (and $s_1$) given in
Eq.~(\ref{landau_extreme}). The first equation in
Eq.~(\ref{extr-detq4}) with the help of Eq.~(\ref{landau_extreme})
leads to
\beqn
\det Q_3(s_1,M_H^2)\det Q_3(s_1,s)= 0. \label{extr-detq4_1}
\eeqn
The second equation, using again the same quadratic form in
Eq.~(\ref{landau_extreme}) leads to
\beqn
\frac{\partial \det Q_3(s_1,M_H^2)}{\partial s_1} \det
Q_3(s_1,s)+\frac{\partial \det Q_3(s_1,s)}{\partial s_1} \det
Q_3(s_1,M_H^2)=0. \label{extr-detq4_2}
\eeqn
The constraints of Eqs.~(\ref{extr-detq4_1},\ref{extr-detq4_2}) then
require either i) \underline{both} sub-determinants in
Eq.~(\ref{landau_extreme}) to vanish, $\det Q_3(s_1,M_H^2)=\det
Q_3(s_1,s)=0$. The latter requirement is exactly the condition
given in Eq.~(\ref{curve_s_MH_max}). The other solutions of
Eqs.~(\ref{extr-detq4_1},\ref{extr-detq4_2}) give the boundaries
related to the \underline{normal thresholds}, ii) $\det
Q_3(s_1,M_H^2)=\frac{\partial \det Q_3(s_1,M_H^2)}{\partial s_1}=0$
which implies see Eq.~(\ref{ex-eqc81}) the normal threshold
$M_H=2M_W$ is reached , while the third solution iii) $\det
Q_3(s_1,s)=\frac{\partial \det Q_3(s_1,s)}{\partial s_1}=0$
corresponds to the normal threshold $s=(2m_t)^2$. These equations
for the boundary define the LLS region presented in Fig.~\ref{landau_range}.
Note that ii) and iii) can also be derived from i) if one insists on
finding the extrema of the curve $\det Q_3(s_1,M_H^2)=0$ for
example. This is the same argument that is used in
Appendix~\ref{appendix_3pt} for the three-point function. Here we
can carry this argument  one step further starting from the fact
that $\det Q_3=0$ is a condition for the Landau singularity of a
3-point function. The extrema and tangent argument applied at this
level will show that the range in $M_H$ and $s$ are given by the
vanishing of the corresponding $\det Q_2$ which give the normal
thresholds, $M_H=2M_W$ and $s=(2 m_t)^2$. This derivation shows that
when the normal threshold is met all singularities of the 2-, 3- and
4-point function coalesce. Observe that in Fig.~\ref{landau_range}
the lines given by $M_H=2M_W$ and $\sqrt{s}=2 m_t$ are not only
boundaries of the LLS region but also tangents to the extremum
bounding curve given by Eq.~(\ref{curve_s_MH_max}).

The arguments given above can be applied to derive the
bounding curve and the range of the LLS's in the $(s_1,s_2)$ plane
after elimination of the variables $(M_H^2,s)$ and taking into
account the normal threshold condition, $s_{1,2}> (m_t+M_W)^2$ as
the lower bound. The starting point in this case is to express
$\det Q_4$ as a quadratic polynomial in $M_H^2$ for example. The
solution of the bounding curve is given by
\beqn
s_1-s_2=\frac{m_t^2-M_W^2}{m_t^2+M_W^2} \biggl(
\sqrt{\lambda(s_1,m_t^2,M_W^2)}  - \sqrt{\lambda(s_2,m_t^2,M_W^2)}
\biggr).
\eeqn
This translates into the bounds

\beqn
(m_t+M_W)^2\le s_{1,2}& \le &
(m_t+M_W)^2+\fr{(m_t^2-M_W^2)^2}{m_tM_W}, \quad {\rm numerically}
\nonumber \\
254.38\text{GeV} & \le & \sqrt{s_{1,2}}\le 324.44\text{GeV}.
\label{bounds_s1_s2}
\eeqn
\section{The width as a regulator of the Landau singularity}
\label{width-reg} As we have seen the leading Landau singularity
requires all internal particles to be on their mass shell, see for
example Eq.~(\ref{landau_cond0}). This is akin to the usual
singularity that occurs on resonance for a massive particle. These
equations also show that if any parameter $m_i$ is complex with a
non zero imaginary part, the singularity is avoided. For an
unstable particle the width provides this imaginary part. As can
be inferred from Eq.~(\ref{landau_cond0}), mathematically, the
width effect is to move the Landau singularities into the complex
plane, so they do not occur in the physical region (the real
axis). For our problem, the Landau condition in the interpretation
of Coleman and Norton through Eq.~\ref{thresh_mtgmw}, $m_t > M_W$,
clearly shows that the singularity develops because of the
instability of the top quark. Therefore, in principle, one should
only include the width of the top as a regulator. Including the
width of an unstable particle, whereby the mass of the internal
particle becomes complex effectively sums a subset of higher order
Feynman diagrams thereby taming the Landau
singularity\cite{stuart}. On the other hand, if one goes to higher
order to implement the width then we would not only induce a width
for the top but also for the $W$. Therefore to be realistic one
should include the widths of both the top quark, $\Gamma_t$,  as
well as the width of the $W$, $\Gamma_W$.

\begin{figure}[httb!]
\centering \mbox{
\includegraphics[width=0.45\textwidth]{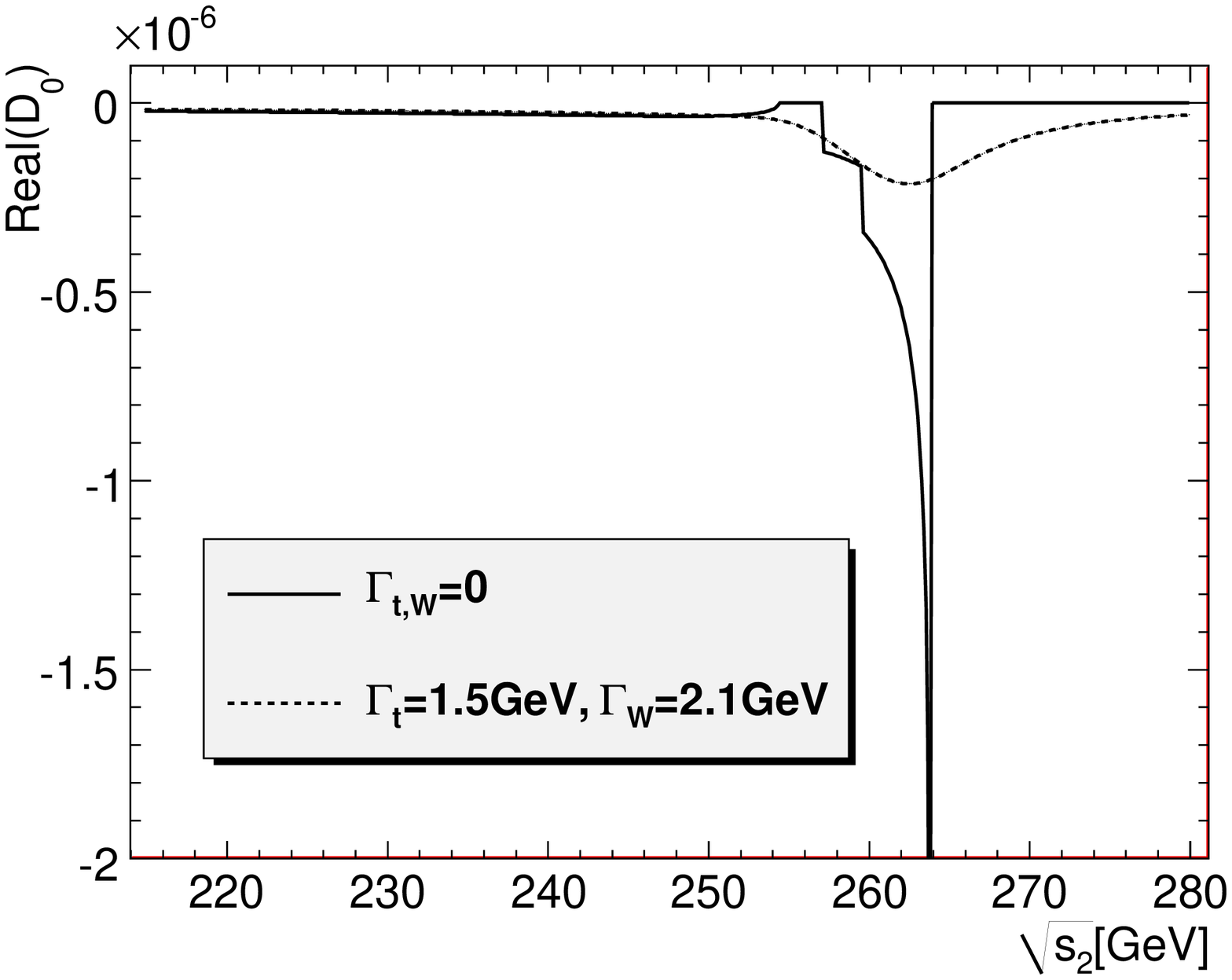}
\includegraphics[width=0.45\textwidth]{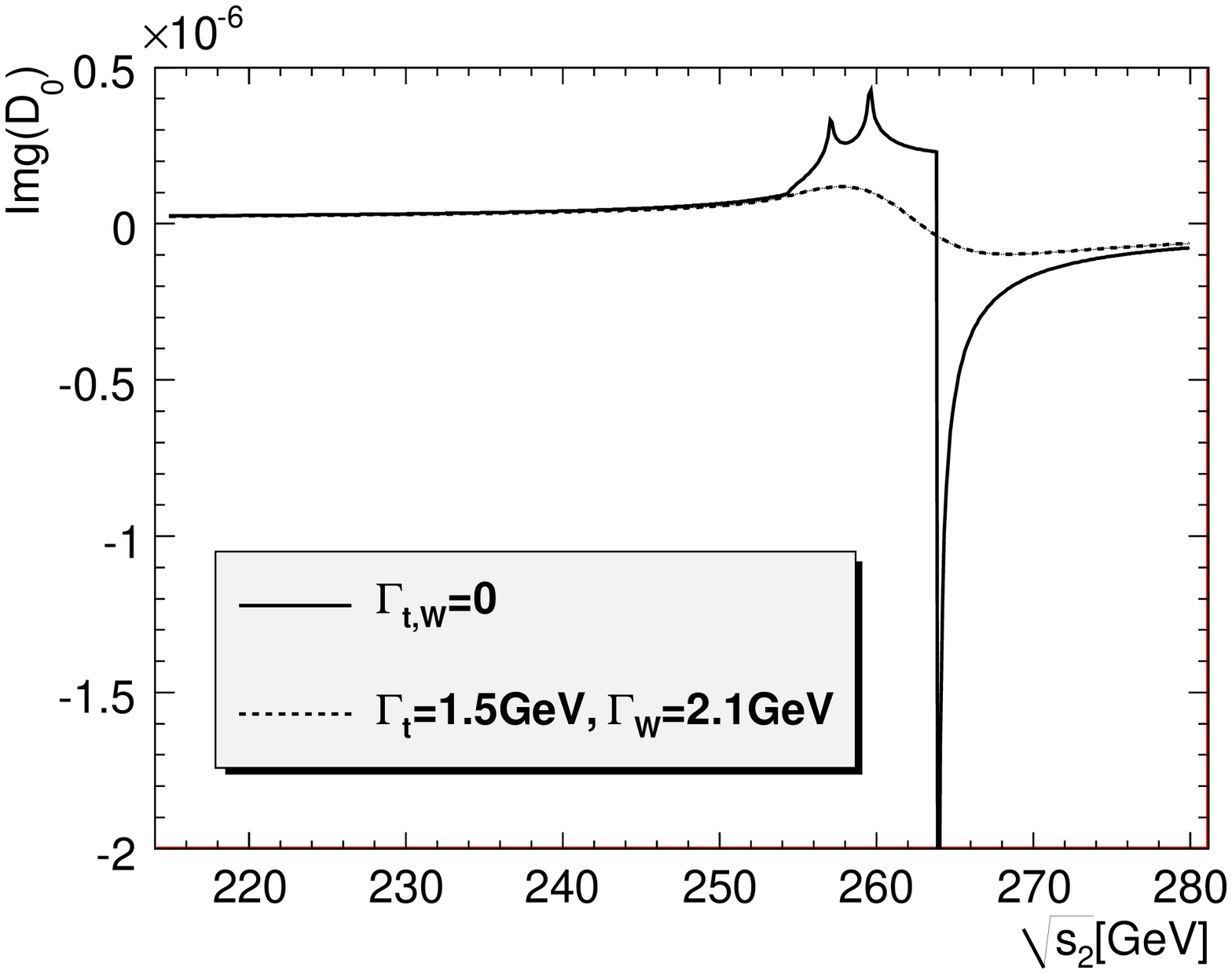}}
\caption{\textit{Effect of the  width of the $W$, $\Gamma_W$ and
of the top, $\Gamma_t$, on the real and imaginary part of the
four-point scalar function.}} \label{dessin-width-smooth}
\end{figure}
We take the simple prescription of a fixed width and make the
substitution
\bea
m_t^2\to m_t^2-im_t\Gamma_t,\hs M_W^2\to
M_W^2-iM_W\Gamma_W.\label{mass_real_complex}
\eea
Applied to the case of our four-point function one sees in
Fig.~\ref{dessin-width-smooth} that indeed the width regulates the
LLS and gives a smooth result that nicely interpolates with the
result at zero width away from the singularity. The normal
threshold and the 3-point sub-leading singularity are also
softened. The real part of the 4-point function still shows a
smooth valley at the location of the LLS after regularisation. For
the imaginary part we note that after introducing the width the
LLS singularity is drastically reduced with a contribution of the
order of the sub-leading singularity.  \\
\noi As we will explain in the next section and in more detail in
Appendix~B the introduction of the width in a four-point function
requires careful extension of the usual 4-point function
libraries. In the case at hand, as will be shown, the four-point function with complex internal masses can be written in an
analytical form, albeit with a larger number of Spence functions
compared to the case of real masses.

In our calculation of Yukawa corrections where all the
relevant couplings depend only on the top-quark mass, the Higgs mass
and the vacuum expectation value $\upsilon$, we will keep $m_t$,
$M_H$ and $\upsilon$ real while applying rules (\ref{mass_real_complex}) to all the loop
integrals.

One might ask whether the same prescription as in
Eq.~(\ref{mass_real_complex}) for the Higgs mass can be of any
relevance. A justification for this will require to consider the
corresponding process including the Higgs decays with among other
contributions, ``resonant contributions" with an integration over
the propagator of the Higgs. At least on a diagram by diagram
basis this will not solve the problem since for example one still
has to deal with the same 4-point function but with $M_H^2$
replaced by a certain $p_H^2$, taking into account the fact the
leading Landau singularity occurs for a wide range of Higgs masses
and values of the invariant $p_H^2$. On the practical side, recall
that compared to the top and $W$ width of about $2$GeV, for Higgs
masses of about $2M_W$ the width of the Higgs is $0.1$GeV, more
than an order of magnitude smaller\footnote{Our calculation of the
leading Yukawa effects involves the charged Goldstone boson in the
Feynman gauge through which the $W$ mass enters. One may question
whether it is appropriate to introduce a width here for a
Goldstone boson considering that a Goldstone is defined as a
massless state. Independently of the width one should first
question why the Goldstone has a mass here. The point is in any
other gauge than the Feynman gauge we would have had to consider
the effect of the Goldstone and $W$ exchange to derive the leading
Yukawa effects. The physical thresholds are therefore captured in
the Feynman gauge.}.
\section{Implementation of complex masses in the loop integrals}
We have implemented complex masses in all the loop integrals we
encounter in calculating the cross section in the limit
$\la_{bbH}=0$ where the tree-level prediction vanishes. In this
limit we can also set the mass of the bottom quark to zero. In the
$SU(3)$-gauge invariant classification of Fig~\ref{diag_3group},
class (a) vanishes in this approximation. In fact we had shown
\cite{fawzi_bbH} that even with $m_b=4.62$GeV class (a) is totally
negligible. Although it is only class (c) that shows severe
numerical instabilities due to the presence of a leading Landau
singularity in the 4-point box function and non-leading
singularity in the 5-point function we introduce the width in all
diagrams of both classes (b) and (c).

For the tensorial and scalar loop integrals with up-to three legs
we rely on {\tt LoopTools} \cite{looptools} which handles complex
masses in up to $3$-point functions. The $5$-point  functions are
reduced to $4$-point functions according to \cite{denner_5p,
looptools_5p}. The tensorial $4$-point functions are reduced to
the scalar $4$-point function and $3$-point functions. We
therefore have to calculate only the scalar $4$-point function
with complex masses. The analytical calculation of $4$-point
function with complex masses in the most general case is
practically intractable. If one of the external particles is
lightlike, the standard technique of 't Hooft and Veltman
\cite{hooft_velt} brings some light although the result writes in
terms of $72$ Spence functions. In our example, $gg\ra b \bar b H$
with massless bottom quarks, there are at least $2$ lightlike
external momenta in all boxes, including the ones derived from the
pentagon diagrams. If the positions, in the box, of two lightlike
momenta are opposite then we can write the result in terms of $32$
Spence functions. If the two lightlike momenta are adjacent, the
result contains $60$ Spence functions. The detailed derivation and
results are given in Appendix~\ref{appendix-box-integral}. We have
implemented those analytical formulae for the case of two massless
external momenta into a code and added this into {\tt LoopTools}
\footnote{The implementation for the case of one massless external
momentum is straightforward. However, we have not done this yet
since it is not necessary for our present calculation.}.

We have performed a variety of checks on the new loop integrals
with complex internal masses. First of all,  for {\em all} the
tensorial and scalar loop integrals ($4$- and $5$- point
functions), we have performed a trivial numerical consistency
check making sure that as the numerical value of the widths is
negligibly small, $widths\to 0^+$, one recovers the well tested
result with real internal masses. For the scalar loop integrals,
the results are compared to the ones calculated numerically in the
limit of large widths, {\it e.g.} $\Gamma_{t,W}=100$GeV, we find
an excellent agreement.  Furthermore, for the scalar box integrals
the results can be checked by using the segmentation technique
described in \cite{Boudjema:2005hb}. The idea is the following. At
the boundary of phase space where the Gram determinant vanishes,
the $4-$point function can be written as a sum of four 3-point
functions. The $3$-point functions with complex masses can be
calculated by using {\tt LoopTools}. In this way, we have verified
with excellent precision that the results of the scalar $4$-point
functions are correct at the boundary of phase space. We have also
carried out a comparison with a dedicated purely numerical
approach based on an extension of the extrapolation
technique\cite{grace-extrapolation}. We have found perfect
agreement\footnote{We thank F.~Yuasa for sending us the results of
the extrapolation technique.}.

In a second stage we have performed checks at the amplitude level.
A very trivial one was to check that the results with the new loop
library exactly match the ones with the standard loop library with
real masses in the limit  $widths\to 0^+$.  Another important
check was to verify that the results calculated with complex
internal masses are QCD gauge invariant, see \cite{fawzi_bbH} for
this check.

Since the leading Landau singularity is integrable at interference
level, the NLO calculation with $\la_{bbH}\neq 0$ performed in
\cite{fawzi_bbH} can be trivially extended to the region of
$M_H\ge 2M_W$ by using the same method without introducing widths
for unstable internal particles. However, there is a small problem
related to the universal correction $(\delta
Z_H^{1/2}-\delta\upsilon)$ where the wave function renormalisation
of the Higgs $\delta Z_H^{1/2}$ related to the derivative of the
Higgs two-point function becomes singular when $M_H$ equal to
$2M_W$ or $2M_Z$\cite{eezhsmrc}. We regularise this singularity by
separately introducing the widths of the $W$ and the $Z$. This
singularity, contrary to the leading Landau singularity, is due to
the Higgs being an external one-shell particle. Other ways for
dealing with this problem have been discussed\cite{Kniehl-wfrh}.
\section{Inputs parameters and kinematical cuts}
\label{section_inputs} The input parameters are the same as given
in \cite{fawzi_bbH}. We rewrite them here together with  new
inputs which are  the widths of the unstable particles appearing
in the calculation.
\bea
\alpha(0)&=&1/137.03599911, \hs \alpha_s(M_Z)=0.118, \crn
M_W&=&80.3766\GeV(G_{\mu}=1.16639\times 10^{-5}\GeV^{-2}), \hs
M_Z=91.1876\GeV, \crn
m_t&=&174.0\GeV,\hs \Gamma_{W}=2.1\GeV, \hs \Gamma_{Z}=2.4952\GeV,
\eea
the top-quark width is calculated at the tree level in the SM  as
\bea
\Gamma_t=\fr{G_\mu(m_t^2-M_W^2)^2(m_t^2+2M_W^2)}{8\pi\sqrt{2}m_t^3}\approx
1.5\GeV
\eea
where the bottom-quark mass has been neglected.  The
Cabibbo-Kobayashi-Maskawa parameter $V_{tb}$ is set to be $1$.
Most of our discussion concerns the most interesting case of the
limit $\lambda_{bbH} \ra 0$ where as we have discussed at length,
see also \cite{fawzi_bbH}, the effect of the $b$-quark mass other
than in the Higgs coupling is totally negligible. Therefore we set
this mass to zero when discussing this limit in
section~\ref{section_nnlo}. For completeness we will  also give
results for the NLO corrections in section~\ref{section_nlo} which
require $\lambda_{bbH} \neq 0$. There we will set $m_b=4.62$GeV.
When we refer to the leading order contribution we will have in
mind the cross section at the Born level calculated with
$m_b=4.62$GeV. The cross section from the one-loop amplitude
squared with $\lambda_{bbH} \ra 0$ will, in a few instances, be
normalised to  this Born cross section to give a measure of the
new electroweak effect and so as to allow comparison with the NLO
corrections.

We consider the case at the LHC where the $pp$ center of mass
energy  is $\sqrt{s}=14$TeV. Neglecting the small light quark
initiated contribution, see \cite{fawzi_bbH}, we use
CTEQ6L\cite{cteq6,cteq6_1,cteq6_2,cteq6_3} for the gluon density function in the proton.
The factorisation scale for this density and the energy scale for
the strong coupling constant are both chosen to be $Q=M_Z$ for
simplicity.

As has been done in previous
analyses~\cite{dawson_bbH,LH03_bbh,fawzi_bbH}, for the exclusive
$b\bar{b}H$ final state, we require the outgoing $b$ and $\bar{b}$
to have high transverse momenta $|\textbf{p}_{T}^{b,\bar{b}}|\ge
20$GeV and pseudo-rapidity $|\eta^{b,\bar{b}}|<2.5$. These
kinematical cuts reduce the total rate of the signal but also
greatly reduce the QCD background. As pointed
in~\cite{dittmaier_bbH} these cuts also stabilise the scale
dependence of the QCD NLO corrections compared to the case where
no cut is applied. In the following, these kinematical cuts are
always applied.
\section{Results in the limit of vanishing $\la_{bbH}$}
\label{section_nnlo}
\subsection{Total cross section}

We start with the cross section in the case where $\la_{bbH}=0$.
In \cite{fawzi_bbH} we reported on  results up to $M_H=150$GeV
that showed that this cross section was  rising fast as one
approached the threshold $M_H=2M_W$. Beyond this threshold our
integrated cross sections showed large instabilities. As we
discussed in section~\ref{section_landau} this is due to the
appearance of a leading singularity which as we have advocated can
be cured by the introduction of a width for the unstable top quark and
$W$ gauge boson. We also showed in section~\ref{section_landau} that the
region of Landau singularity spans the region $2M_W\le M_H\le
211$GeV with $2 m_t <\sqrt{s_{gg}}=\sqrt{s}\le 457$\;GeV, see
Fig.~\ref{landau_range}. Before convoluting with the gluon
distribution let us briefly look at the behaviour of the partonic
cross section $gg \ra b \bar b H$ paying a particular attention to
this leading Landau singularity region.

\begin{figure}[h]
\begin{center}
\mbox{\includegraphics[width=0.45\textwidth]{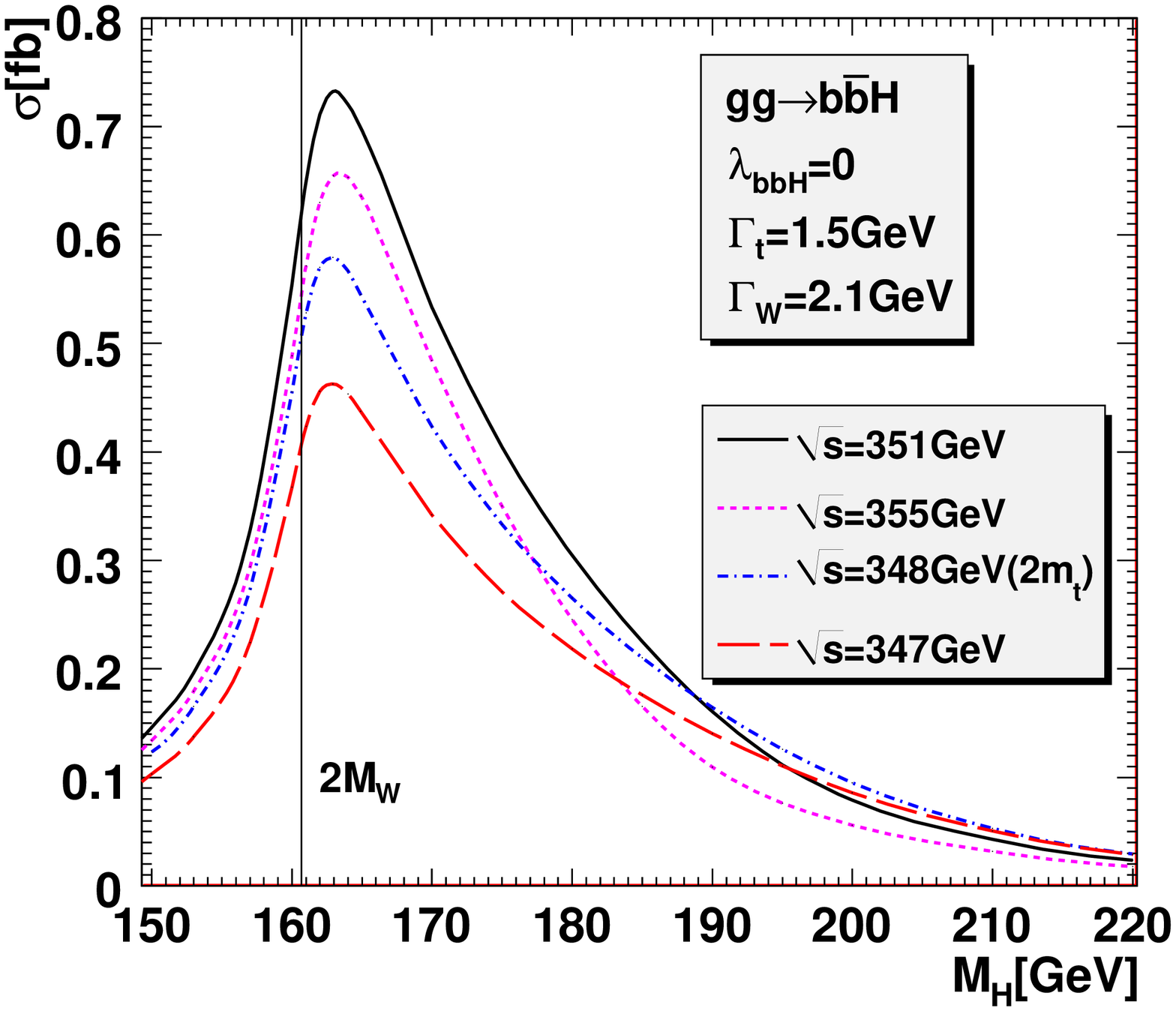}
\hspace*{0.075\textwidth}
\includegraphics[width=0.45\textwidth]{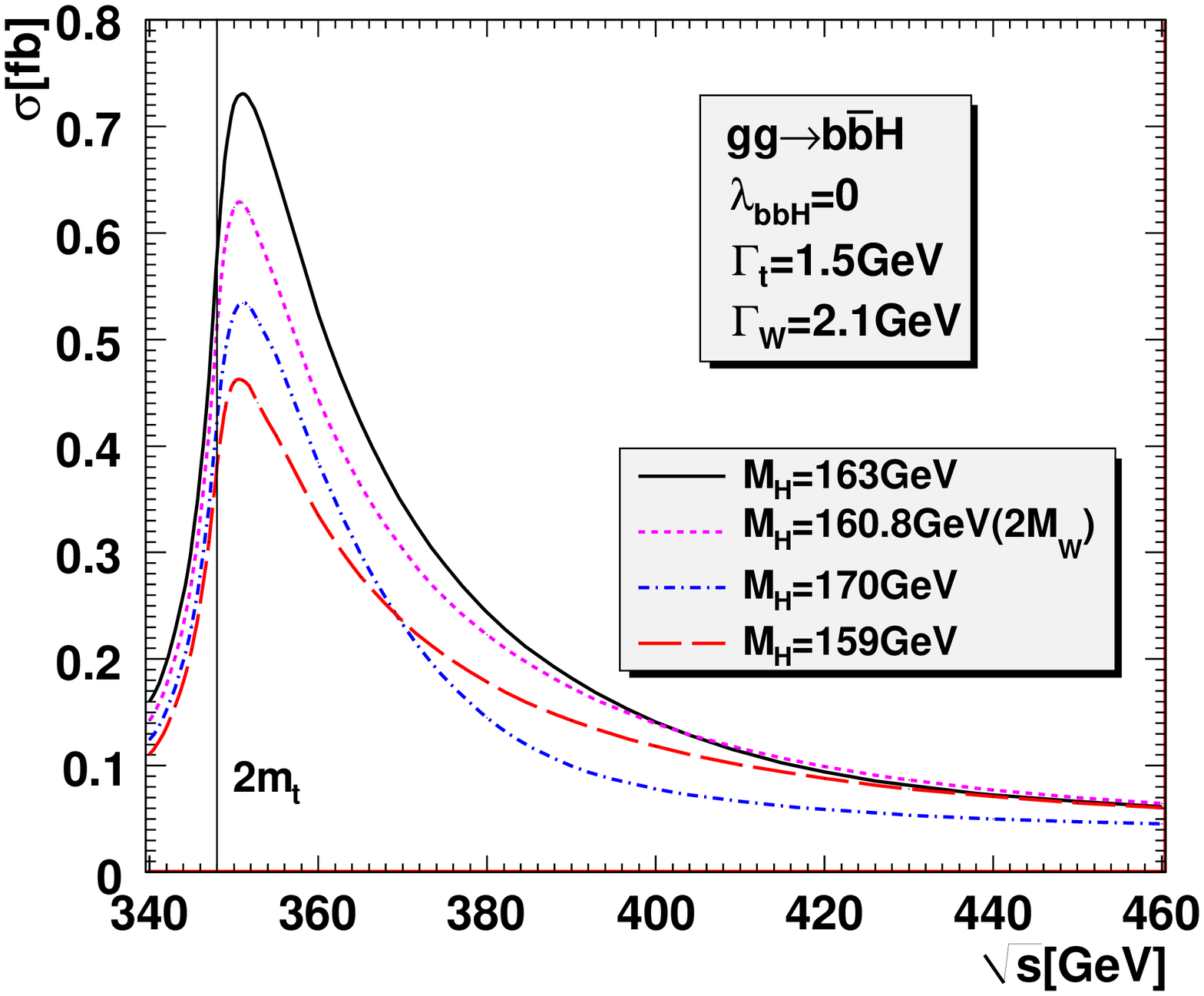}}
\caption{\label{gg_LL_rs}{\em Left: the cross section for the
subprocess $gg\to b\bar{b}H$ as functions of $M_H$ for various
values of $\sqrt{s}$ including the case $\sqrt{s}=2m_t=348GeV$.
Right: the cross section for the subprocess $gg\to b\bar{b}H$ as
functions of $\sqrt{s}$ for various values of $M_H$ including the
case $M_H=2M_W=160.7532GeV$.}}
\end{center}
\end{figure}

Figs~\ref{gg_LL_rs} show that  indeed the widths do regulate the
cross section. Moreover it is within this range that the cross
section is largest even after being regulated. The (highest) peak
of the cross section occurs for a Higgs mass of $163$GeV about
$\Gamma_W$ above the $M_H=2M_W$ threshold and for
$\sqrt{s}=351$GeV about $2\Gamma_t$ above the $\sqrt{s}=2m_t$
threshold. Figs.~\ref{gg_LL_rs} show that the cross section
exhibits a peak structure close to the onset of the normal
thresholds in $M_H, \sqrt{s}$ even when one is slightly outside
the leading Landau singularity region of the 4-point function. In
fact, this enhancement at the normal threshold is far from being
totally due the 4-point LLS especially after the latter has been
regularised by the introduction of the width. At the normal
threshold there is an enhancement from the accumulation of all the
$2$-point, $3$-point and of course the $4$-point function.
Moreover as we noted in section~\ref{width-reg}, see
Fig.~\ref{dessin-width-smooth}, the introduction of the widths
brings the contribution of the LLS to the level of a sub-leading
singularity.

\begin{figure}[h]
\begin{center}
\mbox{\includegraphics[width=0.45\textwidth]{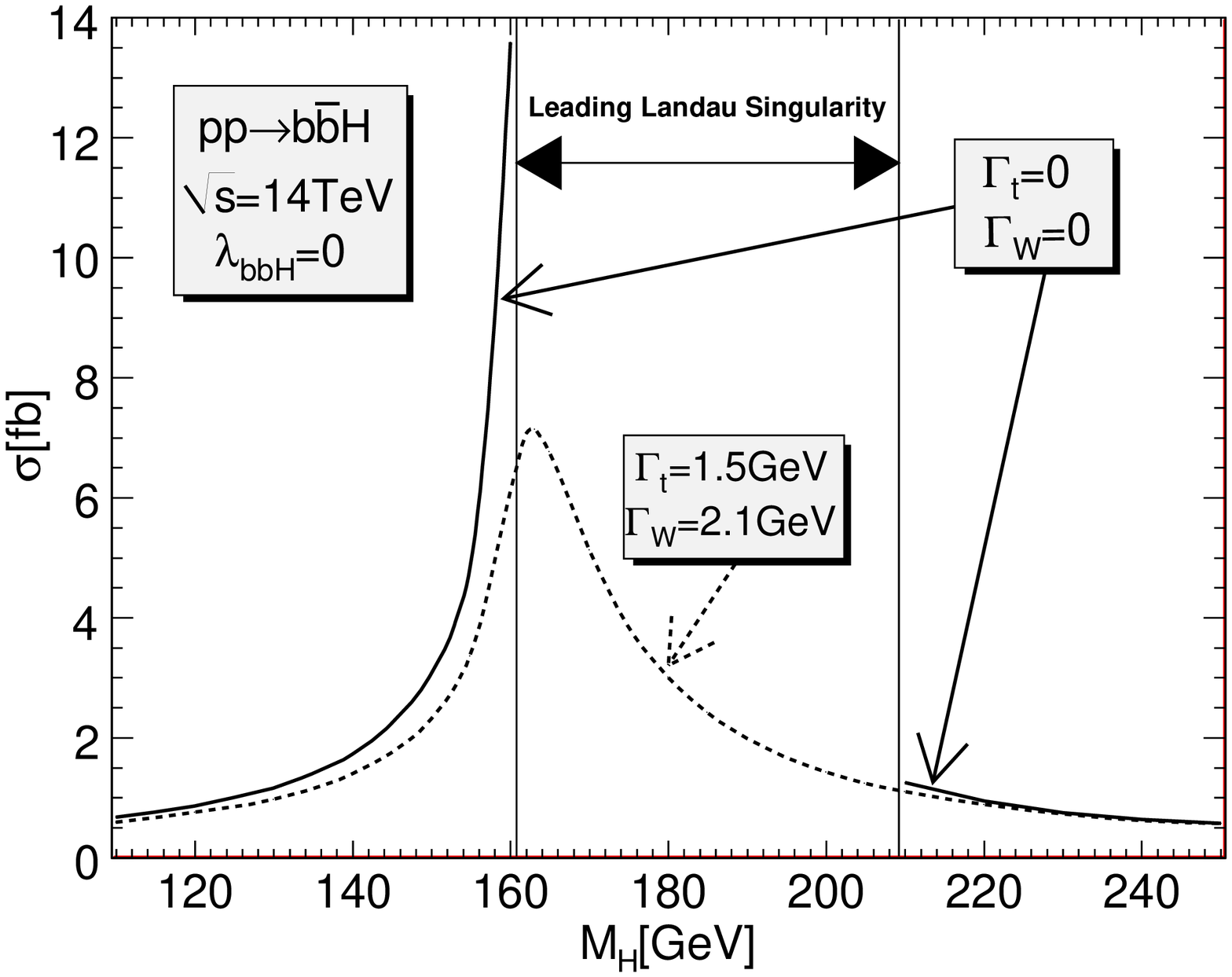}
\hspace*{0.075\textwidth}
\includegraphics[width=0.45\textwidth]{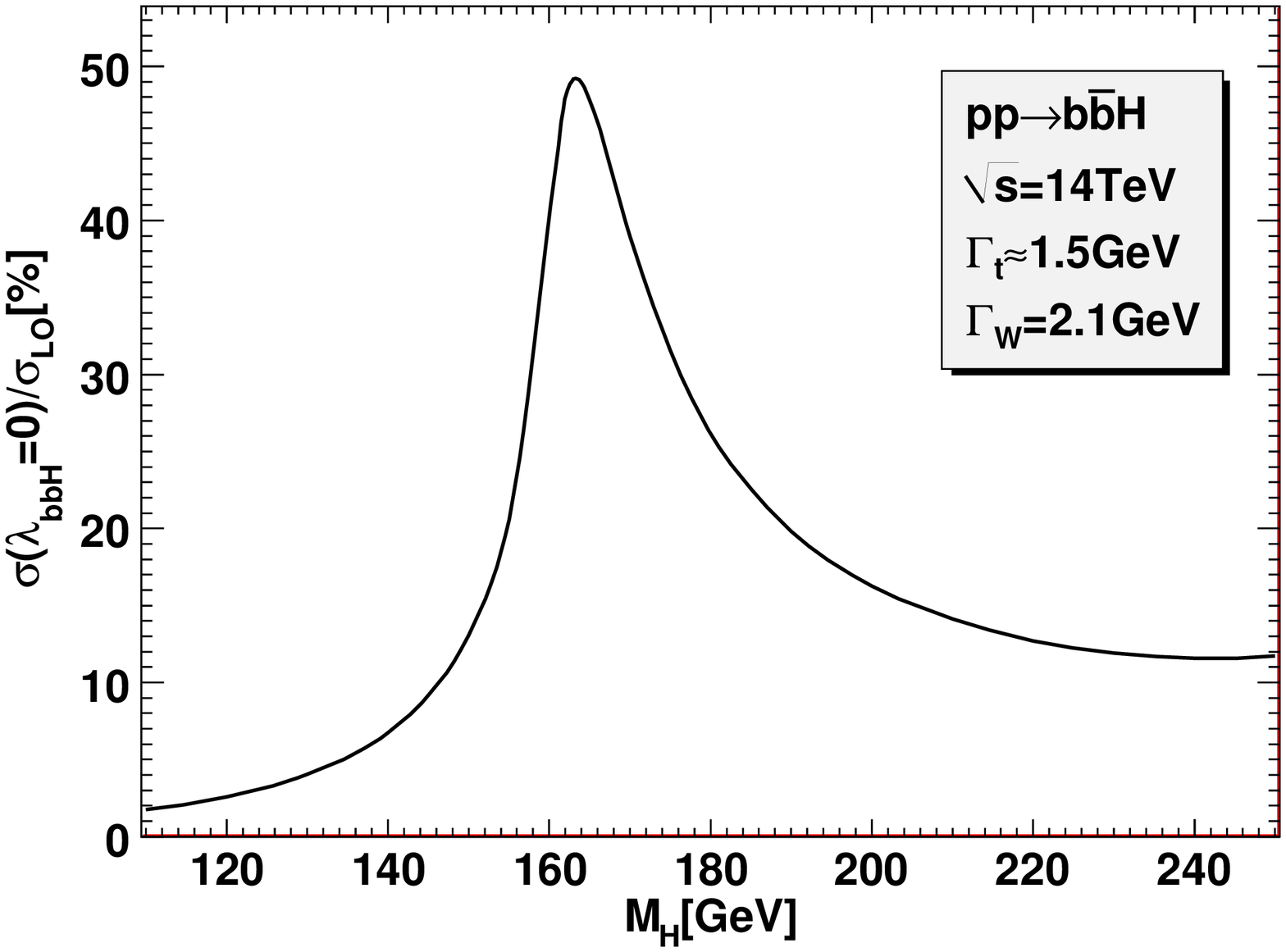}}
\caption{\label{p_LL_mH}{\em Left: the one-loop induced cross
section as a function of $M_H$ in the limit of vanishing
bottom-Higgs Yukawa coupling for two cases: with and without
widths. Right: the percentage correction of the contribution with
widths relative to the tree level cross section calculated with
$\la_{bbH}\neq 0$.}}
\end{center}
\end{figure}

The cross section at the $pp$ level for the $14$TeV centre of
mass energy at the LHC as a function of the Higgs mass is shown in
Fig.~\ref{p_LL_mH} taking into account the width of the top quark and
the $W$ gauge boson.
For comparison we also show the cross section without the width
effect outside the leading Landau singularity range of $M_H$. The
sharp rise above $M_H>150$GeV is nicely tamed. On the other hand
note that on leaving the leading Landau singularity region around
$M_H=211$GeV, the width effect is much smaller and the figures
suggest that one could have ``entered this region from the right"
without having recourse to introducing a width. Indeed our
numerical integration routine over phase space  with the default
{\tt LoopTools} library does not show as bad behaviour until we
venture around values of $2M_W\le M_H<200$GeV. The reason for this
can be understood by taking a glance at Fig.~\ref{landau_range}.
For $200 {\rm GeV}<M_H<211 {\rm GeV}$ the singularity region is
considerably shrunk to a line so that one is integrating over an almost zero
measure. The effect of the widths outside the singularity region
is to reduce the cross section for $M_H=$ $120$GeV, $140$GeV and
$150$GeV by respectively $15\%$, $24\%$ and $33\%$ while for
$M_H=$ $210$GeV, $230$GeV and $250$GeV the reduction is
comparatively more modest with respectively
$15\%$, $5\%$ and $2\%$.

Normalised to the Born cross section the new contribution
represents a mere $2.6\%$ for $M_H=120$GeV. It increases however
to as much as $49\%$ for $M_H=163$GeV before stabilizing to about
$10\%$ for larger Higgs masses.

\subsection{Distributions}
\begin{figure}[hp]
\begin{center}
\mbox{\includegraphics[width=0.45\textwidth]{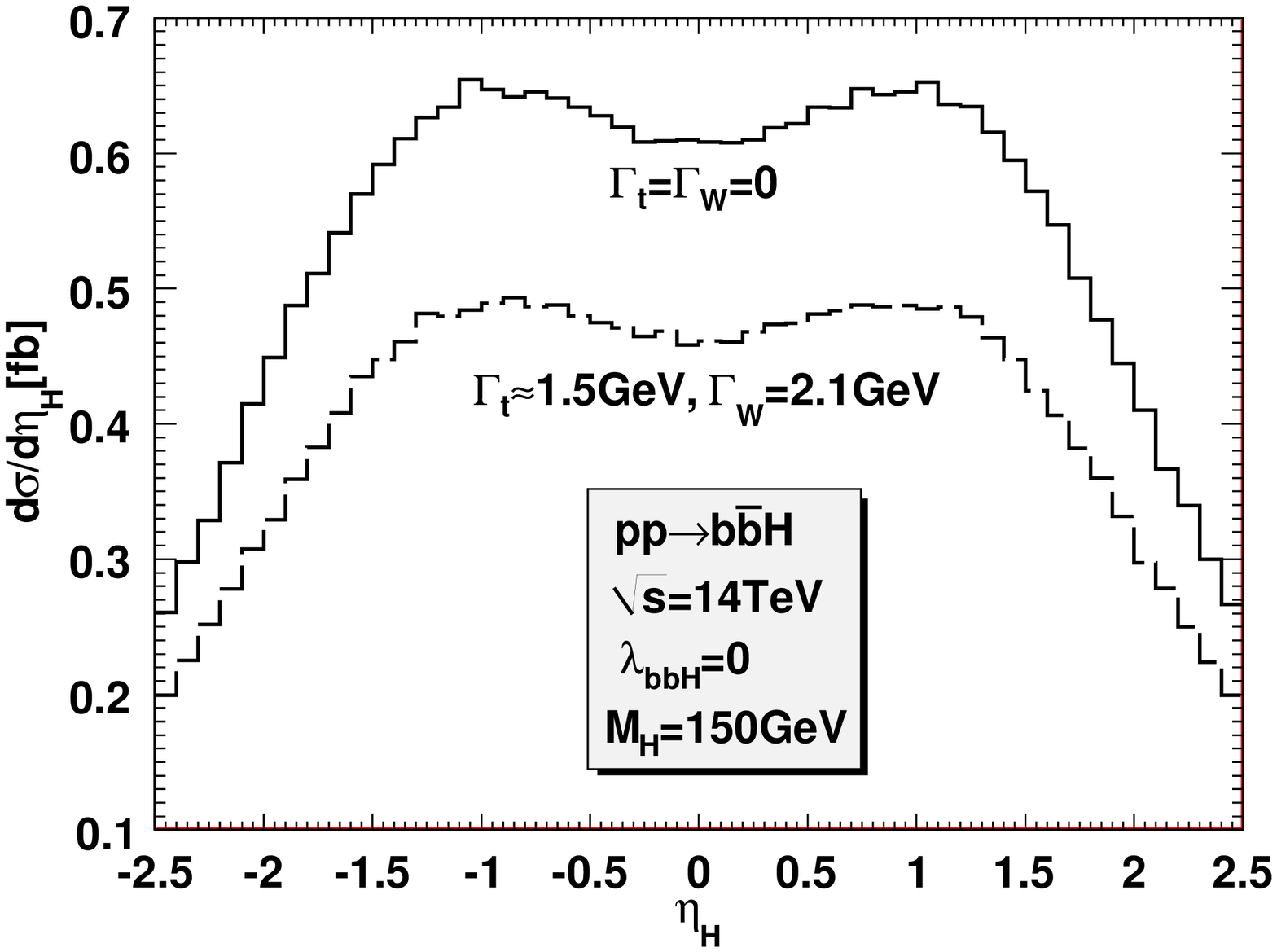}
\hspace*{0.075\textwidth}
\includegraphics[width=0.45\textwidth]{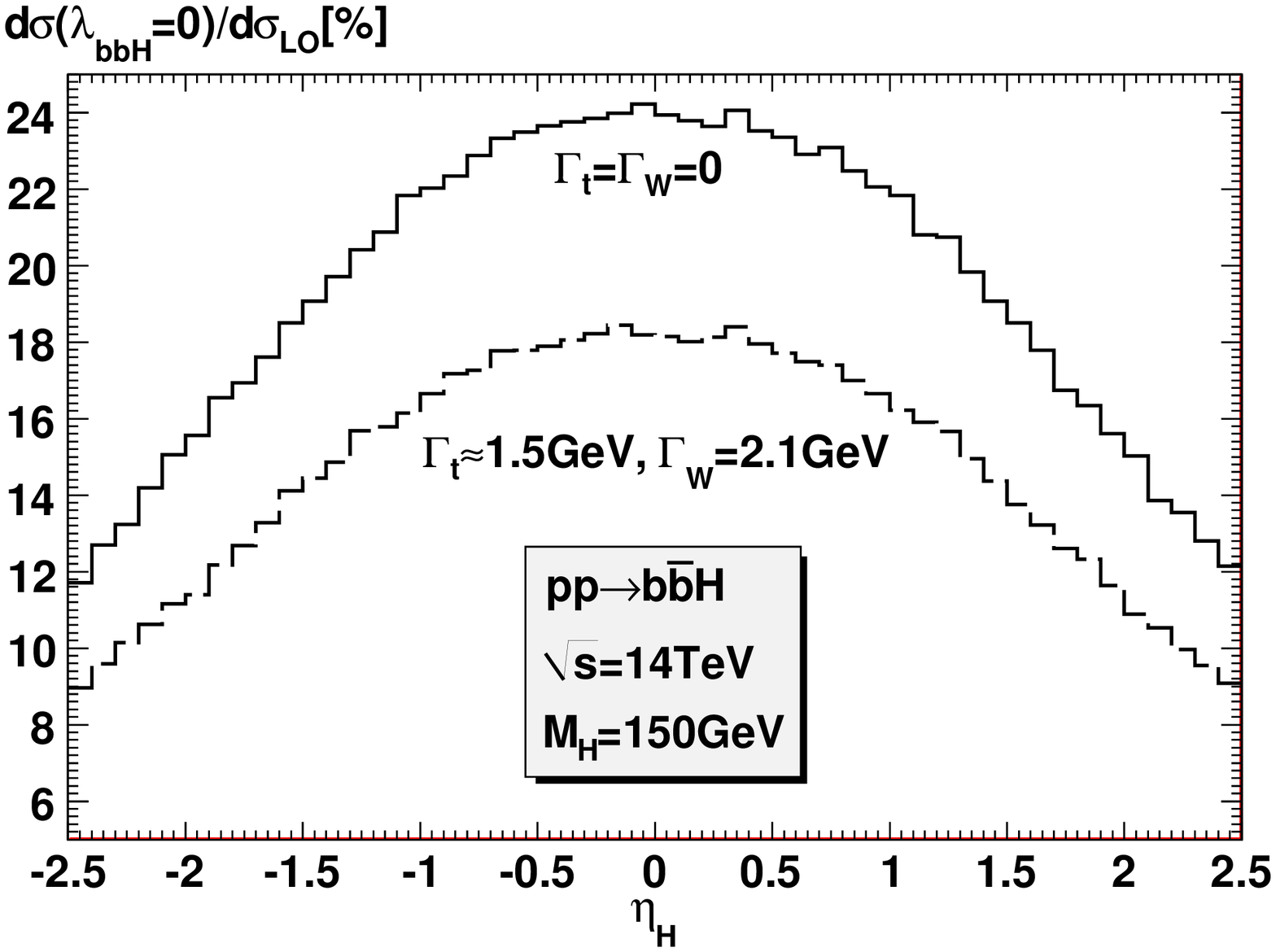}}
\mbox{\includegraphics[width=0.45\textwidth]{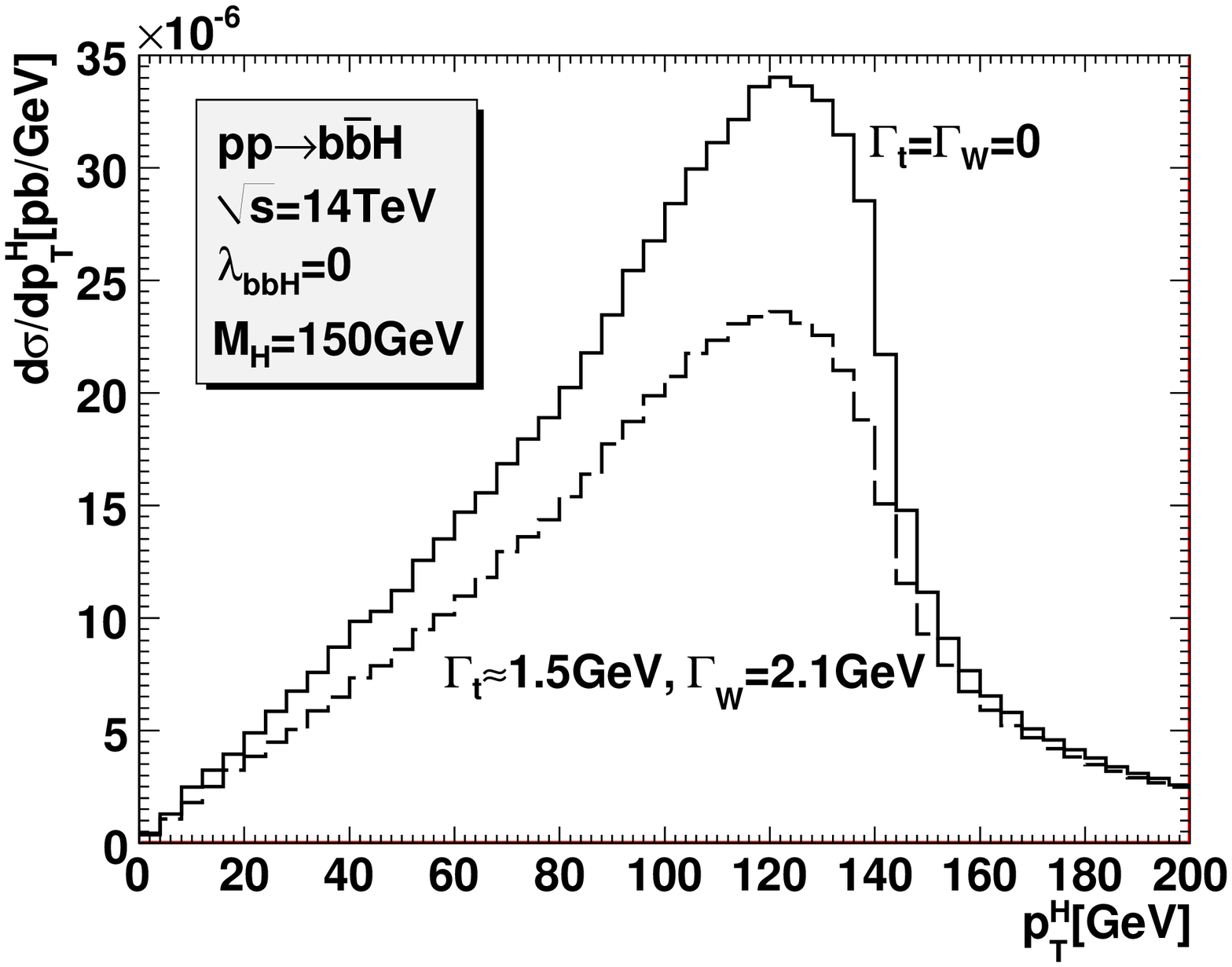}
\hspace*{0.075\textwidth}
\includegraphics[width=0.45\textwidth]{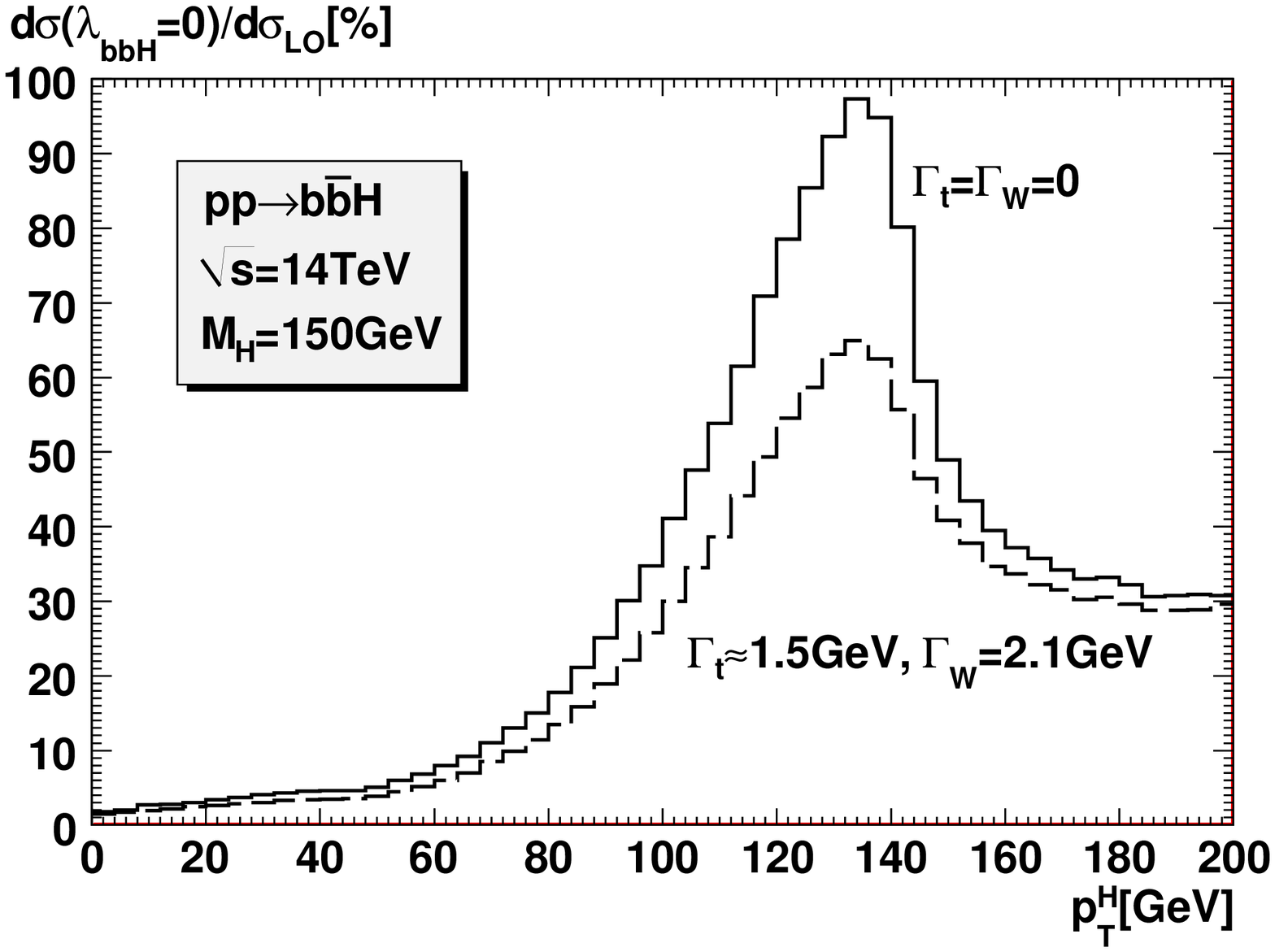}}
\mbox{\includegraphics[width=0.45\textwidth]{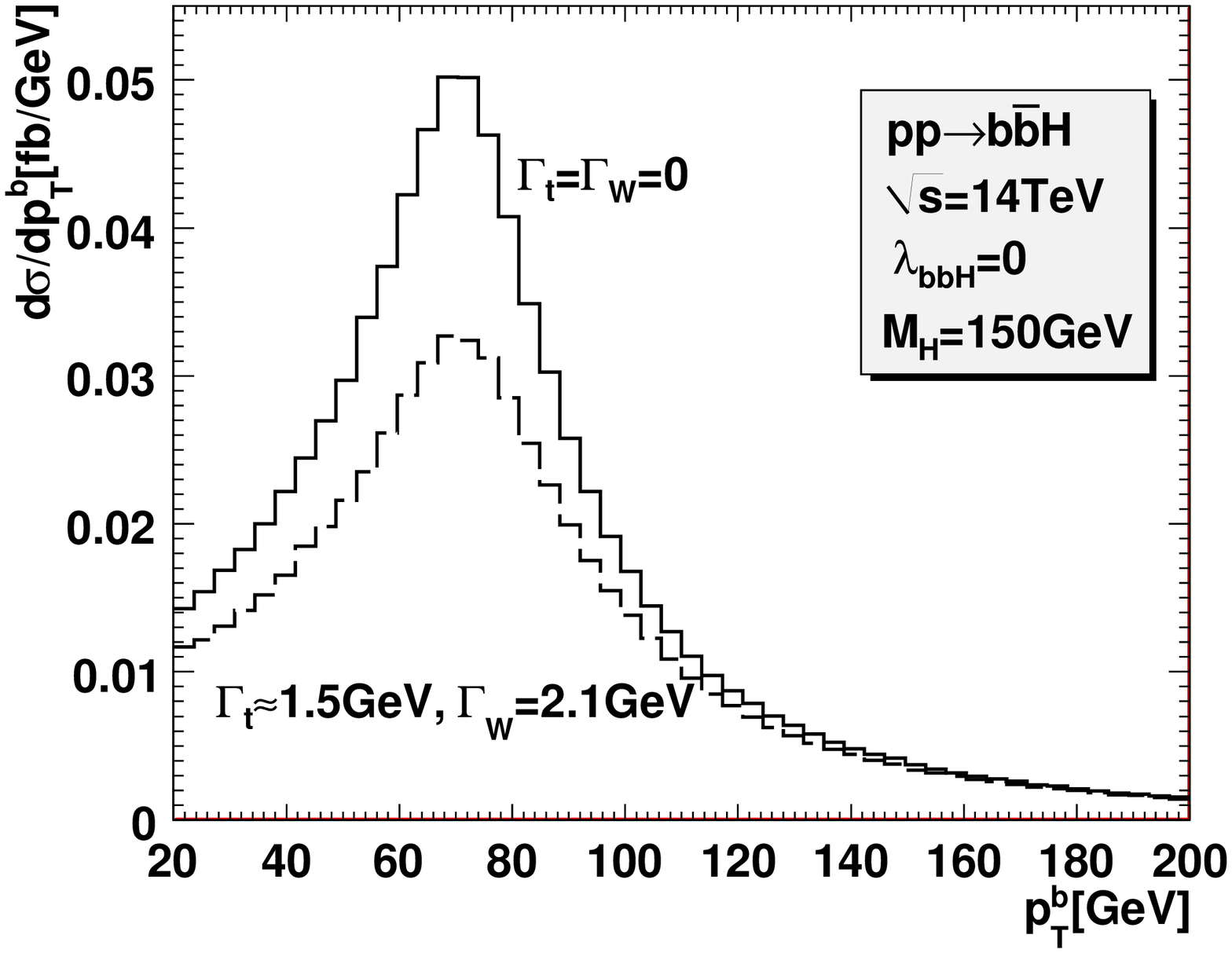}
\hspace*{0.075\textwidth}
\includegraphics[width=0.45\textwidth]{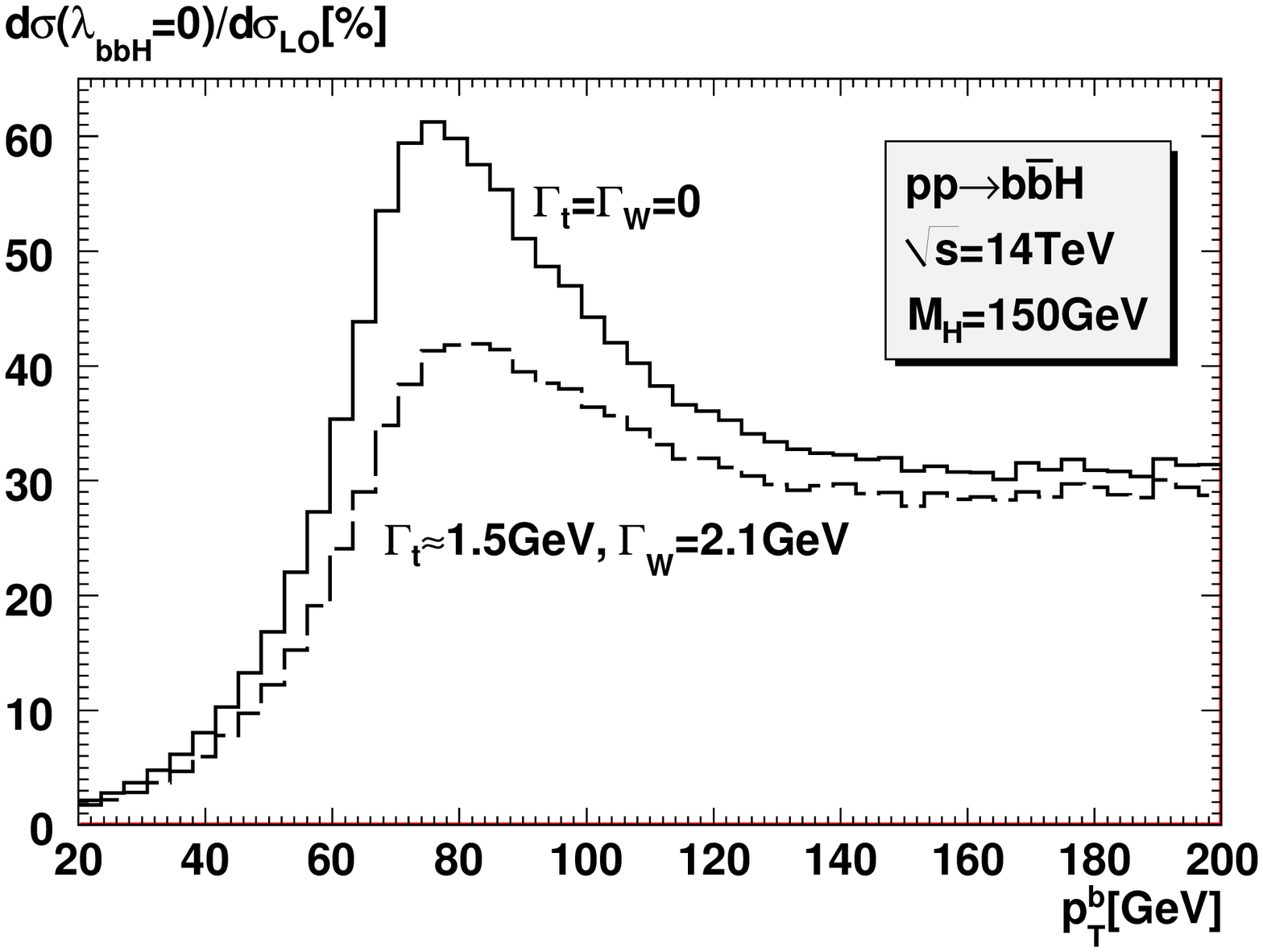}}
\caption{\label{p_LL_histo_mH150}{\em The pseudo-rapidity of the
Higgs and transverse momentum distributions of the Higgs and the
bottom for $M_H=150$GeV arising from the purely one-loop
contribution in the limit of vanishing LO ($\la_{bbH}=0$) for two
cases: with and without widths. The relative percentage
contribution $d\sigma(\lambda_{bbH}=0)/d\sigma_{LO}$ is also
shown.}}
\end{center}
\end{figure}

Effects of the new purely one-loop contribution being as large as
$\sim 50\%$, compared to the Born cross section even after being
regulated through the introduction of the widths, it is essential
that one looks at different distributions to see if this new
effect can be described as a simple $K$-factor. The two examples
we show for $M_H=150$GeV (before the onset of the leading Landau
singularity) and for the $M_H=163$GeV where the effect on the
total cross section are largest show that the corrections are not
uniformly distributed for all distributions. Figures
\ref{p_LL_histo_mH150} for  $M_H=150$GeV show the effect of the
width. While the relative difference is rather uniform, about
$33\%$, on the Higgs pseudorapidity, $\eta_H$, distribution, the
transverse momentum distributions of the Higgs, $p_T^H$, and the
bottom $p_T^b$ are strongly affected in particular for values
which in the absence of the width showed a peak structure. There
is still some peak structure in the $p_T$ distributions but the
width effect reduces this by as much as $50\%$, while in the tails
it is about $10\%$.

Let us now turn to $M_H=163$GeV.  The correction, normalised to
the Born cross section, for the Higgs pseudorapidity distribution
is about $60\%$ around the center region. The corrections to the
$p_T$ distributions can be enormous in some regions of phase
space, up to $200\%$ for the Higgs and about $170\%$ for the
bottom quark case. These huge corrections to the distributions in
some region of phase space are again due to the effect of Landau
singularities.

\begin{figure}[hp]
\begin{center}
\mbox{\includegraphics[width=0.45\textwidth]{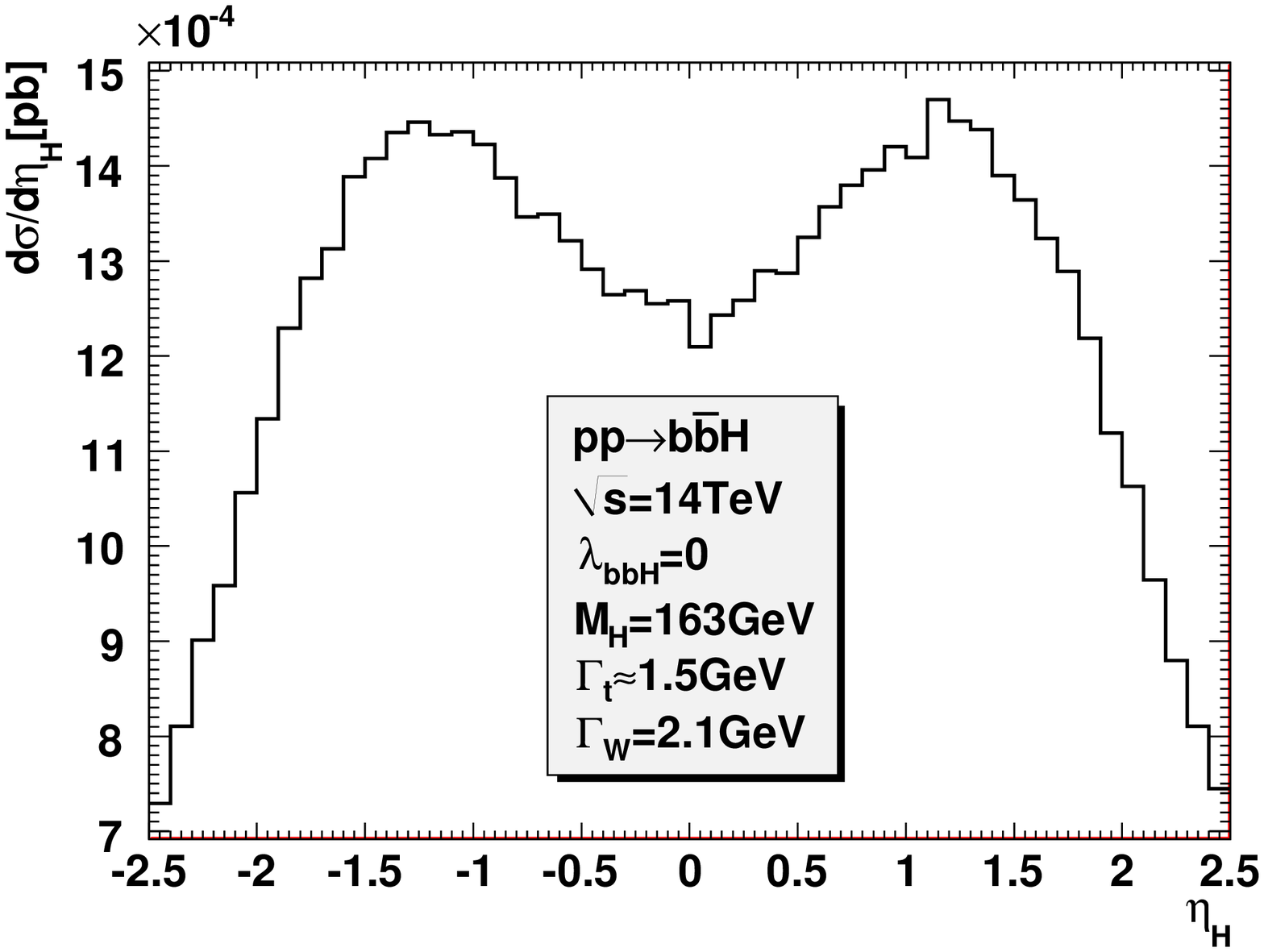}
\hspace*{0.075\textwidth}
\includegraphics[width=0.45\textwidth]{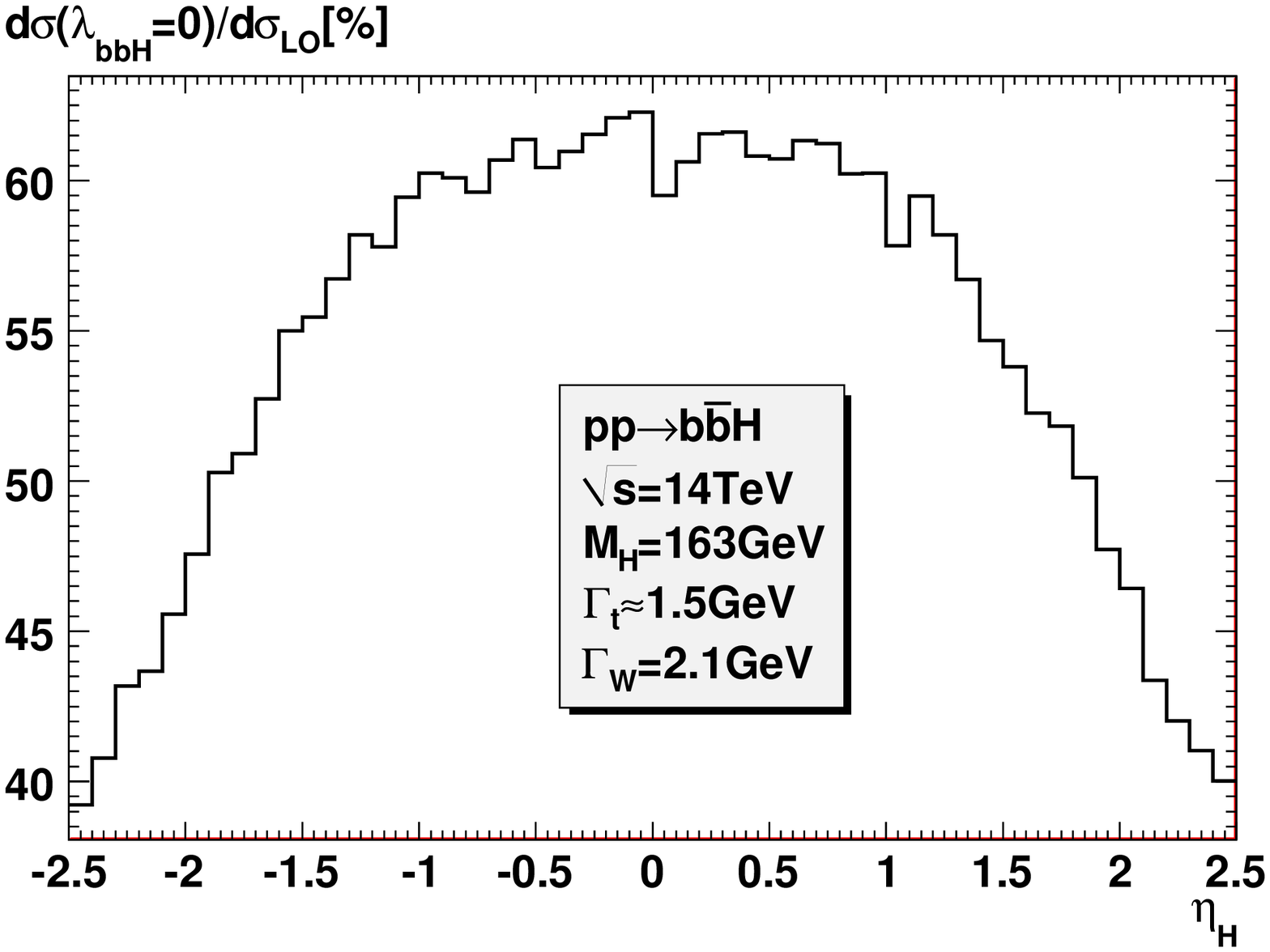}}
\mbox{\includegraphics[width=0.45\textwidth]{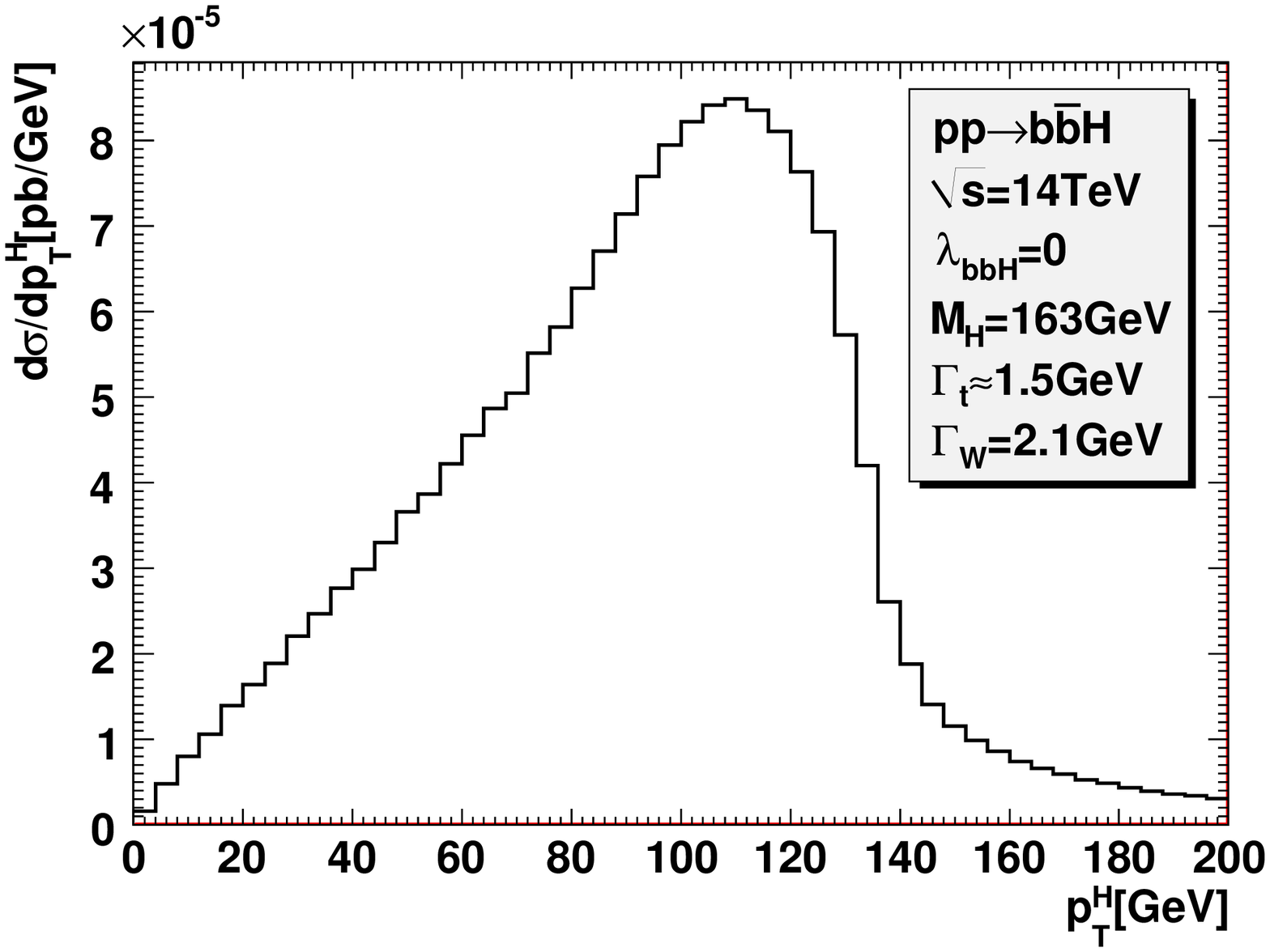}
\hspace*{0.075\textwidth}
\includegraphics[width=0.45\textwidth]{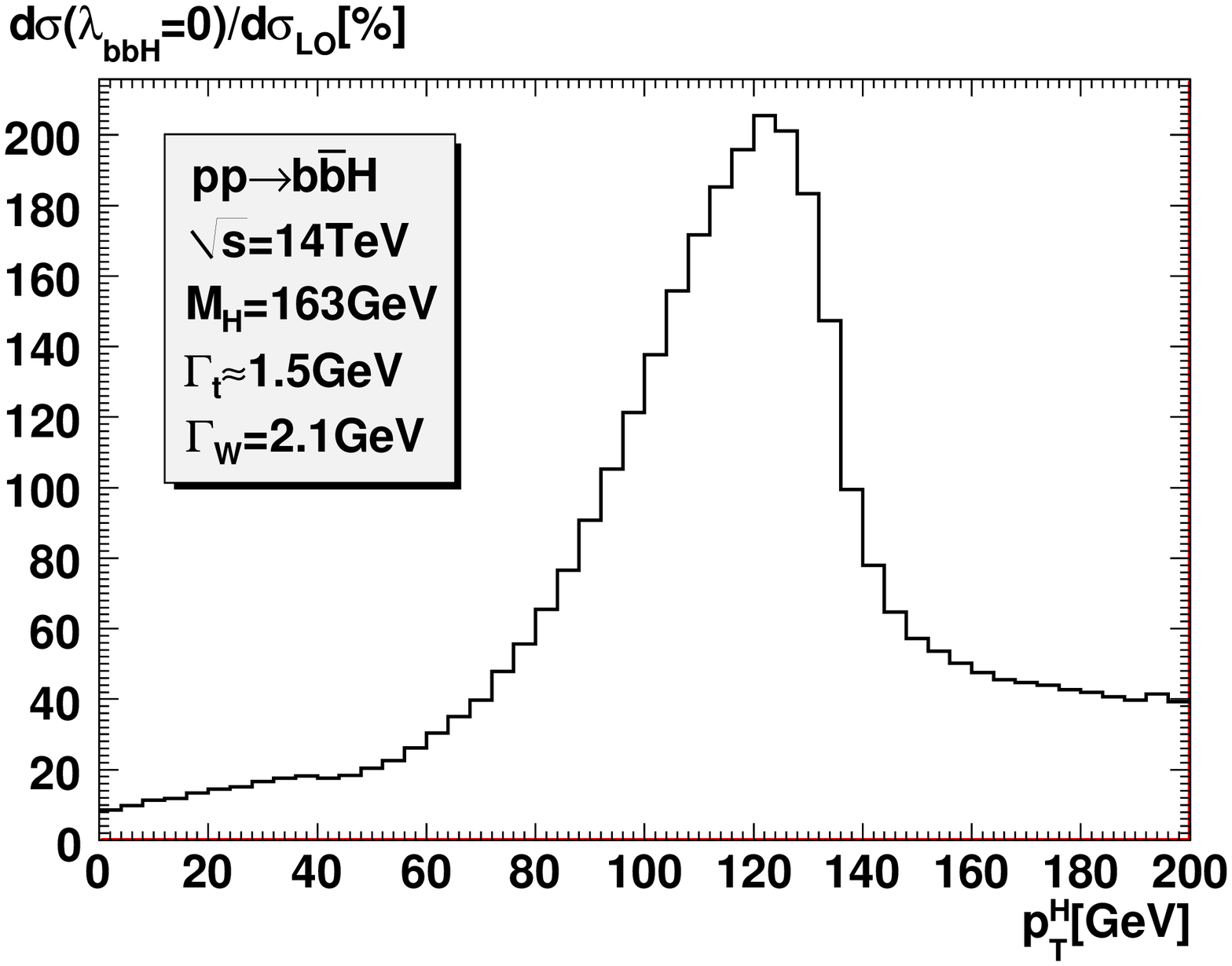}}
\mbox{\includegraphics[width=0.45\textwidth]{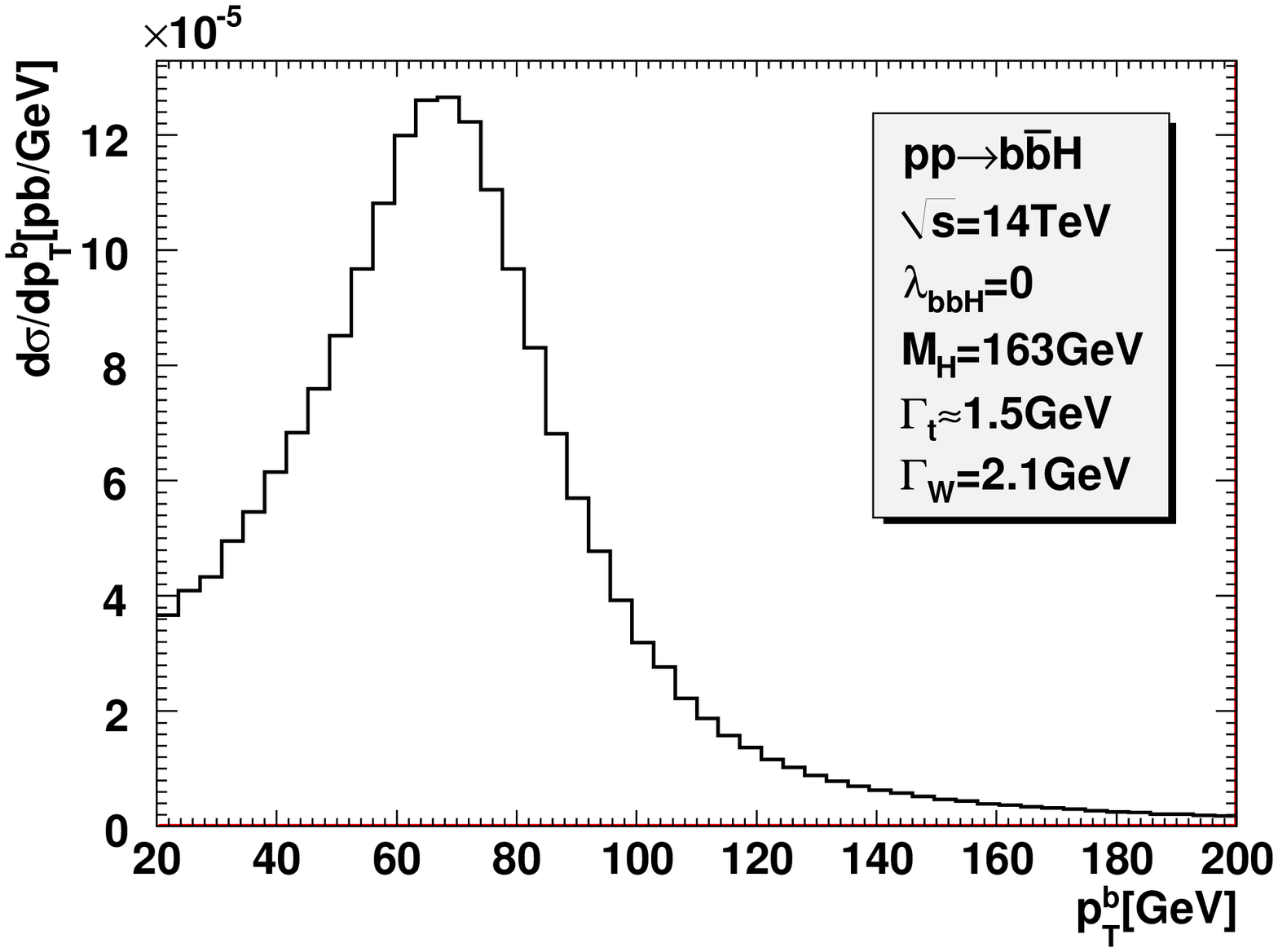}
\hspace*{0.075\textwidth}
\includegraphics[width=0.45\textwidth]{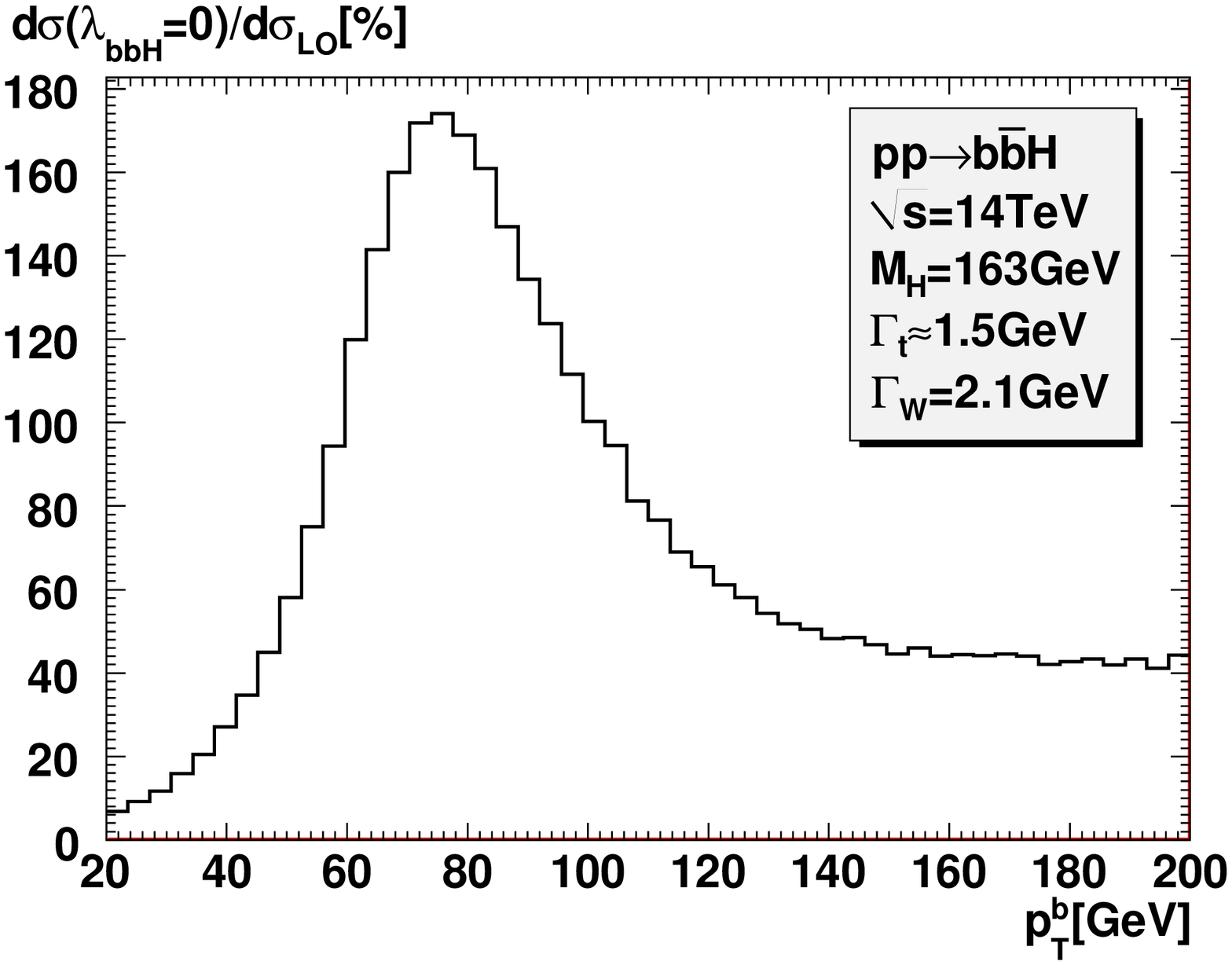}}
\caption{\label{p_LL_histo_mH163}{\em The pseudo-rapidity of the
Higgs and transverse momentum distributions of the Higgs and the
bottom for $M_H=163$GeV arising from the purely one-loop
contribution in the limit of vanishing LO ($\la_{bbH}=0$). Its
relative percentage contribution
$d\sigma(\lambda_{bbH}=0)/d\sigma_{LO}$ is also shown.}}
\end{center}
\end{figure}

One may question whether these large corrections signal the
breakdown of perturbation theory and whether one expects (even)
higher order effects to be large. We do not think so. First of all
the relative large corrections have to do with the fact that for
vanishing $\lambda_{bbH}$ the tree-level cross section vanishes. Second,
higher order effects have been captured in the introduction of the
width and there is no reason to suspect that the leading Landau
singularity we have encountered is affected by higher order
effects.
\section{Results at NLO  with $\la_{bbH}\neq 0$}
\label{section_nlo}

The results of the electroweak corrections at NLO which represent
the interference contribution between the Born and the one-loop
amplitude are much less interesting and numerically quite small, a
trend that we had found already when studying at some length the
electroweak NLO for $M_H<150$GeV\cite{fawzi_bbH}. Moreover
although some one-loop diagrams contain a leading Landau
singularity at the interference level this singularity as we have
shown in section~\ref{section_landau} is integrable, see
Eq.~(\ref{eq_T04h}). The NLO contribution, apart from the Higgs
wave-function renormalisation  effect, is numerically stable even
if one does not implement widths of the internal particles. The
purpose of this section is to briefly present the results for the
NLO. We first show that the effect of introducing the width is
very small then show the NLO result without the internal widths
being implemented hence these results are genuinely NLO results.
These results thus complement the study we made for
$M_H<150$GeV\cite{fawzi_bbH}.

As discussed in section~\ref{section_general} the NLO Yukawa
corrections consist of 3 QCD gauge invariant classes, see
Fig.~\ref{diag_3group}. Class (a) gives a totally negligible
correction below $0.1\%$. We will not discuss this contribution
any further here. Moreover, the leading Landau singularity we have
discussed only shows up in class (c). As a first step we therefore
study the NLO correction due to class (c) and weigh the effect of
implementing the width of the internal particles. Class (b) does
not develop a leading Landau singularity and therefore the widths
effects will be marginal. \\
Another correction with enhanced Yukawa coupling is  the universal
correction, $(\delta Z_{H}^{1/2}-\delta\upsilon)$ where $\delta
Z_{H}^{1/2}$, the Higgs wave-function renormalisation constant
involving the derivative of the two-point function Higgs
self-energy. The latter is ill-defined when $M_H$ is equal to
$2M_W$ or $2M_Z$. Here the width of all unstable particles,
$W,Z,t$, will be kept \footnote{Note that $\delta Z_{H}^{1/2}$
does not diverge when $M_H=2m_t$ and the top-quark width thus has
a marginal effect on $\delta Z_{H}^{1/2}$.}.

\subsection{Width effect at NLO}
\begin{figure}[th]
\begin{center}
\includegraphics[width=0.75\textwidth]{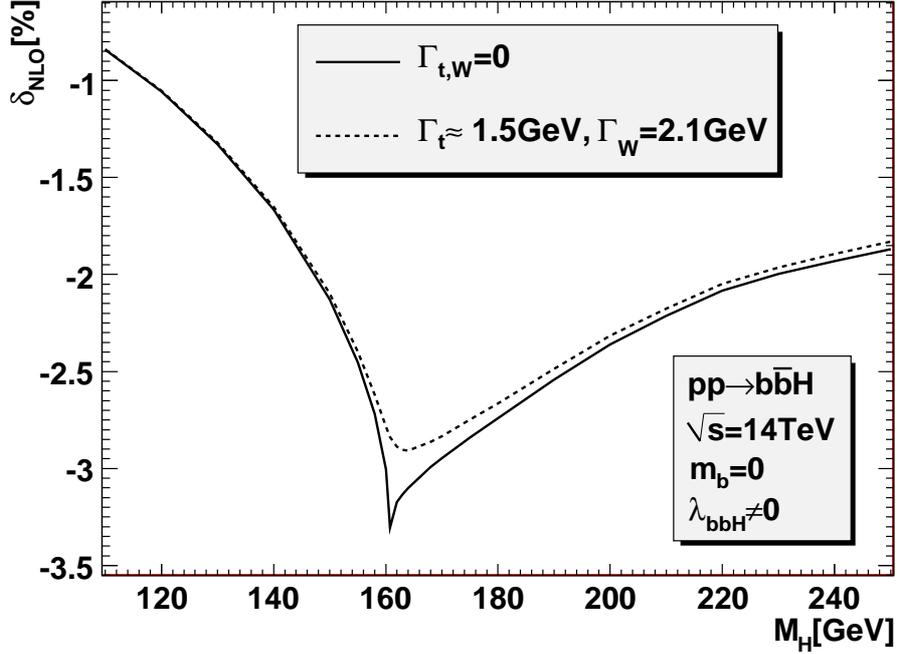}
\caption{\label{pp_Loop_mb0_mH}{\em Contribution of class (c) at
NLO. The one-loop amplitude is calculated by keeping $m_b$ only in
the $\la_{bbH}$ coupling.}}
\end{center}
\end{figure}

Our implementation of the width in the four-point function has
been done in the limit of massless external quarks. To be fully
consistent in the calculation of the one-loop amplitude with
widths using the modified 4-point function we switch off the
bottom mass in the spinors and propagators but keep
$\lambda_{bbH}\neq 0$ as an independent parameter. Our results for
the NLO contribution of class (c) is shown  in
Fig.~\ref{pp_Loop_mb0_mH}. First of all as we can see the overall
correction is quite small, even at the onset of the (integrable)
leading Landau singularity, the correction to the Born term is
below $3.5\%$. The existence of a dip at the expected location is
noticeable. Width effect softens the dip behaviour somehow but the
effect is not as dramatic as what we have seen in the previous
section for the loop squared results. We find that if $M_H<
158$GeV or $M_H>165$GeV then the width effect change the NLO
result but not more than $5\%$ and are therefore totally
negligible especially if one takes into account the smallness of
the NLO result itself.
Therefore the full NLO results
can be studied by safely neglecting the width effect in classes
(b) and (c).

\subsection{NLO corrections with $m_b \neq 0$}
\begin{figure}[htbp]
\begin{center}
\mbox{\includegraphics[width=0.45\textwidth]{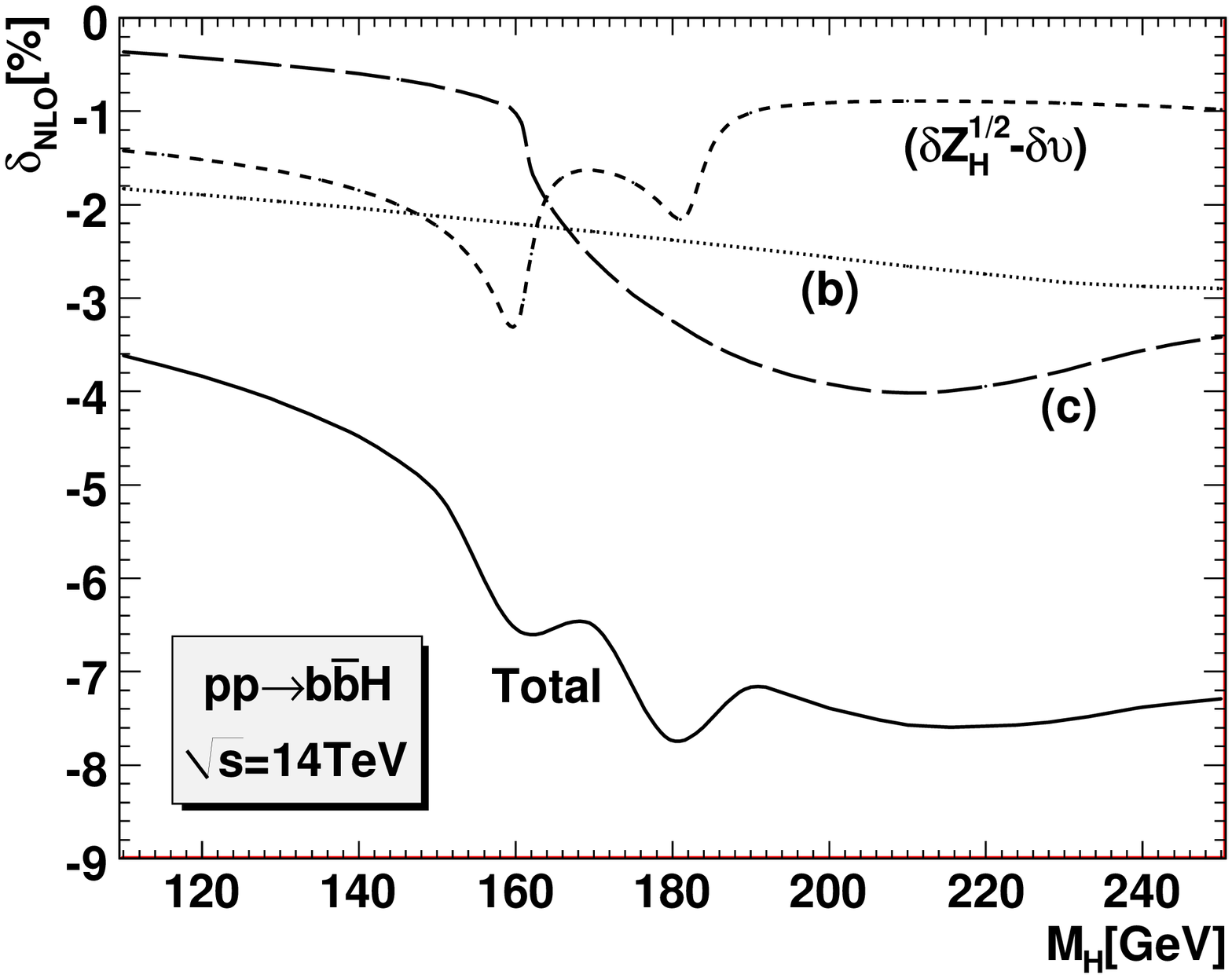}
\hspace*{0.075\textwidth}
\includegraphics[width=0.45\textwidth]{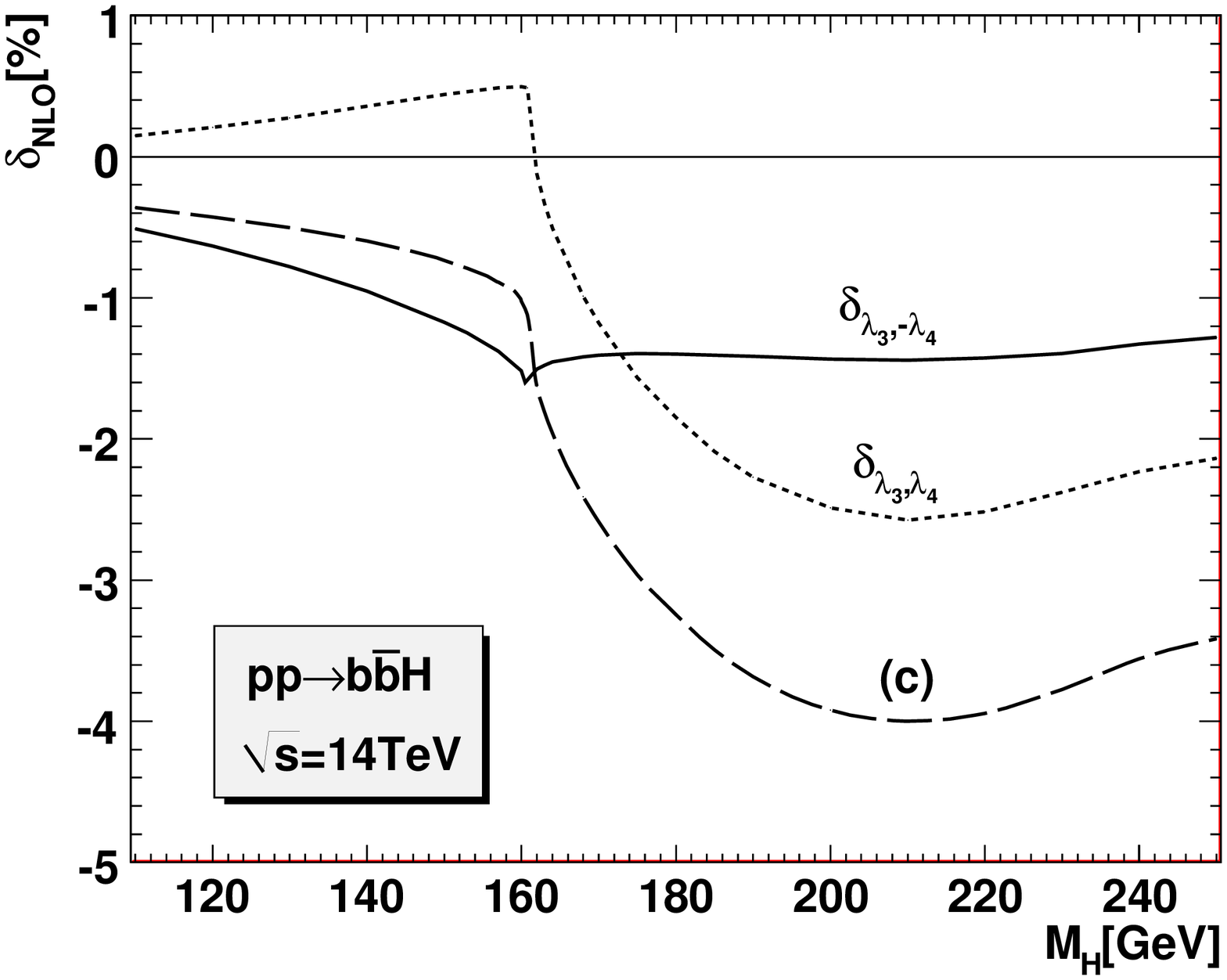}}
\caption{\label{pp_Loop_mb_mH}{\em Left: The relative NLO EW
corrections normalized to the tree-level cross section. (b) and
(c) correspond to the two classes of diagrams displayed in
Fig.~\ref{diag_3group}. $(\delta Z_{H}^{1/2}-\delta\upsilon)$ is
the universal correction contained in the renormalization of the
$b\bar{b}H$ vertex. "Total" refers to the sum of those $3$
corrections. $\delta Z_{H}^{1/2}$ is calculated by taking into
account the widths of $W$, $Z$ and the top quark. Right: the
structure of (c)-correction which is a sum of two independent
helicity configurations.}}
\end{center}
\end{figure}
The results for the NLO corrections are shown in
Fig.~\ref{pp_Loop_mb_mH} as a function of the Higgs mass. We
implement widths only in the two-point function wave function
renormalisation of the Higgs. The latter contributes an almost
constant $-1\%$ correction apart from oscillations in the range
$2M_W$ to $2M_Z$ due to the dips at $2M_W$ and $2M_Z$ where the
Higgs wave function is not analytic at those value. The effect of
the widths of the $W$ and $Z$ smooths the behaviour and the
corrections is never larger than $3.5\%$ in this range of Higgs
masses. The contribution from class (b) where the Higgs couples to
the internal top decreases very slowly as the Higgs mass increases
from $110$GeV to $250$GeV, as expected there is no structure as
would be the case if this contribution were sensitive to any
threshold or singularity. Class (c) on the other hand does, as
expected, reveal some structure around $M_H=2M_W$ where we see a
fall in the relative correction. The correction is however,
despite this fall, quite modest ranging from $\sim -1\%$ for
$M_H=160$GeV to $-4\%$ for $M_H=210$GeV. When we studied the
effect of the width of the internal particles on class (c) at NLO,
we did so  in the massless quark limit. In that limit the outgoing
quarks have opposite helicity so that only the
$\delta_{\la_3,-\la_4}$ helicity amplitude survives,
$\lambda_{3,4}$ are the quark helicities. In our case here when
the quark mass is reinstated, the $\delta_{\la_3,\la_4}$ helicity
amplitude switches on. Fig.~\ref{pp_Loop_mb_mH}(right) shows that
these two helicity amplitudes behave differently as a function of
the Higgs mass. The effect of the $b$-quark mass makes the dip in
the $\delta_{\la_3,-\la_4}$ much softer that in the massless case
displayed in Fig.~~\ref{pp_Loop_mb0_mH}. In the
$\delta_{\la_3,\la_4}$ the fall of the correction around
$M_H=2M_W$ is more apparent. This is another manifestation of how
the dynamics can affect  the structure of a singularity.

Adding the effect of all the contributions at NLO the total
correction changes from $-4\%$ for $M_H=110$GeV to $-8\%$ at $M_H
\sim 2M_Z$ stabilising to around $-7\%$ past this value up to
$M_H=250$GeV.
\section{Conclusions}
\label{section_conclusions}

At tree-level Higgs production in association with a $b$-quark
pair at the LHC is dominated by $gg \ra b \bar b H$ where the
Higgs is radiated from the $b$-quark with a strength proportional
to the bottom-Higgs Yukawa coupling. Unfortunately in the Standard
Model this coupling is extremely small and therefore this
mechanism is not a Higgs discovery channel although once the Higgs
has been found the study of the Higgs coupling to the $b$-quark
through this reaction could probe interesting phenomena having to
do with the the mechanism of symmetry breaking and the role played
by the third generation fermions. Electroweak one-loop effects are
usually small compared to the QCD corrections, however processes
involving the bottom quark, electroweak one-loop corrections
involve the top quark whose Yukawa coupling is of the order the
QCD strength. More interesting for $b \bar b H$ production is that
even in the limit where the bottom-Higgs Yukawa coupling vanishes
and therefore the Born tree-level cross section vanishes,
electroweak one-loop effects, through the top-Higgs Yukawa
coupling in particular, can still trigger this reaction. We
studied these effects in some detail in a previous
publication\cite{fawzi_bbH} but presented results for Higgs masses
below $2M_W$. We remarked that for the one-loop contribution in
the limit of vanishing bottom-Higgs Yukawa coupling, the cross
section was growing as the Higgs mass increased and that numerical
results started showing instabilities past $M_H\ge 2M_W$. The aim
of this paper was to extend the study performed in
\cite{fawzi_bbH} to the mass range where numerical instabilities
occurred. The origin of the numerical instabilities is due to the
fact that some one-loop contributions, contained in some box
diagrams, develop a leading Landau singularity. We have here
reviewed in some detail the problem of the occurrence of the
leading Landau singularity and investigated in more details the
conditions and dynamics as concerns $b \bar b H$ production. Since
this singularity is not integrable when the one-loop amplitude is
squared, we regulate the cross section by taking into account the
width of the internal top and $W$ particles. This requires that we
extend the usual box one-loop function to the case of complex
masses. We show how this can be implemented analytically in our
case. We study in some detail the cross section at the LHC as a
function of the Higgs mass and show how some distributions can be
drastically affected compared to the tree-level result. For
completeness we have also extended our study of the NLO Yukawa
electroweak corrections which represent the interference between
the one-loop amplitude and the tree-level amplitude. At this level
the Landau singularity is integrable and therefore does not
require that one endows the internal particle with a width. The
NLO correction is found to be small.\\[1cm]
\noi {\bf Acknowledgments} \\
LDN expresses his gratitude and thanks to P.~Aurenche for his support, most helpful discussions and
comments. We benefited a lot from discussions with G.~Altarelli, C.~Bernicot,
DO Hoang Son, J.~Fujimoto, J.P.~Guillet, K.~Kato, Y.~Kurihara, E.~Pilon
and F.~Yuasa. Special thanks go to F.~Yuasa for comparisons
between her numerical code and our code for the four-point
function with complex masses.
We thank A.~Denner for his many useful comments related to the manuscript.
LDN acknowledges the financial
support of {\em Rencontres du Vietnam}.

\newpage

\renewcommand{\thesection}{\Alph{section}}
\setcounter{section}{0}
\renewcommand{\theequation}{\thesection.\arabic{equation}}
\setcounter{equation}{0}
\noi {\Large {\bf Appendices}}

\section{Nature of the leading Landau singularity}
\label{appendix_landau}

 We give in this section more detail about our derivation of Eq.~(\ref{ccnature_landau_N}). One can rewrite Eq.~(\ref{eq_TN0})
in the form
\bea
T^{N}_{0}=\Gamma(N)\int_0^\infty dx_1\cdots dx_N\delta(\sum_{i=1}^{N}x_i-1)\int\fr{d^Dq}{(2\pi)^Di}\fr{1}{(q^2-\Delta)^N},\label{eq_TN0_1}
\eea
where
\bea
\Delta=\fr{1}{2}\sum_{i,j=1}^{N}x_ix_jQ_{ij}-i\eps\,
\eea
with $Q_{ij}$ given in Eq.~(\ref{def_Qij}).
Integrating over $q$ gives
\bea
T^{N}_{0}=\fr{(-1)^N\Gamma(N-D/2)}{(4\pi)^{D/2}}\int_0^1dx_1\cdots dx_N\fr{\delta(\sum_{i=1}^{N}x_i-1)}{\Delta^{N-D/2}}.\label{eq_TN0_2}
\eea
The Landau equations for the representation (\ref{eq_TN0_2}) are \cite{book_eden}
\bea
\left\{
\begin{array}{ll}
\Delta=0,\\
\fr{\partial \Delta}{\partial x_i}=0. \label{landau_cond_3rd}
\end{array}
\right.
\eea
Since $\Delta$ is a homogeneous function of $x_i$, the first
equation in (\ref{landau_cond_3rd}) is automatically satisfied
when the second is. Eq.~(\ref{landau_cond_3rd}) is equivalent to
Eq.~(\ref{landau_Meqs}), which means that the solution of
Eq.~(\ref{landau_cond_3rd}) is an eigenvector of $Q$ with zero
eigenvalue. In general, $Q$ has $N$ real eigenvalues  $\la_1$,
\ldots, $\la_N$. The characteristic equation of $Q$ is given by
\bea
f(\la)&=&\la^N+(-1)a_{N-1}\la^{N-1}+(-1)^2a_{N-2}\la^{N-2}-\ldots (-1)^{N-1}a_1\la+(-1)^{N}a_0\crn
&=&(\la-\la_1)(\la-\la_2)\ldots (\la-\la_n)=0.
\eea
For the case $N=4$ we have
\bea
a_0&=&\la_1\la_2\la_3\la_4=\det(Q_4),\crn
a_1&=&\la_1\la_2\la_3+\la_1\la_2\la_4+\la_1\la_3\la_4+\la_2\la_3\la_4,\crn
a_2&=&\la_1\la_2+\la_1\la_3+\la_1\la_4+\la_2\la_3+\la_2\la_4+\la_3\la_4,\crn
a_3&=&\la_1+\la_2+\la_3+\la_4=\Trace(Q_4),
\eea
Consider the case where $Q$ has only one very small eigenvalue
$\la_N\ll 1$, then to a very good approximation
\bea
\la_N\simeq \fr{a_0}{a_1},\,\,\, a_{1}=\la_1\la_2\ldots
\la_{N-1}\neq 0.
\eea
Let $V=\{x_1^0,x_2^0,\ldots,x_N^0\}$ be the eigenvector
corresponding to the eigenvalue $\la_N$. $V$ is normalised to
\bea
\sum_{i=1}^Nx_i^0=1.
\eea
For latter use, we define
\bea
\upsilon^2=V.V.
\eea
The expansion of $\Delta$ around $V$ reads
\bea
\Delta=\fr{1}{2}\sum_{i,j=1}^NQ_{ij}y_iy_j+\la_{N}\sum_{i=1}^Nx_i^0y_i+\fr{1}{2}\la_N\upsilon^2-i\eps,
\eea
where $y_i=x_i-x_i^0$. In order to find the leading singularity, it will be sufficient to neglect the
linear term in the rhs. The $Q$-matrix can be diagonalised by rotating the $y$-vector
\bea
y_i=\sum_{j=1}^NA_{ij}z_j,
\eea
where $A$ is an orthogonal matrix whose columns are the normalised
eigenvectors of $Q$. Thus we have
\bea
\det(A)=1,\,\,\, \sum_{j=1}^NA_{Nj}=\fr{\sum_{i=1}^Nx_i^0}{\sqrt{V.V}}=\fr{1}{\upsilon},\crn
\Delta=\fr{1}{2}\sum_{i=1}^{N-1}\la_iz_i^2+\fr{1}{2}\la_N\upsilon^2-i\eps.
\eea
Note that the term $\la_Nz_N^2$ in the rhs has been neglected as
this term would give a contribution of the order
$\mathcal{O}(\la_N^2)$ to the final result. Eq.~(\ref{eq_TN0_2}) can now be re-written in the form
\bea
T^{N}_{0}=\fr{(-1)^N\Gamma(N-D/2)}{\pi^{D/2}2^{3D/2-N}}\int_{-\infty}^{+\infty} dz_1\cdots dz_N\fr{\delta(\sum_{i,j=1}^{N}A_{ij}z_j)}
{(\sum_{i=1}^{N-1}\la_iz_i^2+\la_N\upsilon^2-i\eps)^{N-D/2}}.\label{eq_TN0_2a}
\eea
Although the original integration contour is some segment around the singular point $z_i=0$ with $i=1,\ldots,N$, the singular part
will not be changed if we extend the integration contour to infinity, provided the power $(N-D/2)$ of the denominator in
Eq.~(\ref{eq_TN0_2a}) is sufficiently large.
Integrating over $z_N$ gives
\bea
T^{N}_{0}=\fr{(-1)^N\Gamma(N-D/2)\upsilon}{\pi^{D/2}2^{3D/2-N}}\int_{-\infty}^{+\infty} dz_1\cdots dz_{N-1}\fr{1}
{(\sum_{i=1}^{N-1}\la_iz_i^2+\la_N\upsilon^2-i\eps)^{N-D/2}},
\eea
where the factor $\upsilon$ comes from the $\delta$-fuction.
Asumming that $\la_i>0$ for $i=1,\ldots,K$ and $\la_j<0$ for
$j=K+1,\ldots,N-1$ with $0\le K\le N-1$,  we change the
integration variables as follows
\bea
\left\{
\begin{array}{ll}
t_i=\sqrt{\la_i}z_i\,\,\, \text{for} \,\,\, i=1,\ldots K,\\
t_j=\sqrt{-\la_j}z_j\,\,\, \text{for} \,\,\, j=K+1,\ldots N-1.
\end{array}
\right.
\eea
This makes sure that all $t_i$ are real. We get
\bea
T^{N}_{0}&=&\fr{(-1)^N\Gamma(N-D/2)\upsilon}{\pi^{D/2}2^{3D/2-N}\sqrt{(-1)^{N-K-1}a_1}}\crn
&\times&\int_{-\infty}^{+\infty}
dt_1\cdots dt_{K}\int_{-\infty}^{+\infty} dt_{K+1}\cdots dt_{N-1}\fr{1}
{(-\sum_{i=K+1}^{N-1}t_i^2+b^2)^{N-D/2}},
\eea
where
\bea
b^2=\sum_{i=1}^{K}t_i^2+\la_N\upsilon^2-i\eps,\,\,\, \RE(b^2)>0.
\eea
Changing to spherical coordinates and using the following formulae
for the volume
\bea
\int_{-\infty}^{+\infty} dt_1\cdots dt_{K}=\int_{0}^{\infty} r^{K-1}drd\Omega_{K-1},\,\,\, \int d\Omega_{K-1}=\fr{2\pi^{K/2}}{\Gamma(K/2)},
\eea
we arrive at
\bea
T^{N}_{0}&=&\fr{(-1)^N\Gamma(N-D/2)\upsilon}{\pi^{D/2}2^{3D/2-N}\sqrt{(-1)^{N-K-1}a_1}}
\fr{2\pi^{(N-K-1)/2}}{\Gamma((N-K-1)/2)}\crn
&\times&\int_{-\infty}^{+\infty} dt_1\cdots
dt_{K}\int_0^{\infty}dr\fr{r^{N-K-2}} {(b^2-r^2)^{N-D/2}}.
\eea
Note that $(b^2-r^2)^{N-D/2}=e^{-i\pi(N-D/2)}(r^2-b^2)^{N-D/2}$
due to the fact that $\eps>0$. Using
\bea
\int_0^{\infty}ds\fr{s^{\alpha-1}}{(z+s)^\beta}=z^{(\alpha-\beta)}
\frac{\Gamma(\beta-\alpha)\Gamma(\alpha)}{\Gamma(\beta)},
\eea
gives
\bea
T^{N}_{0}&=&\fr{(-1)^Ne^{i\pi(N-K-1)/2}\upsilon}{\pi^{D/2}2^{3D/2-N}\sqrt{(-1)^{N-K-1}a_1}}
\pi^{(N-K-1)/2}\Gamma((N-D+K+1)/2)\crn
&\times&\int_{-\infty}^{+\infty} dt_1\cdots dt_{K}\fr{1}{(\sum_{i=1}^{K}t_i^2+\la_N\upsilon^2-i\eps)^{(N-D+K+1)/2}}.
\eea
Repeat the above steps to write
\bea
T^{N}_{0}&=&\fr{(-1)^Ne^{i\pi(N-K-1)/2}\upsilon}{2^{3D/2-N}\sqrt{(-1)^{N-K-1}a_1}}
\fr{\pi^{(N-D-1)/2}\Gamma((N-D+1)/2)}{(\la_N\upsilon^2-i\eps)^{(N-D+1)/2}}.
\eea
This result was derived with the condition
\bea
a_1\neq 0 \,\,\, \text{and} \,\,\, N-D+1>0.
\eea
However it can be trivially analytically continued if we work in
$D=4-2\epsilon$ so that it applies to $N\leq 3$ in $D=4$ by taking
the limit $\epsilon \ra 0$. \\
\noi Alternatively, with $D=4$ and $N=3$ the scalar function
\bea
T^{3}_{0}=\fr{-\upsilon}{8\pi^2}\int dz_1 dz_{2}\fr{1}
{(\la_1z_1^2+\la_2z_2^2+\la_3\upsilon^2-i\eps)}.
\eea
one first needs to dispose of the ultraviolet divergent first. To
that effect we  differentiate the above equation with respect to
$\eta=\la_3\upsilon^2$ with the result
\bea
\fr{dT^{3}_{0}}{d\eta}&=&\fr{\upsilon}{8\pi^2}\int_{-\infty}^{\infty}
dz_1 dz_{2}\fr{1} {(\la_1z_1^2+\la_2z_2^2+\eta-i\eps)^2}\crn
&=&\fr{e^{i\pi(2-K)/2}\upsilon}{8\pi\sqrt{(-1)^{2-K}\la_1\la_2}}\fr{1}{\eta-i\eps}.
\eea
Integrating back (with respect to $\eta$) we get
\bea
T^{3}_{0}=\fr{e^{i\pi(2-K)/2}\upsilon}{8\pi\sqrt{(-1)^{2-K}\la_1\la_2}}\ln(\la_3\upsilon^2-i\eps)+C,
\eea
where $C$ is a constant independent of $\eta$. This result
coincides with Eq.~(\ref{nature_landau_vertex}).
\section{Scalar box integrals with complex masses }
\label{appendix-box-integral}

The derivation of the analytical expression of the scalar one-loop
function for the box ($N=4$) with complex internal masses in the
most general case with no restriction on the external invariants
is not tractable. However, if at least one of the invariant masses
of the external legs is light-like one can derive an analytical
formula in closed form starting from the the standard approach of
't Hooft and Veltman \cite{hooft_velt} (see also
\cite{Denner:1991qq}). For our application there are at least $2$
lightlike external momenta in all boxes. We explain here our
derivation based on the method given in \cite{hooft_velt} for this
special case.

The scalar box integral is deduced from Eq.~(\ref{eq_TN0_2}) with
$x_4$ integrated out with the result
\bea
D_0&\equiv& (4\pi)^2T_0^4\crn
&=&\int_0^1dx\int_0^xdy\int_0^ydz\fr{1}{(ax^2+by^2+gz^2+cxy+hxz+jyz+dx+ey+kz+f)^2},\crn\label{d0_xyz}
\eea
where we have changed the integration variables as $t=\sum_{i=1}^4x_i$, $x=\sum_{i=1}^3x_i$, $y=x_1+x_2$, $z=x_1$; and
\bea
a&=&\fr{1}{2}(Q_{33}+Q_{44}-2Q_{34})=p_3^2,\hspace*{3mm} b=\fr{1}{2}(Q_{22}+Q_{33}-2Q_{23})=p_2^2,\crn
g&=&\fr{1}{2}(Q_{11}+Q_{22}-2Q_{12})=p_1^2,\hspace*{3mm} c=Q_{23}+Q_{34}-Q_{33}-Q_{24}=2p_2.p_3,\crn
h&=&Q_{13}+Q_{24}-Q_{14}-Q_{23}=2p_1.p_3,\hspace*{3mm} j=Q_{12}+Q_{23}-Q_{22}-Q_{13}=2p_1.p_2,\crn
d&=&Q_{34}-Q_{44}=m_3^2-m_4^2-p_3^2,\hspace*{3mm} e=Q_{24}-Q_{34}=m_2^2-m_3^2-p_2^2-2p_2.p_3,\crn
k&=&Q_{14}-Q_{24}=m_1^2-m_2^2+p_1^2+2p_1.p_4,\hspace*{3mm} f=\fr{Q_{44}}{2}-i\eps=m_4^2-i\eps,
\eea
with $Q_{ij}$ is defined in Eq.~(\ref{def_Qij}). Our application
will be to complex masses, $m_i^2$, with $i=1,2,3,4$, $d$, $e$,
$k$, $f$ are therefore complex parameters while other parameters
are real. The two light-like external momenta can be either
adjacent or opposite to each other. We consider in each of these
tow cases separately.

\subsection{Integral with two opposite lightlike external momenta}
\begin{figure}[htb]
\begin{center}
\includegraphics[width=0.3\textwidth]{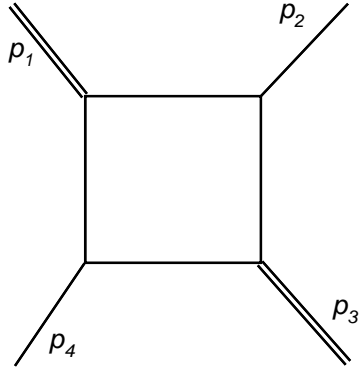}
\caption{\label{box_p1p3}{\em A box diagram with two opposite lightlike external momenta $p_1$ and $p_3$. Double line means massless.}}
\end{center}
\end{figure}
For the box shown in Fig.~\ref{box_p1p3} with $p_1^2=p_3^2=0$ one gets $a=g=0$ and writes
\bea
D_0^{(13)}=\int_0^1dx\int_0^xdy\int_0^ydz\fr{1}{(by^2+cxy+hxz+jyz+dx+ey+kz+f)^2}.\label{d0_xyz_13}
\eea
Integrating over $z$ to get
\bea
D_0^{(13)}=\int_0^1dx\int_0^xdy\fr{y}{(Ax+B)(Cx+D)}.
\eea
with
\bea
A&=&cy+d, \hs B=by^2+ey+f,\crn
C&=&(c+h)y+d, \hs D=(b+j)y^2+(e+k)y+f.\label{ABC_d0_13}
\eea
One changes the integration order as
\bea
\int_0^1dx\int_0^xdy=\int_0^1dy\int_y^1dx.
\eea
We get
\bea
D_0^{(13)}=\int_0^1dyy\int_y^1dx\fr{1}{(Ax+B)(Cx+D)},\label{d0_xy_13}
\eea
where $A$, $B$, $C$, $D$ are complex. Integrating over $x$ as follows
\bea
\int_y^1dx\fr{1}{(Ax+B)(Cx+D)}&=&\fr{1}{AC}\int_y^1\fr{dx}{(x+\fr{B}{A})(x+\fr{D}{C})}\crn
&=&\fr{1}{AD-BC}\int_y^1\left(\fr{1}{x+\fr{B}{A}}-\fr{1}{x+\fr{D}{C}}\right)dx\crn
&=&\fr{1}{AD-BC}\left(\ln\fr{1+\fr{B}{A}}{y+\fr{B}{A}}-\ln\fr{1+\fr{D}{C}}{y+\fr{D}{C}}\right)\crn
&=&\fr{1}{AD-BC}\left(\ln\fr{A+B}{Ay+B}-\ln\fr{C+D}{Cy+D}\right),\label{int_x_ABCD}
\eea
where we have made sure that the arguments of the logarithms never cross the cut along the negative real axis. One easily gets
\bea
D_0^{(13)}=\int_0^1\fr{dy}{(cj-bh)y^2+(dj+ck-eh)y+dk-fh}\left(\ln\fr{A+B}{Ay+B}-\ln\fr{C+D}{Cy+D}\right).
\eea
The discriminant of the quadratic function in the denominator of
the prefactor is nothing but the Landau determinant. Indeed,
\bea
\det Q_4=(dj+ck-eh)^2-4(cj-bh)(dk-fh).
\eea
We write
\bea
D_0^{(13)}=\fr{1}{(cj-bh)(y_{2}-y_{1})}\int_0^1\left(\fr{1}{y-y_2}-\fr{1}{y-y_1}\right)\left(\ln\fr{A+B}{Ay+B}-\ln\fr{C+D}{Cy+D}\right),\label{d0_y_13}
\eea
with
\bea
{\large
y}_{\stackrel{1}{2}}=\fr{-(dj+ck-eh)\mp\sqrt{\det Q_4}}{2(cj-bh)}.
\eea

Now we have to look at the imaginary parts of the arguments of the logarithms in (\ref{d0_y_13}). We write them explicitly
\bea
A+B&=&by^2+(c+e)y+d+f,\crn
Ay+B&=&(b+c)y^2+(e+d)y+f,\crn
C+D&=&(b+j)y^2+(e+k+c+h)y+d+f,\crn
Cy+D&=&(b+j+c+h)y^2+(e+k+d)y+f.
\eea
Imaginary parts read
\bea
\Img(A+B)&=&\Img(ey+d+f)=\Img[ym_2^2+(1-y)m_3^2-i\eps]<0,\crn
\Img(Ay+B)&=&\Img(ey+dy+f)=\Img[ym_2^2+(1-y)m_4^2-i\eps]<0,\crn
\Img(C+D)&=&\Img[(e+k)y+d+f]=\Img[ym_1^2+(1-y)m_3^2-i\eps]<0,\crn
\Img(Cy+D)&=&\Img[(e+k)y+dy+f]=\Img[ym_1^2+(1-y)m_4^2-i\eps]<0.\label{sign_img_deno}
\eea
Using formula $\ln(a/b)=\ln a-\ln b$ for $\Img(a)\Img(b)>0$, we rewrite (\ref{d0_y_13}) as
\bea
D_0^{(13)}=\fr{1}{\sqrt{\det Q_4}}\sum_{i=1}^2\sum_{j=1}^4(-1)^{i+j}\int_0^1dy\fr{1}{y-y_i}\ln(A_jy^2+B_jy+C_j),\label{d0_y_13_sumij}
\eea
with
\bea
A_1&=&b+c, \hs B_1=e+d, \hs C_1=f,\crn
A_2&=&b, \hs B_2=c+e, \hs C_2=d+f,\crn
A_3&=&b+j, \hs B_3=e+k+c+h, \hs C_3=d+f,\crn
A_4&=&b+j+c+h, \hs B_4=e+k+d, \hs C_4=f.
\eea
We would like to make an important remark here. From Eq. (\ref{sign_img_deno}) we can re-write Eq. (\ref{d0_y_13}) in the form
\bea
D_0^{(13)}=\fr{1}{(cj-bh)(y_{2}-y_{1})}\int_0^1\left(\fr{1}{y-y_2}-\fr{1}{y-y_1}\right)\left(\ln\fr{A+B}{C+D}-\ln\fr{Ay+B}{Cy+D}\right).\hs\hs \label{d0_y_13_ab}
\eea
We notice that if $y=y_{1,2}$ then $AD=BC$ which means
\bea
\fr{A+B}{C+D}\Big\vert_{y=y_{1,2}}=\fr{Ay+B}{Cy+D}\Big\vert_{y=y_{1,2}}=\fr{B}{D}\Big\vert_{y=y_{1,2}}.
\eea
Thus, we get
\bea
\int_0^1\left(\fr{1}{y-y_2}-\fr{1}{y-y_1}\right)\left(\ln\fr{A+B}{C+D}\Big\vert_{y=y_{1,2}}-\ln\fr{Ay+B}{Cy+D}\Big\vert_{y=y_{1,2}}\right)=0.
\eea
Subtracting this zero contribution from Eq. (\ref{d0_y_13_sumij}) we get another form
\bea
D_0^{(13)}&=&\fr{1}{\sqrt{\det(Q_4)}}\sum_{i=1}^2\sum_{j=1}^4(-1)^{i+j}\crn
&\times&\int_0^1dy\fr{\ln(A_jy^2+B_jy+C_j)-\ln(A_jy_i^2+B_jy_i+C_j)}{y-y_i}\label{d0_y_13_sumij_extra}
\eea
which is more convenient for the evaluation in terms of Spence functions.

Each integral in Eq. (\ref{d0_y_13_sumij_extra}) can be written in terms of $4$ Spence functions as given in Appendix B of \cite{hooft_velt}.
Thus $D_0^{(13)}$ can be written in terms of $32$ Spence functions.
\subsection{Integral with two adjacent lightlike external momenta}
\begin{figure}[htb]
\begin{center}
\includegraphics[width=0.3\textwidth]{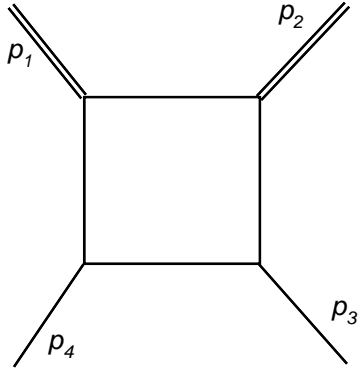}
\caption{\label{box_p1p2}{\em A box diagram with two adjacent lightlike external momenta $p_1$ and $p_2$. Double line means massless.}}
\end{center}
\end{figure}
For the box shown in Fig.~\ref{box_p1p2} with $p_1^2=p_2^2=0$ one gets $b=g=0$ and writes
\bea
D_0^{(12)}=\int_0^1dx\int_0^xdy\int_0^ydz\fr{1}{(ax^2+cxy+hxz+jyz+dx+ey+kz+f)^2}.\label{d0_xyz_12}
\eea
As in the case of $D_0^{(13)}$,
integrating over $z$ gives
\bea
D_0^{(12)}=\underbrace{\int_0^1dx\int_0^xdy\fr{1}{a_1b_1}}_{I_1}+s_k\underbrace{\int_0^1dx\int_0^xdy\fr{1}{-s_ka_1(a_1y+b_1)}}_{I_2},
\eea
with
\bea
s_k&=&\sign(\Img(k)),\hs -s_ka_1=-s_k(hx+jy+k)-i\eps^\prime,\crn
b_1&=&ax^2+cxy+dx+ey+f,\crn
a_1y+b_1&=&ax^2+jy^2+(c+h)xy+dx+(e+k)y+f-i\eps,
\eea
where we have used the fact that
$\Img(a_1y+b_1)=\Img[dx+(e+k)y+f]=\Img[(x-y)m_3^2+(1-x)m_4^2+ym_1^2-i\eps]<0$
because $0\le y\le x \le1$. $\eps$ and $\eps^\prime$ are
infinitesimal positive quantities which carry the sign of the
imaginary parts of $-s_ka_1$ and $a_1y+b_1$. For $I_1$, we
integrate over $y$, similar to (\ref{int_x_ABCD}), to get
\bea
I_1&=&\int_0^1dy\fr{1}{(ja-hc)y^2+(jd-he-kc)y+jf-ke}\crn
&\times&\left[\ln\fr{(j+h)y+k-i\eps^\prime}{hy+k-i\eps^\prime}-\ln\fr{(a+c)y^2+(d+e)y+f}{ay^2+dy+f}\right].
\eea
Consider the prefactor
\bea
\det(Q_4)&=&(jd-he-kc)^2-4(ja-hc)(jf-ke),\crn
y_{11(12)}&=&\fr{(he+kc-jd)\mp\sqrt{\det(Q_4)}}{2(ja-hc)},
\eea
where the indices $11$, $12$ correspond to $-$ and $+$ signs respectively.
We rewrite $I_1$ as
\bea
I_1&=&\fr{1}{\sqrt{\det(Q_4)}}\sum_{i=1}^2(-1)^i\crn
&&\int_0^1dy\fr{1}{y-y_{1i}}\left[\ln\fr{(j+h)y+k-i\eps^\prime}{hy+k-i\eps^\prime}-\ln\fr{(a+c)y^2+(d+e)y+f}{ay^2+dy+f}\right]\crn
&=&\fr{1}{\sqrt{\det(Q_4)}}\sum_{i=1}^2\sum_{j=1}^4(-1)^{i+j}\int_0^1dy\fr{1}{y-y_{1i}}\ln(A_{1j}y^2+B_{1j}y+C_{1j}),\label{I1_result}
\eea
with
\bea
A_{11}&=&0,\hs B_{11}=h,\hs C_{11}=k,\crn
A_{12}&=&0,\hs B_{12}=j+h,\hs C_{12}=k,\crn
A_{13}&=&a+c,\hs B_{13}=d+e,\hs C_{13}=f,\crn
A_{14}&=&a,\hs B_{14}=d,\hs C_{14}=f.
\eea
Thus $I_1$ can be written in terms of $24$ Spence functions. For $I_2$ we shift $y=y+\alpha x$, $\alpha$ such that
\bea
j\alpha^2+(c+h)\alpha+a=0.
\eea
There are, in general, two values of $\alpha$. The final result does not depend on which value of $\alpha$ we take.
We have used this freedom to find bugs in the numerical calculation and it turns out to be a very powerful
method to check the correctness of the imaginary part which can be very tricky for the case of equal masses.
One gets
\bea
I_2=\int_0^1dx\int_{-\alpha x}^{(1-\alpha)x}dy\fr{1}{(Gx+H-i\eps^\prime)(Ex+F-i\eps)},\label{int_xy_EF}
\eea
with
\bea
G&=&-s_kh-s_kj\alpha,\hs H=-s_kjy-s_kk,\crn
E&=&(2j\alpha+c+h)y+d+\alpha(e+k),\hs F=jy^2+(e+k)y+f.
\eea
For real $\alpha$ we have
\bea
\int_0^1dx\int_{-\alpha x}^{(1-\alpha)x}dy&=&\int_0^1dx\int_0^{(1-\alpha)x}dy-\int_0^1dx\int_{0}^{-\alpha x}dy\crn
&=&\int_0^{1-\alpha}dy\int_{y/(1-\alpha)}^{1}dx-\int_0^{-\alpha}dy\int_{-y/\alpha}^{1}dx.
\eea
We write
\bea
\fr{1}{(Gx+H-i\eps^\prime)(Ex+F-i\eps)}=\fr{1}{GF-HE}\left(\fr{G}{Gx+H-i\eps^\prime}-\fr{E}{Ex+F-i\eps}\right).
\eea
Integrating over $x$, we get
\bea
I_2&=&\int_{-\alpha}^{1-\alpha}\fr{dy}{GF-HE}\ln\fr{G+H}{E+F}-\int_0^{1-\alpha}\fr{dy}{GF-HE}\ln\fr{\fr{Gy}{1-\alpha}+H}{\fr{Ey}{1-\alpha}+F}\crn
&+&\int_0^{-\alpha}\fr{dy}{GF-HE}\ln\fr{\fr{Gy}{-\alpha}+H}{\fr{Ey}{-\alpha}+F}.
\eea
The prefactor
\bea
\fr{GF-HE}{s_k}&=&j(j\alpha+c)y^2+(2\alpha jk+jd-he+kc)y+\alpha(ke+k^2-jf)+kd-hf\crn
&=&j(j\alpha+c)(y-y_{21})(y-y_{22}),
\eea
with
\bea
y_{21(22)}=\fr{-(2\alpha jk+jd-he+kc)\mp\sqrt{\det(Q_4)}}{2j(j\alpha+c)},
\eea
where the indices $21$, $22$ correspond to $-$ and $+$ signs respectively.
We rewrite $I_2$ as
\bea
I_2&=&\fr{1}{s_k\sqrt{\det(Q_4)}}\sum_{i=1}^2(-1)^iI_2^{(i)},\crn
I_2^{(i)}&=&\int_{-\alpha}^{1-\alpha}\fr{dy}{y-y_{2i}}\ln\fr{G+H}{E+F}-\int_0^{1-\alpha}\fr{dy}{y-y_{2i}}\ln\fr{\fr{Gy}{1-\alpha}+H}{\fr{Ey}{1-\alpha}+F}\crn
&+&\int_0^{-\alpha}\fr{dy}{y-y_{2i}}\ln\fr{\fr{Gy}{-\alpha}+H}{\fr{Ey}{-\alpha}+F}.\label{I2_alpha}
\eea
We make the substitutions $y=y-\alpha$ for the first integral, $y=(1-\alpha)y$ for the second integral and $y=-\alpha y$ for the third integral to get
\bea
I_2^{(i)}&=&\int_{0}^{1}\fr{dy}{y-\alpha-y_{2i}}\ln\fr{-s_kjy-s_kh-s_kk-i\eps^\prime}{jy^2+(c+h+e+k)y+a+d+f-i\eps}\crn
&-&\int_0^{1}\fr{(1-\alpha)dy}{(1-\alpha)y-y_{2i}}\ln\fr{-s_k(j+h)y-s_kk-i\eps^\prime}{(a+c+j+h)y^2+(d+e+k)y+f-i\eps}\crn
&+&\int_0^{1}\fr{-\alpha dy}{-\alpha y-y_{2i}}\ln\fr{-s_khy-s_kk-i\eps^\prime}{ay^2+dy+f-i\eps}.\label{I2_01}
\eea
Consider the arguments of the three logarithms, as demonstrated in (\ref{sign_img_deno}), it is easy to see that the sign of the imaginary parts of the denominators is negative as indicated by $-i\eps$. The derivation is for real $\alpha$. However, this result is also correct if $\alpha$ is complex as proven in \cite{hooft_velt}. We can now rewrite $I_2$ as
\bea
I_2=\fr{1}{s_k\sqrt{\det(Q_4)}}\sum_{i=1}^2\sum_{j=1}^6(-1)^{i}\int_0^1dy\fr{c_j}{a_jy-b_j-y_{2i}}\ln(A_{2j}y^2+B_{2j}y+C_{2j}),\label{I2_result}
\eea
with
\bea
c_1&=&1,\hs a_1=1,\hs b_1=\alpha,\crn
c_2&=&-(1-\alpha),\hs a_2=1-\alpha,\hs b_2=0,\crn
c_3&=&-\alpha,\hs a_3=-\alpha,\hs b_3=0,\crn
c_4&=&-1,\hs a_4=1,\hs b_4=\alpha,\crn
c_5&=&1-\alpha,\hs a_5=1-\alpha,\hs b_5=0,\crn
c_6&=&\alpha,\hs a_6=-\alpha,\hs b_6=0,\crn
A_{21}&=&0,\hs B_{21}=-s_kj,\hs C_{21}=-s_kk-s_kh,\crn
A_{22}&=&0,\hs B_{22}=-s_k(j+h),\hs C_{22}=-s_kk,\crn
A_{23}&=&0,\hs B_{23}=-s_kh,\hs C_{23}=-s_kk,\crn
A_{24}&=&j,\hs B_{24}=c+h+e+k,\hs C_{24}=a+d+f,\crn
A_{25}&=&a+c+j+h,\hs B_{25}=d+e+k,\hs C_{25}=f,\crn
A_{26}&=&a,\hs B_{26}=d,\hs C_{26}=f.
\eea
$I_2$ can be written in terms of $36$ Spence functions.
Thus
\bea
D_0^{(12)}=I_1+s_kI_2
\label{box_adj}
\eea
contains $60$ Spence functions. For the evaluation of $D_0^{(12)}$ in terms of Spence functions,
it is better to do the following replacement for each logarithm in $I_{1,2}$:
\bea
\ln(A_{1j}y^2+B_{1j}y+C_{1j})&\to& \ln(A_{1j}y^2+B_{1j}y+C_{1j})-\ln(A_{1j}y_{1i}^2+B_{1j}y_{1i}+C_{1j}),\crn
\ln(A_{2j}y^2+B_{2j}y+C_{2j})&\to& \ln(A_{2j}y^2+B_{2j}y+C_{2j})-\ln(A_{2j}\hat{y}_{2i}^2+B_{2j}\hat{y}_{2i}+C_{2j}),\crn
\eea
with $\hat{y}_{2i}=(y_{2i}+b_j)/a_j$. The argument for this is similar to that explained in the previous section, see Eq. (\ref{d0_y_13_sumij_extra}).

For the boxes with one lightlike external momentum, the result is written in terms of $72$ Spence functions by
using exactly the same method.
\section{Singularities of the three point function}
\label{appendix_3pt}

\begin{figure}[htb]
\begin{center}
\includegraphics[width=0.4\textwidth]{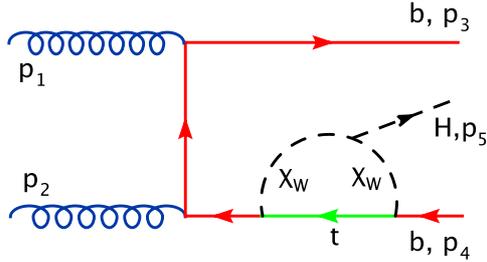}
\caption{\label{3pt-LLS}{\em A triangle diagram contributing to
$gg \ra b \bar b H$  that can develop a leading Landau singularity for
$M_H\ge 2M_W$ and $\sqrt{s_2}\ge m_t+M_W$, i.e. all the three particles
in the loop can be simultaneously on-shell.}}
\end{center}
\end{figure}

In the main text we concentrated on the properties of the 4-point
one-loop function especially as concerns the occurrence of the
leading Landau singularity which in that case is not integrable.
Although a leading singularity in the  3-point function is
integrable, it is instructive to study the case of the 3-point
function in some detail as it sheds light on some properties we
unravelled in the 4-point function. Moreover the three-point
function appears also when shrinking or collapsing one of the
internal lines into a point and therefore its singularities are
part of the singularities of the corresponding 4-point function.
The study of the 3-point scalar integral is easier to handle as it
involves less parameters. We take as an example, the 3-point loop
integral shown in Fig.~\ref{3pt-LLS} that is part of the diagrams
contributing to class (c).

\noi In terms of the Passarino-Veltman appellation, this scalar
integral writes
\bea
T_0^3(s_2)=C_0(s_2,M_H^2,0,m_t^2,M_W^2,M_W^2), \quad
s_2=(p_4+p_5)^2. \label{c0-s2}
\eea
\noi The bottom-quark mass has been neglected by setting it to $0$. The phase-space
constraint on $s_2$ is $M_H^2\le s_2\le s$, see
Eq.~(\ref{phys_region_ggbbH}).

We will define the Landau determinant, $\det Q_3$, corresponding
to a 3-point function, \newline $C_0(p_1^2,p_2^2,p_3^3,
m_1^2,m_2^2,m_3^2)$ according to the Passarino-Veltman notation
\beqn
\det Q_3 (p_1^2,p_2^2,p_3^3; m_1^2,m_2^2,m_3^2) \label{def-q3}
\eeqn
with $m_i$ the internal masses and $p_i^2$ the invariants of the
external momenta. In the same spirit the determinant of a 2-point
function will be defined as
\beqn
\det Q_2 (p^2; m_1^2,m_2^2)=-\lambda(p^2;
m_1^2,m_2^2)=-\biggl(p^2-(m_1+m_2)^2\biggr) \biggl(
p^2-(m_1-m_2)^2\biggr) \label{def-q3}
\eeqn
where $\lambda(a,b,c)$ is the usual kinematic function, see
Eq.~(\ref{cond_real}). For completeness and later reference, the
determinant of the 1-point function is defined as
\beqn
\det Q_1(m^2)= 2 m^2.
\eeqn
\noi A necessary condition for a three point function to have a
LLS is that it has exactly two cuts which can produce physical
on-shell particles. The diagram in Fig.~\ref{3pt-LLS} satisfies
this condition when
\bea
M_H\ge 2M_W\hs \textrm{and}\hs \sqrt{s_2}\ge m_t+M_W.
\label{cond_mass_3pt}
\eea
\noi These conditions are part of the conditions  for our 4-point
function (that we studied in section~\ref{section_landau}) to have
an LLS. In fact this three-point function is a reduced diagram
from the point of view of our 4-point function where it is
considered as a subleading Landau singularity. These conditions
Eq.~(\ref{cond_mass_3pt}) represent the opening up of normal
thresholds. We will refer to the first threshold $M_H \ge 2 M_W$
as the Higgs threshold ($H \ra W^+W^-$), while
the second condition will be referred to as the $s_2$ threshold ($Hb\ra W t$).\\
\noi   The sign condition ($x_i>0$), Eq.~(\ref{landau_cond1}) for
the case at hand are particularly simple here. For example,
\bea
\det \hat{Q}_{13}&=&-M_H^2(m_t^2+M_W^2)+2s_2M_W^2\le 0,\crn \det
\hat{Q}_{23} &=&-M_H^2(m_t^2+M_W^2)+s_2(M_H^2-2M_W^2)\le 0,
\eea
which together with Eq.~(\ref{cond_mass_3pt}) give
\bea
s_2\le \fr{M_W^2+m_t^2}{M_H^2-2M_W^2}M_H^2\le 2(m_t^2+M_W^2).
\label{s2_max}
\eea
These inequalities are supplemented by the condition of vanishing
Landau determinant in order for the appearance of the LLS. The
Landau determinant in our case is
\bea
\det Q_3(s_2,M_H^2)&\equiv&\det(s_2,M_H^2,0;m_t^2,M_W^2,M_W^2) \nonumber \\
&=&-2M_W^2s_2^2+2M_H^2(m_t^2 + M_W^2)s_2-2M_H^2\biggl(
M_H^2m_t^2+(m_t^2-M_W^2)^2\biggr). \nonumber \\
\eea
We have chosen to pick up $s_2$ as the variable in which to study
the location of the LLS, hence our notation $\det Q_3(s_2,M_H^2)$.
It is very rewarding to express this determinant in terms of a
perfect square in   $s_2$ plus a remainder which is the
discriminant of the quadratic equation. We can then write
\beqn
\det Q_3(s_2,M_H^2)&=&-\det Q_1(m_t^2) \biggl( (s_2-s_2^{0})^2
-\frac{\det Q_2^0}{\det Q_1(m_t^2)}  \frac{\det
Q_2^{M_H^2}}{Q_1(m_t^2)} \biggr) \quad {\rm with} \nonumber \\
\det Q_2^{M_H^2}&=&\det Q_2 (M_H^2; M_W^2,M_W^2) \quad \det
Q_2^{0}=\det Q_2 (0; m_t^2,M_W^2)  \\
s_2^0&=&2 (m_t^2+M_W^2) +(M_H^2-4
M_W^2)\biggl(1+\frac{m_t^2-M_W^2}{2M_W^2} \biggr).
\eeqn
It is important to note that the discriminant is the product of
two sub-determiants, independent of $s_2$, corresponding to two
two-point functions each one obtained by collapsing or shrinking
one of the internal lines bringing one vertex of the original
3-point function to coincide with the ``$s_2$ vertex", $s_2$ in
which we write the perfect square. This is a general
theorem\cite{Tarski} that applies to symmetric matrices based on
the Jacobi ratio theorem for
determinants\cite{determinants_wonders}.

The roots $s_{2,\pm}$  (from $\det Q_3(s_2,M_H^2) =0$) give the
position of the LLS as a function of $M_H$, for fixed $m_t,M_W$.
In view of the constraint Eq.~(\ref{s2_max}) only one solution is
possible. It is given by
\bea
s_{2}^H&=&s_{2}^{\rm{LLS}}=\fr{1}{2M_W^2}\left(M_H^2(M_W^2+m_t^2)-(m_t^2-M_W^2)M_H\sqrt{M_H^2-4M_W^2}\right)
\nonumber \\
&=&s_2^0-\frac{m_t^2-M_W^2}{2 M_W^2} M_H^2 \sqrt{1-4M_W^2/M_H^2}.
\label{s2_3pt}
\eea
The surface that defines Eq.~(\ref{s2_3pt}) is the surface of the
LLS region. This surface is bounded however due to the constraint
from the inequalities due to the normal thresholds and the sign
condition. This is what defines the region of the LLS singularity.
In fact the normal thresholds are directly related to the range of
the LLS region. First of all, if $M_H<2M_W$ there is no LLS.  At
exactly the Higgs threshold, $M_H=2M_W$ and $\det Q_2^{M_H^2}=0$,
the LLS according to Eq.~(\ref{s2_3pt}) occurs at $s_{2}^{{\rm
LLS}}=2(m_t^2+M_W^2)$ which is the {\em maximum} value of $s_2$
given by Eq.~(\ref{s2_max}). When  $M_H$ increases, the value of
$s_{2}^{{\rm LLS}}$ decreases until $s_2$ reaches the $s_2$
threshold, $(m_t+M_W)^2$, below which the LLS disappears.
Therefore the $s_2$ threshold, via the vanishing of the Landau
determinant will give the maximum value of $M_H$ for the
appearance of the LLS. We therefore find that the region of the
LLS is delimited as
\bea
&&4M_W^2\le M_H^2\le 4M_W^2+\fr{M_W}{m_t}(m_t-M_W)^2,\crn
&&(m_t+M_W)^2\le s_2\le 2(m_t^2+M_W^2).
\label{range_MH_s2_3pt}
\eea
Numerically, this corresponds to
\bea
&&160.75\text{GeV}\le M_H\le 172.89\text{GeV},\crn
&&254.38\text{GeV}\le \sqrt{s_2}\le 271.06\text{GeV}.
\eea
This range in the variables $M_H,s_2$  can be derived in a much
simpler way. The Landau constraint of vanishing determinant $\det
Q_3 (s_2,M_H^2)$ is a surface. This is bounded by tangents
parallel to the coordinate variables\cite{Tarski}, $s_2,M_H^2$ so
that with $m_t$ and $M_W$ fixed, these extrema are given by
\beqn
\frac{\partial \det Q_3(s_2,M_H^2)}{\partial s_2}=0 &\implies &
s_2^{{\rm ext}}=s_2^0 \implies  \det Q_2^{M_H^2}=0 \;\; ({\rm
since} \det Q_3(s_2,M_H^2)=0) \nonumber \\
&\implies & M_H=2M_W \implies s_2^{{\rm extr.1}}=2 (m_t^2+M_W^2).
\label{ex-eqc81}
\eeqn
The other extrema are derived in a similar way by considering
\beqn
\frac{\partial \det Q_3 (s_2,M_H^2)}{\partial M_H^2}&=&0 \implies
\det Q_2^{s_2}=0 \implies \\
 s_2^{{\rm extr.2}}&=&(m_t+M_W)^2 \implies
M_H^2=4M_W^2+\fr{M_W}{m_t}(m_t-M_W)^2.    \nonumber
\eeqn

It is crucially important to observe that these extrema do
correspond to normal thresholds where a leading singularity and a
sub-leading singularity coincide. This feature will be carried
through to the case of the 4-point function.

\begin{figure}[ht]
\begin{center}
\mbox{\includegraphics[width=0.49\textwidth]{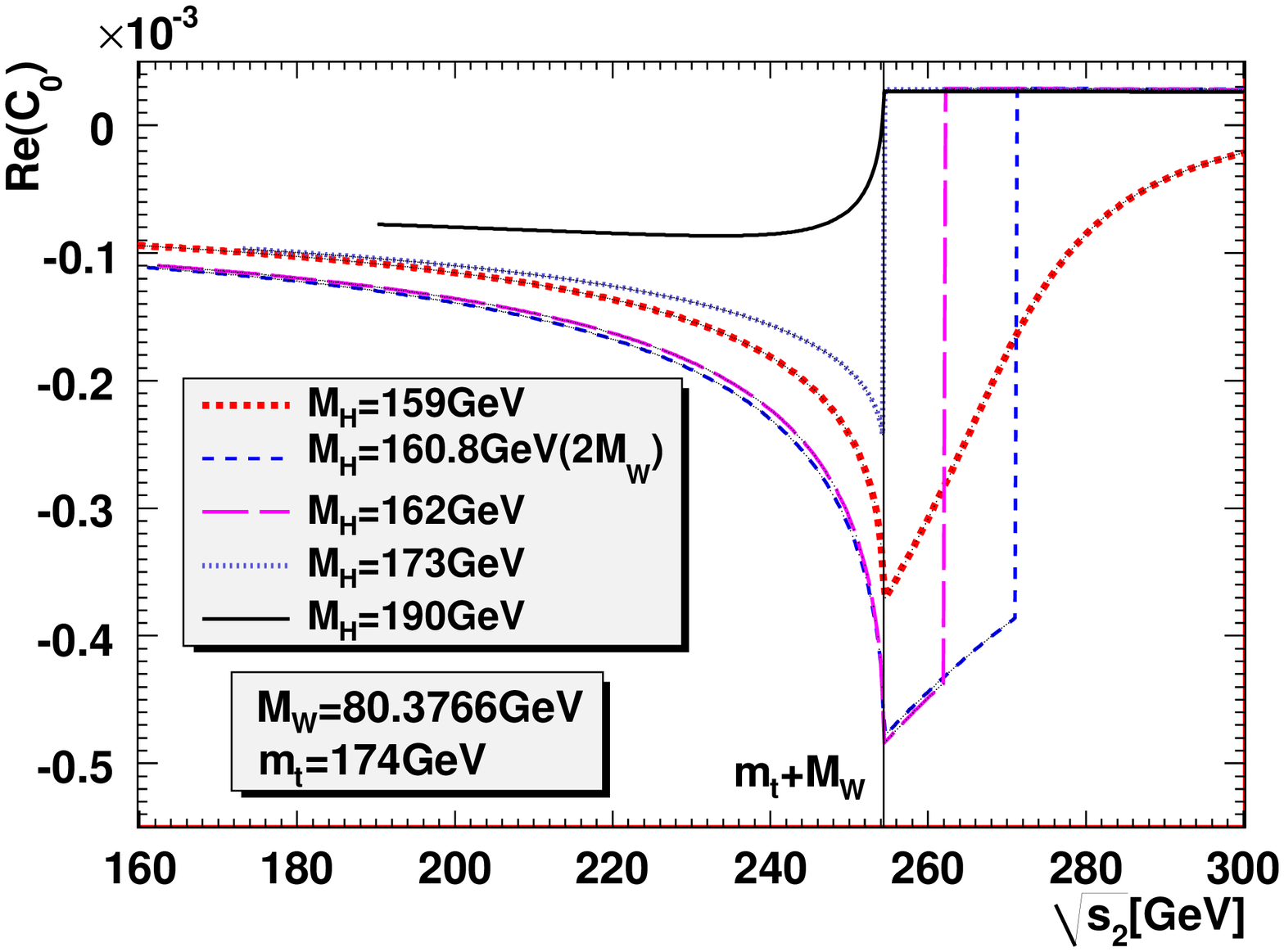}
\hspace*{0.075\textwidth}
\includegraphics[width=0.459\textwidth]{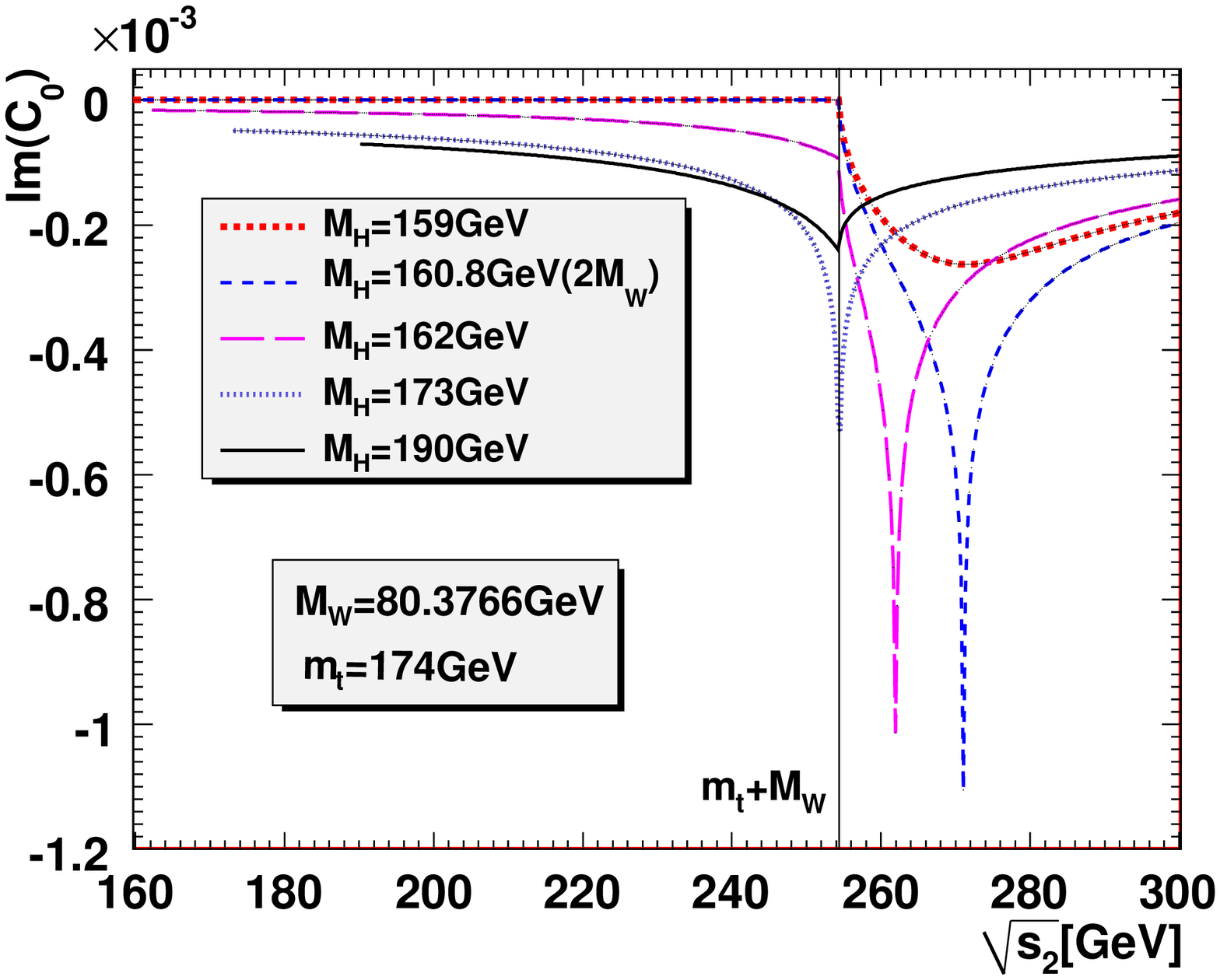}}
\caption{\label{fig_c0_width0}{\em Left: the real part of $C_0$ as
a function of $\sqrt{s_2}$ with various values of $M_H$. Right:
the same plots for the imaginary part.}}
\end{center}
\end{figure}

The location of the singularity, as well as its range, is well
rendered in Fig.~\ref{fig_c0_width0} which shows how the location
of the LLS moves as we vary the Higgs mass.
Fig.~\ref{fig_c0_width0} shows both the real and imaginary part of
scalar 3-point function. Note that as shown in
section~\ref{section_landau}, here the LLS is of a logarithm type.
This explains why one observes a jump, a step function
discontinuity, in the real part and a logarithmic singularity in
the imaginary part or vice-versa. We see that for $M_H=159{\rm
GeV} < 2M_W$, a funnel develops at the normal $s_2$ threshold for
the real part while the imaginary part develops a non-zero value
past this threshold with a rather smooth and broad structure. For
$M_H=2M_W$, at the Higgs threshold, the imaginary part develops
are very sharp dip at $s_2=2(m_t^2+M_W^2)$ which is furthest from
the normal $s_2$ threshold at $s_2=(m_t+M_W)^2$. As the Higgs mass
increases, this sharp dip moves to the left towards the normal
$s_2$ threshold beyond which the sharp peaks signalling the LLS
disappear leaving only a {\em dent} at the normal ($s_2$)
threshold.

\end{document}